\documentclass[iop]{emulateapj}
\usepackage{comment}
\usepackage{color}
\usepackage{xspace}
\usepackage{rotating}
\usepackage[caption=false]{subfig}

\newcommand{\unitlpco}    {K\,km\,s$^{-1}$\,pc$^2$\xspace}
\newcommand{\unitxco}    {$M_{\sun}$ (\unitlpco)$^{-1}$\xspace}
\newcommand{\unitsfr}    {$M_{\sun}$ yr$^{-1}$\xspace}
\newcommand{\kms}        {\,km\,s$^{-1}$\xspace}

\newcommand\um      {\ifmmode\mu{\rm m}\else$\mu${\rm m}\fi\xspace}

\newcommand\lpco    {${L^{\prime}}_{\rm\!\!CO}$\xspace}

\newcommand\msun    {$M_{\sun}$\xspace}

\newcommand\mhtwo   {$M({\rm H_2})$\xspace}

\newcommand\alphaco {$\alpha_{\rm CO}$\xspace}

\newcommand{\dfour}{$D_n(4000)$\xspace}

\slugcomment{Submitted to ApJ 5/30/2014, Accepted 12/30/2014}
\shorttitle{Molecular Gas in Post-Starbursts}
\shortauthors{}

\begin{document}

\title{Discovery of Large Molecular Gas Reservoirs in Post-Starburst Galaxies
}

\author{
K. Decker French     \altaffilmark{1},
Yujin Yang          \altaffilmark{2},
Ann Zabludoff      \altaffilmark{1},
Desika Narayanan   \altaffilmark{1,3},
Yancy Shirley      \altaffilmark{1},
Fabian Walter    \altaffilmark{4}, 
John-David Smith   \altaffilmark{5},
Christy A. Tremonti      \altaffilmark{6}
}

\altaffiltext{1}{Steward Observatory, University of Arizona, 933 North Cherry Avenue, Tucson AZ 85721}
\altaffiltext{2}{Argelander Institut f\"ur Astronomie, Universit\"at Bonn, Auf dem H\"ugel 71, 53121 Bonn, Germany}
\altaffiltext{3}{Department of Physics and Astronomy, Haverford College, 370 Lancaster Avenue
Haverford, PA 19041}
\altaffiltext{4}{Max-Planck-Institut f\"ur Astronomie, K\"onigstuhl 17, Heidelberg, Germany}
\altaffiltext{5}{Department of Physics and Astronomy, University of Toledo, Ritter Obs., MS \#113, Toledo, OH 43606}
\altaffiltext{6}{Department of Astronomy, University of Wisconsin-Madison, Madison WI, 53706}

\begin{abstract}
Post-starburst (or ``E+A'') galaxies are characterized by low H$\alpha$ emission and strong Balmer absorption, suggesting a recent starburst, but little current star formation. 
Although many of these galaxies show evidence of recent mergers, the mechanism for ending the starburst is not yet understood.
To study the fate of the molecular gas, we search for CO (1--0) and (2--1) emission with the IRAM 30m and SMT 10m telescopes in 32 nearby ($0.01<z<0.12$) post-starburst galaxies drawn from the Sloan Digital Sky Survey. We detect CO in 17 (53\%). Using CO as a tracer for molecular hydrogen, and a Galactic conversion factor, we obtain molecular gas masses of $M(H_2)=10^{8.6}$--$10^{9.8} M_\sun$ and molecular gas mass to stellar mass fractions of $\sim10^{-2}$--$10^{-0.5}$, comparable to those of star-forming galaxies.
The large amounts of molecular gas rule out complete gas consumption, expulsion, or starvation as the primary mechanism that ends the starburst in these galaxies.
The upper limits on $M(H_2)$ for the 15 undetected galaxies range from $10^{7.7} M_\sun$ to $10^{9.7} M_\sun$, with the median more consistent with early-type galaxies than with star-forming galaxies.
Upper limits on the post-starburst star formation rates (SFRs) are lower by $\sim10\times$ than for star-forming galaxies with the same $M(H_2)$.
We also compare the molecular gas surface densities ($\Sigma_{\rm H_2}$) to upper limits on the SFR surface densities ($\Sigma_{\rm SFR}$), finding a significant offset, with lower $\Sigma_{\rm SFR}$ for a given $\Sigma_{\rm H_2}$ than is typical for star-forming galaxies.
This offset from the Kennicutt-Schmidt relation suggests that post-starbursts have lower star formation efficiency, a low CO-to-H$_2$ conversion factor characteristic of ULIRGs, and/or a bottom-heavy initial mass function, although uncertainties in the rate and distribution of current star formation remain.

\end{abstract}


\section{Introduction}
\setcounter{footnote}{0}

Post-starburst (or ``E+A'') galaxies show signs of being caught in the middle of a dramatic, but brief, stage in their evolution. Emission line indicators suggest little-to-no current star formation, but strong Balmer absorption lines indicate a population of A stars that formed in a substantial burst of star formation before a sudden stop $\sim1$ Gyr ago \citep{Dressler1983,Couch1987}. 

Post-starburst galaxies are likely in transition between star-forming gas-rich disk galaxies and passively evolving gas-poor early types. 
Their disturbed morphologies indicate that many are post-merger, and most have spheroid-dominated kinematics \citep{Zabludoff1996,Norton2001,Yang2004,Yang2008}. Many have blue cores, which can fade into the color gradients observed in early type galaxies \citep{Yamauchi2005,Yang2006,Yang2008}, and many lie in the ``green valley'' of the color magnitude diagram \citep{Wong2012}.
Although only $\sim0.2$\% of local galaxies are post-starbursts, the short duration of this phase suggests that 
$\sim40$\% of galaxies could have passed through it \citep{Zabludoff1996,Snyder2011}. 

A critical part of galaxy evolution is the end, or possible ``quenching,'' of star formation.
As transitional objects, post-starburst galaxies serve as a unique laboratory for understanding the processes that drive this cessation. Explanations for the end of the starburst fall into two general categories: elimination of the molecular gas or suppression of star formation.

One possibility is that the starburst uses up the dense molecular clouds in forming stars \citep{Kennicutt1998, Gao2004}. Molecular gas could also be removed from the galaxy in outflows \citep{Narayanan2008}. Evidence of LINER activity and large outflows are seen in post-starbursts \citep{Yan2006,Yang2006,Tremonti2007}, and AGN are observed to eject molecular gas in outflows \citep{Feruglio2010}, although the driver of the outflows in post-starbursts may be due to star formation, not AGN activity \citep{Sell2014}.  Some environmental effects, such as starvation \citep[e.g.,][]{Larson1980,Boselli2006}, are thought to eliminate molecular gas reservoirs in galaxies.
The molecular gas mass is several orders of magnitude lower in early types than in late types \citep[e.g.,][]{Young2011,Crocker2011}. If post-starbursts are becoming early types, they must lose or repurpose most of their gas.

Feedback mechanisms could be responsible for suppressing star formation, resulting in the end of the starburst.
Molecular gas heating and suppressed star formation efficiency have been claimed in galaxies with AGN \citep{Nesvadba2010}, resulting in higher observed molecular gas surface densities than the Kennicutt-Schmidt relation \citep{Kennicutt1998} would predict for their star formation rate (SFR) densities. Observations of cold gas in early type galaxies with AGN and recent bursts of star formation reveal little molecular gas ($<10^9 M_\sun$), which declines steeply with the age since the last period of star formation \citep{Schawinski2009}.  In our sample, the timescales necessary for outflows to expel the molecular gas from the galaxy are less than the time elapsed since the starburst ended (about 0.3-1 Gyr), so if AGN feedback has significantly reduced the molecular gas reservoirs, we should observe the galaxies in their depleted state. 

A lower star formation efficiency is suggested in gas-rich, fast-rotating early type galaxies by \citet{Davis2014}, who observe lower SFR surface densities than the molecular gas surface densities would predict by a factor of $\sim2.5$. These authors favor dynamical methods of lowering star formation efficiency in this sample of galaxies, such as morphological quenching \citep{Martig2009}, where the gravitational stability of the gas prevents it from collapsing and forming stars. Although post-starburst galaxies are likely to evolve into early types, it is not clear that the gas-rich sample studied by \citet{Davis2014} are on the same evolutionary sequence as post-starbursts. 

We aim to test these explanations for the starbursts' end by constraining the properties of molecular gas within post-starbursts. 
Reservoirs of HI have been observed in post-starburst galaxies \citep{Chang2001,Buyle2006,Zwaan2013}. In six of the eleven post-starbursts targeted in these samples, HI 21 cm emission is detected, with atomic gas to stellar mass fractions typically between those of early and late type galaxies. However, HI is not a good tracer of star formation fuel \citep{KennicuttJr.2007}, and we must look at molecular gas signatures to understand the starbursts' end.

Detailed CO maps have been measured for only a handful of local post-starburst galaxies \citep{Kohno2002,Alatalo2013}. Even then, the two galaxies studied, NGC5195 and NGC1266, are not universally agreed-upon as post-starbursts due to their H$\alpha$ emission. The molecular gas in these galaxies is centrally concentrated, reaching starburst-like gas surface densities. Their kinematics led these authors to suggest morphological quenching \citep{Martig2009}, where the gravitational stability of the gas prevents it from collapsing and forming stars. There is a need for a survey of the molecular gas content in a representative sample of post-starburst galaxies.

We set out here to determine how much molecular gas remains in a sample of 32 post-starburst galaxies drawn from the Sloan Digital Sky Survey \citep[SDSS,][]{York2000}, and to determine whether the molecular gas densities are consistent with the small or negligible levels of current star formation. 
We observe the CO (1--0) and CO (2--1) lines with the IRAM 30m telescope, and observe a subset of 13 galaxies in CO (2--1) with the SMT 10m telescope. 
By assuming that the CO traces H$_2$, we test whether the cessation of star formation was due to a lack of molecular gas, or to the gas being consumed by the burst, expelled in outflows, or prevented from entering the galaxy (starvation of HI \citep{Larson1980}). By comparing to the molecular gas vs SFR surface density relation for other galaxies, we will be able to determine if the star formation efficiency in post-starbursts is reduced by either gas heating, morphological quenching, or some other mechanism.

We discuss our sample and observations in \S2. 
Measurements of molecular gas masses and comparisons to the SFRs are presented in \S3.
We test these results and consider their implications for galaxy evolution in \S4, presenting our conclusions in \S5.
When needed, we assume a cosmology of $\Omega_m=0.3$, $\Omega_\Lambda=0.7$, and $h=0.7$.

\section{Observations and Data Analysis}
\label{sec:observation}

\subsection{Sample Selection}
\label{sampleselection}

Our parent sample is drawn from the SDSS main galaxy spectroscopic sample \citep{Strauss2002}, which is selected to have a limiting magnitude of $r<$ 17.77 mag.  The initial sample was selected from the SDSS DR7 \citep{Abazajian2009}, using the line fluxes and indices from the MPA-JHU catalogs \citep{Aihara2011}. We exclude galaxies with $z <0.01$ to eliminate those that are very large on the sky relative to the 3\arcsec \ diameter of the SDSS fibers.  We also exclude galaxies with unreliable \footnote{We require \texttt{h\_alpha\_eqw\_err > -1}} H$\alpha$
equivalent widths (EW), or median signal-to-noise (S/N)
values of less than 10 per pixel. These cuts ensure that the line index
measurements are reliable.  Our final parent sample from DR10 is composed of 595,268 galaxies.

We select post-starburst galaxies from our parent sample by identifying galaxies with
strong stellar Balmer absorption lines signifying a recent ($\lesssim$
Gyr) starburst but little nebular emission indicative of on-going star
formation.  We use the Lick H$\delta$ index to characterize the stellar Balmer absorption. We require H$\delta_{\rm A}$ $-$
$\sigma$(H$\delta_{\rm A}$) $>$ 4\,\AA, where $\sigma$(H$\delta_{\rm A}$)
is the measurement error of the H$\delta_{\rm A}$ index. We ensure that
the galaxies have little on-going star formation by requiring H$\alpha$
EW $<$ 3\,\AA\ in the rest frame. These selection criteria result in a sub-sample of 1207 galaxies from the parent sample (0.20\%). 

We have chosen two sub-samples for {\it HST}, {\it Spitzer}, and {\it Herschel} imaging, which we follow-up here. 15 galaxies designated ``S'' throughout were selected to represent a variety of ages since the end of the burst and based on their projected 8$\mu$m flux from SDSS spectra and serendipitous {\it Spitzer} observations. Galaxies with nearby companions and large [OIII] equivalent widths indicative of AGN activity were excluded. The post-burst ages are determined by fitting stellar population synthesis (SPS) models to the galaxy spectrum, assuming a combination of a young and old single burst stellar populations (French et al. in prep). 17 galaxies designated ``H'' throughout were selected from their bright WISE 12$\mu m$ fluxes and again for a range of post-burst ages (although without the [OIII] equivalent width cut). More details on the ``H'' and ``S'' selection processes are available in Smercina et al. (in prep). The effect of these selection criteria on properties of the resultant sample is studied in \S\ref{samplebias}. Basic parameters of this sample are listed in Table 1.

\subsection{IRAM 30m CO Observations}

Observations were carried out with the IRAM 30m telescope over
two observing campaigns in January 2012 (project ID: {\tt 218-11})
and in August -- September 2012 (ID: {\tt 074-12}).  We use the Eight
Mixer Receiver (EMIR) to observe both CO(1--0) and CO(2--1) lines (rest
frequency: 115.271 and 230.538 GHz).  For each target, we tuned the 3mm
band (E090) and 1.3mm band (E230) receivers to the redshifted CO(1--0)
and CO(2--1) frequencies, $\nu_{\rm obs}$ = 103.5 -- 113.5\,GHz and 207.1
-- 227.1\,GHz, respectively. EMIR provides a bandwidth of 4\,GHz
in dual polarization corresponding to $\sim$11000 and  5500\,\kms for
CO(1--0) and CO(2--1) lines, respectively.  The Wideband Line Multiple
Autocorrelator (WILMA) was used as the backend, with a resolution of
2\,MHz corresponding to $\sim$5\,\kms in the 3mm band.  Data were taken
with a wobbler-switching mode with a frequency of 0.5\,Hz or 1\,Hz
with a throw distance of 120\arcsec\ in azimuth. The weather varied
significantly: the precipitable water vapor (PWV) ranged from  1mm to
10mm with medians of 3mm (winter) and 6.8mm (summer).  Calibration was
performed every 15 min with standard hot/cold load absorbers. The pointing
was checked every 2 hours and was found to be stable within 3\arcsec.
The FWHMs of beam are $\approx$22\arcsec\ and 11\arcsec\ for CO(1--0)
and CO(2--1) lines, respectively.

We reduced the data with {\tt CLASS} within the {\tt GILDAS} software
package{\kern-0.05em}\footnote{\tt http://www.iram.fr/IRAMFR/GILDAS}
and IDL routines.  
We use the velocity intervals [-1200, -400] and [400, 1200] \kms to fit first order polynomials for baseline subtraction. 
The spectra are coadded weighted by the rms noise of each scan.
The on-source time ($T_{\rm ON}$) ranges from 12 to 100 min depending on the strength of the line toward the targets.  If the
source was not detected within 3 hours at the telescope ($T_{\rm ON}$
$\approx$ 1hr), we moved on to the next target.  The resulting rms
noise per 5\kms bin are 1.1 -- 4.4\,mK and 1.9 -- 9.7\,mK for the CO(1--0)
and CO(2--1) observations, respectively (${T_{\rm A}}^{\!\!*}$ scale).  The conversion factors from K
(${T_{\rm A}}^{\!\!*}$ scale) to Jy at our observed frequencies are
$\sim$7.7\,Jy\,K$^{-1}$ and $\sim$6.0\,Jy\,K$^{-1}$ for the 1.3\,mm and
3\,mm bands, respectively.  We summarize the IRAM 30m CO observations in the
Tables \ref{table:iram10} and \ref{table:all21}.

\subsection{SMT CO Observations}
Observations at the SMT 10-m telescope were performed over four runs in May 2011, February 2012, December 2012, and February 2013. We used the 1mm ALMA Band 6 dual polarization sideband separating SIS receiver and 1MHz filterbank to measure the CO(2--1) 230.5 GHz (redshifted to 207.1 -- 227.1 GHz for our sample) line for 13 post-starburst galaxies. The beam size of the SMT for this line is $\approx$33\arcsec. Beam switching was done with the secondary at 2.5 Hz and a throw of 120 \arcsec. Calibration using a hot load and the standard chopper wheel method was performed every 6 minutes. Calibration using a cold load was performed at every tuning.

To reduce the data, we again use {\tt CLASS}. The main beam efficiency $\eta_{mb}$ is calculated using Jupiter in each polarization. We subtract a first-order polynomial baseline from the spectrum using data between [$-$500, 500]\kms, excluding the central region of [$-$300, 300]\kms. The spectra are scaled using $\eta_{mb}$, and coadded, weighting each spectra by the rms noise. We rebin the spectra by a factor of 10, to achieve $\approx$14\,\kms velocity bins. Typical rms per 14 \kms channel is 1 mK. These observations are summarized in Table \ref{table:all21}.

\subsection{Galaxy Properties from the SDSS}
\label{sec:sdss}
We use a variety of data products from the SDSS to study properties of the post-starburst sample, including emission line fluxes, stellar masses, SFRs, and BPT classifications from the MPA-JHU group catalogs \citep[described in][]{Aihara2011}. We use Petrosian \citep{Petrosian1976} optical sizes measured in the $r$ band from the SDSS photometric catalogs, and redshifts from DR7.

We use the stellar masses calculated from the SDSS spectra \citep[method described in][]{Kauffmann2003,Salim2007} and included in the MPA-JHU data products. Because the star formation histories of post-starbursts may not be well represented by the templates assumed in the spectral fitting,  we estimate the systematic error by comparing stellar masses from several different algorithms run on SDSS data. We compare the stellar masses from the MPA-JHU data products to those calculated by \citet{Chen2012}, who use both the \citet{Bruzual2003} and \citet{Maraston2011} SPS models. All three stellar mass calculations use a Kroupa initial mass function (IMF). The systematic error from this method slightly exceeds the formal errors on the MPA-JHU measurements and is typically $\sim30$\%.

\subsection{Star Formation Rate Upper Limits}
\label{sec:sdsssfr}

We use two different methods to calculate SFRs for the post-starburst sample, one employing the H$\alpha$ luminosity and the other the \dfour break. Both are contaminated by other effects (principally LINER and A-stellar emission, see below), and serve as upper limits on the actual current SFR.

Using the emission line fluxes from the MPA-JHU dataset \citep{Aihara2011}, we calculate SFR limits from H$\alpha$ luminosities using the relation from \citet{Kennicutt1994}. We use the Balmer decrement of H$\alpha$/H$\beta$ to calculate dust extinction, assuming the standard case B recombination at T$=10^4$ K and an intrinsic value of 2.86. We use the reddening curve of \citet{O'Donnell1994}. For the cases where the H$\beta$ line flux is uncertain, we use the mean value of $E(B-V)$ of the other post-starburst galaxies. The mean attenuation is then $A_V=0.92$ mag, or $A_{H\alpha}=0.77$ mag.

\begin{figure}
\epsscale{1.2}
\includegraphics[width=0.5\textwidth]{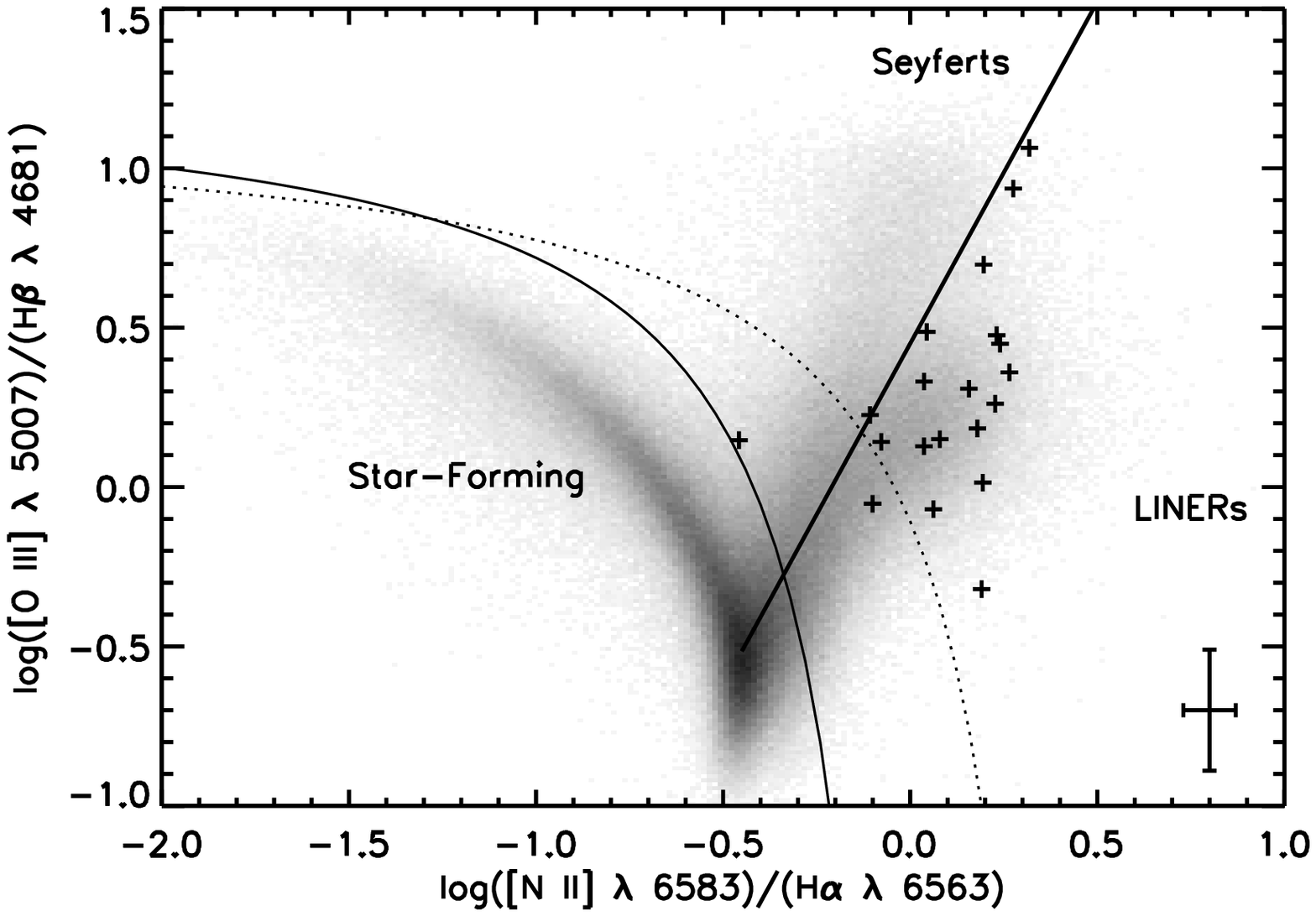}
\caption{BPT \citep{Baldwin1981} diagram for post-starburst sample, measured from SDSS spectra. Galaxies from the SDSS DR7 \citep{Abazajian2009} with well-measured lines are shown as a shaded background. The lines separating star-forming and AGN-like activity from \citet{Kewley2001} and \citet{Kauffmann2003b} are shown as dotted and solid lines, respectively. The line at $\Phi=25$ degrees separates Seyferts from LINERs. The post-starburst sample is plotted as individual points, for galaxies with all lines detected at $>3\sigma$, with a characteristic errorbar shown in the bottom right. Most of the post-starburst sample, except S12, is solidly in the LINER category. The presence of LINERs complicates our calculation of the current SFR, as the nebular emission lines will be contaminated.}
\label{fig:bpt}
\end{figure}

A complicating factor in determining the SFRs from H$\alpha$ for the post-starburst sample is the high incidence of LINER spectra. A BPT diagram for the post-starburst sample is shown in Figure \ref{fig:bpt}. Two galaxies lie in the transition region, and the rest are categorized as LINERs. Although the source of the LINER may not be an AGN (LINER emission is commonly seen in late stage mergers; \citealt{Rich2011} and from post-AGB stars; \citealt{Singh2013}), processes in addition to star formation will contribute to nebular line fluxes here, making the derived SFRs upper limits.

The MPA-JHU group use the \dfour break as a less precise, but less contaminated way to estimate SFRs when galaxies do not lie in the star-forming sequence on the BPT diagram \citep{Brinchmann2004}.
The \dfour break is a measure of the specific SFR (sSFR), and is calculated from regions of the rest-frame spectra bracketing the strong ``break'' observed near 4000\AA.  \dfour is not expected to be influenced by the presence of a Type II AGN \citep{Kauffmann2003b}. The conversion between \dfour and sSFR is calibrated from those galaxies in the SDSS categorized as star-forming. The scatter in this relation is large, and the error bars we show on the \dfour SFRs (derived using the MPA-JHU stellar masses) reflect the low precision of this calibration. 

The problem with using the \dfour-based SFRs in post-starburst galaxies is its sensitivity to the bright A-stellar populations produced in the recent burst. During ongoing star formation, the 4000\AA \ break is minimal, so \dfour is low. In passive galaxies, \dfour is large. However, the timescale over which \dfour is affected by a strong burst \citep[$\sim$1 Gyr, see e.g.,][]{Kauffmann2003} is larger than the post-burst ages of our sample, so \dfour here will reflect both previous and current rates of star formation. \dfour will be lower (more like star-forming galaxies) in post-starbursts than expected given their instantaneous SFRs. Lower values of \dfour correspond to higher SFRs, so \dfour will overestimate the current SFR due to the recent burst.
We use \dfour-based SFRs as upper limits on the current SFR.

To use the SDSS fiber spectra to calculate global SFRs, we must account for any star formation outside the 3\arcsec \ fiber aperture. Like \citet{Brinchmann2004}, we see a trend of increasing fiber-based SFR per total stellar mass with redshift, after breaking up our complete SDSS post-starburst sample  (1207 galaxies) into stellar mass bins. Thus, we expect some contribution to the SFR from outside of the fiber, so we require an aperture correction \footnote{We consider the case where star formation is limited to the fiber aperture in \S\ref{size}, see Figure \ref{fig:radius_comp}b.}. We apply the aperture correction used in the MPA-JHU SFRs \citep{Brinchmann2004,Salim2007,Aihara2011}, which is based on galaxy photometry outside the fiber. Although this aperture correction is calibrated on star-forming galaxies, it successfully removes the trend of sSFR with redshift for our complete post-starburst sample. 
While our use of this correction assumes that it also applies to our post-starbursts, the corrected SFRs remain likely upper limits as post-starbursts tend to have more positive color gradients (relatively bluer cores) than star-forming galaxies \citep{Yang2006,Yang2008}.

One case where H$\alpha$ and \dfour would not provide upper limits on the SFRs is if we have significantly underestimated the dust extinction in post-starbursts. Radio continuum emission at 1.4 GHz is often used as an ``extinction-free'' SFR indicator \citep{Condon1992}.
We search the FIRST \citep[Faint Images of the Radio Sky at Twenty centimeters,][]{Becker1995} and NVSS \citep[NRAO VLA Sky Survey,][]{Condon1998} 1.4GHz surveys for matches within 10\arcsec \ of each galaxy in our sample. We find 6 detections in the FIRST survey (S06, H01, H03, H07, H08, and H09). The galaxies H07, H08, and H09 are also detected in the NVSS. Using the conversion found in \citet{Condon1992}, the SFRs suggested by these detections are higher than the H$\alpha$ SFRs. If we were to accept that the standard 1.4GHz - SFR relation is valid for the post-starburst sample, it would require these galaxies to have up to 4.7 magnitudes of additional extinction on top of the $\sim1$ magnitude already accounted for using the Balmer decrement. While dust extinctions of 5-6 magnitudes are not unheard of, especially for starbursts, the dust extinction is consistent with that implied by Balmer decrement in those cases \citep{Choi2006, Kennicutt2009}. The huge difference between the extinction derived from the Balmer decrement and implied by the 1.4Ghz-SFR relation is unprecedented and suggests a problem with the SFRs derived from the 1.4GHz data for our post-starbursts.

The LINER and recent starburst in these galaxies complicates the standard 1.4GHz SFR conversion. Galaxies with LINER spectra have enhanced 1.4GHz fluxes when compared to other measures of their SFRs \citep{deVries2007, Moric2010}. \citet{Moric2010} find that 90\% of the 1.4GHz flux can come from the LINER, not from star formation, and that the scatter in the 1.4GHz-SFR relation for LINERs is large, of order 2 dex.
For our sample, the 1.4GHz-based SFRs and limits scatter evenly about the H$\alpha$-based SFRs after the radio SFRs are reduced by the expected factor of 10. While the LINER will also contribute to the H$\alpha$ flux, its contribution is typically $\lesssim40$\%, with less scatter \citep{Brinchmann2004}, implying that H$\alpha$ is more reliable than 1.4GHz as a SFR upper limit. Additionally, the recent large starburst may boost the amount of 1.4GHz flux on timescales overlapping with the post-burst ages of our sample \citep{Bressan2002}. As discussed above in \S\ref{sampleselection}, the galaxies marked ``S'' were selected with a cut on the [OIII] equivalent widths, intended to exclude galaxies with strong AGN activity from the sample. This cut was not applied to the selection of galaxies marked ``H,''  and the higher incidence of 1.4GHz detections in the ``H'' sample may be tied to a higher incidence of AGN.

Both the TIR luminosity \citep[total IR, from 8-1000 $\mu$m;][]{Hayward2014} and 24 $\mu$m flux \citep{Utomo2014} are strongly affected by dust heating by the substantial A stellar population in post-starbursts, so we do not consider these SFR indicators here. Ongoing analysis of our sample observed in PAH emission and high ionization species (Smercina et al., in prep) will provide further constraints on any current SFR.

In the following analysis, we use the H$\alpha$-derived SFR as an upper limit, as well as showing the effect of assuming the \dfour-based SFRs.

\section{Results}

\begin{figure*}
\includegraphics[width=\textwidth]{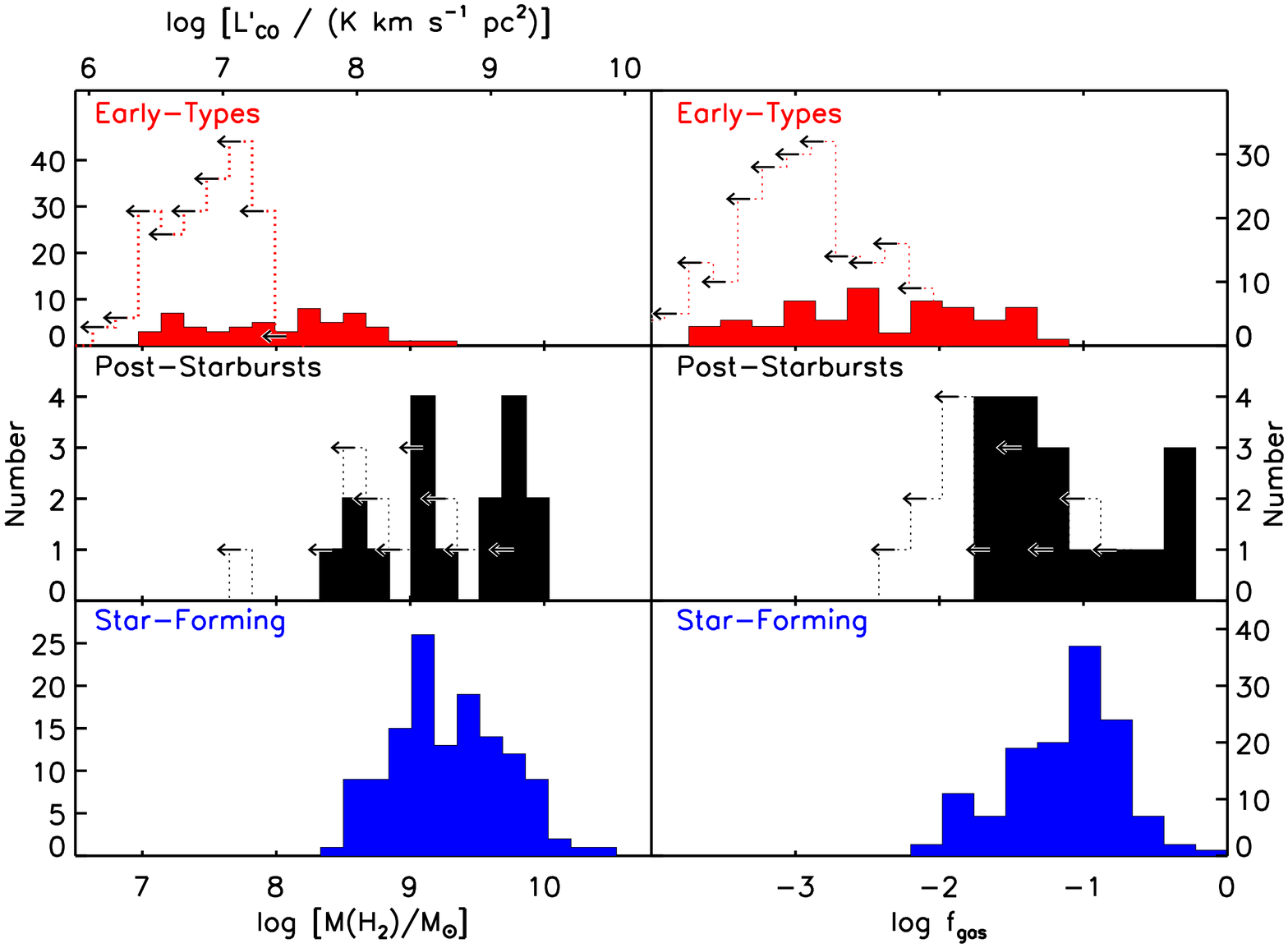}
\caption{Left: Histograms of derived molecular gas masses \mhtwo for a variety of galaxy types: early types \citep[top, from Atlas-3D][]{Young2011}, star-forming \citep[bottom, from COLD GASS;][]{Saintonge2011}, and our post-starburst sample (middle). \alphaco$=4$\unitxco is assumed for all samples.
Bin size represents the mean error in the post-starburst sample, excluding systematic error from uncertainties in \alphaco. A histogram of $3\sigma$ upper limits is overplotted for non-detections.
Right: Histograms of molecular gas normalized to stellar mass ($f_{\rm gas}$) for the same samples. For both \mhtwo and $f_{\rm gas}$, we see considerable overlap between the post-starburst sample and star-forming samples, which is surprising given the difference in SFRs. The lower end and upper limits of the post-starburst sample are consistent with \mhtwo and $f_{\rm gas}$ measured for the early type sample. As seen in Figure \ref{fig:molechist_alpha}, overlap persists even if a ULIRG-type value of \alphaco is assumed for the post-starburst sample.}
\label{fig:molechist}
\end{figure*}

\subsection{Detection of Molecular Gas}
\label{detect}

We detect molecular gas at $>3\sigma$ in 17 of the 32 galaxies observed, using IRAM 30m measurements of the CO (1--0) line. If we increase our detection threshold to $4\sigma$, we detect 14 galaxies, and at $>5\sigma$, we detect 11 galaxies.
To calculate the integrated CO line intensity $I_{CO}$, we fit a Gaussian profile to each line, allowing the center velocity to differ from the optical velocity up to 200 \kms. We use the Gaussian width $\sigma_{\rm gauss}$ to choose integration limits of $\pm 3 \sigma_{\rm gauss}$. Although many of the line shapes are not exactly Gaussian, this method allows us to estimate appropriate velocity intervals for integration in a systematic way. FWHMs given by these fits are listed in Tables \ref{table:iram10} and \ref{table:all21}. If the signal to noise ratio for $\sigma_{\rm gauss}$ is $<3$, we use the interval [-260, 260] \kms\. These velocity limits were chosen to be the median of those of the well-fit sample, and are centered around the optical velocity. The velocity intervals fit from the CO (1--0) data are used for the CO (2--1) data, though we note that fitting the CO (2--1) data separately does not change our results by $>1\sigma$. We calculate the error in the integrated CO line intensity as
\begin{equation}
\sigma_I^2 = (\Delta v)^2 \: \sigma^2 \: N_l \: (1+\frac{N_l}{N_b}),
\end{equation}
where $\Delta v$ is the channel velocity width, $\sigma$ is the channel rms noise, $N_l$ is the number of channels used to integrate over the line, and $N_b$ is the number of channels used to fit the baseline. We also take into account an estimated flux calibration error of 10\%. We calculate upper limits on $I_{CO}$ as $<3 \sigma_I$. 
Following \citet{Solomon1997}, the CO line luminosity $L^\prime_{\rm CO}$
(in \unitlpco) is
\begin{equation}
L^\prime_{\rm CO} =   23.5 \: \Omega_{s*b} \:  D_L^2 \: I_{CO}\:  (1+z)^{-3} \:\: ,
\label{eq:lpco}
\end{equation}
where $I_{CO}=\int T_{mb} \:  dV$ is the integrated line intensity (in K\,km\,s$^{-1}$) as described above, $z$ is the SDSS redshift, and $D_L$ is the luminosity distance (in Mpc). $\Omega_{s*b}$ is the solid angle of the source convolved with the beam, 
\begin{equation}
\Omega_{s*b} = \frac{\pi (\theta_s^2+\theta_b^2)}{4 \ln{2}},
\end{equation}
where $\theta_s$ and $\theta_b$ are the half power beam widths of the source and beam, respectively.
Because the CO emitting size estimates (see \S\ref{size}) are not available for all the sources, we adopt a simple approximation such that the beam is much larger than the source, so $\Omega_{s*b} \approx \Omega_b$. Note that depending on the actual size estimates in \S\ref{size}, we could be underestimating \lpco by $\sim1.1-2.2\times$, with a median of 1.4, but this does not affect our conclusions throughout the paper.

The molecular gas mass can be calculated from $L^\prime_{\rm CO}$ by assuming a conversion factor $\alpha_{\rm CO}$, as
\begin{equation}
M(\rm H_2) = \alpha_{\rm CO} L^\prime_{\rm CO}.
\end{equation}
For now, we assume an \alphaco comparable to that in Galactic molecular clouds and the Local Group \citep[aside from the SMC; see recent reviews by][]{Bolatto2013,Carilli2013,Casey2014}: $\alpha_{\rm CO}=4$ \msun (\unitlpco)$^{-1}$ (units omitted hereafter). This choice is examined below and in \S\ref{alphaks}.

\begin{figure*}
\includegraphics[width=1.0\textwidth]{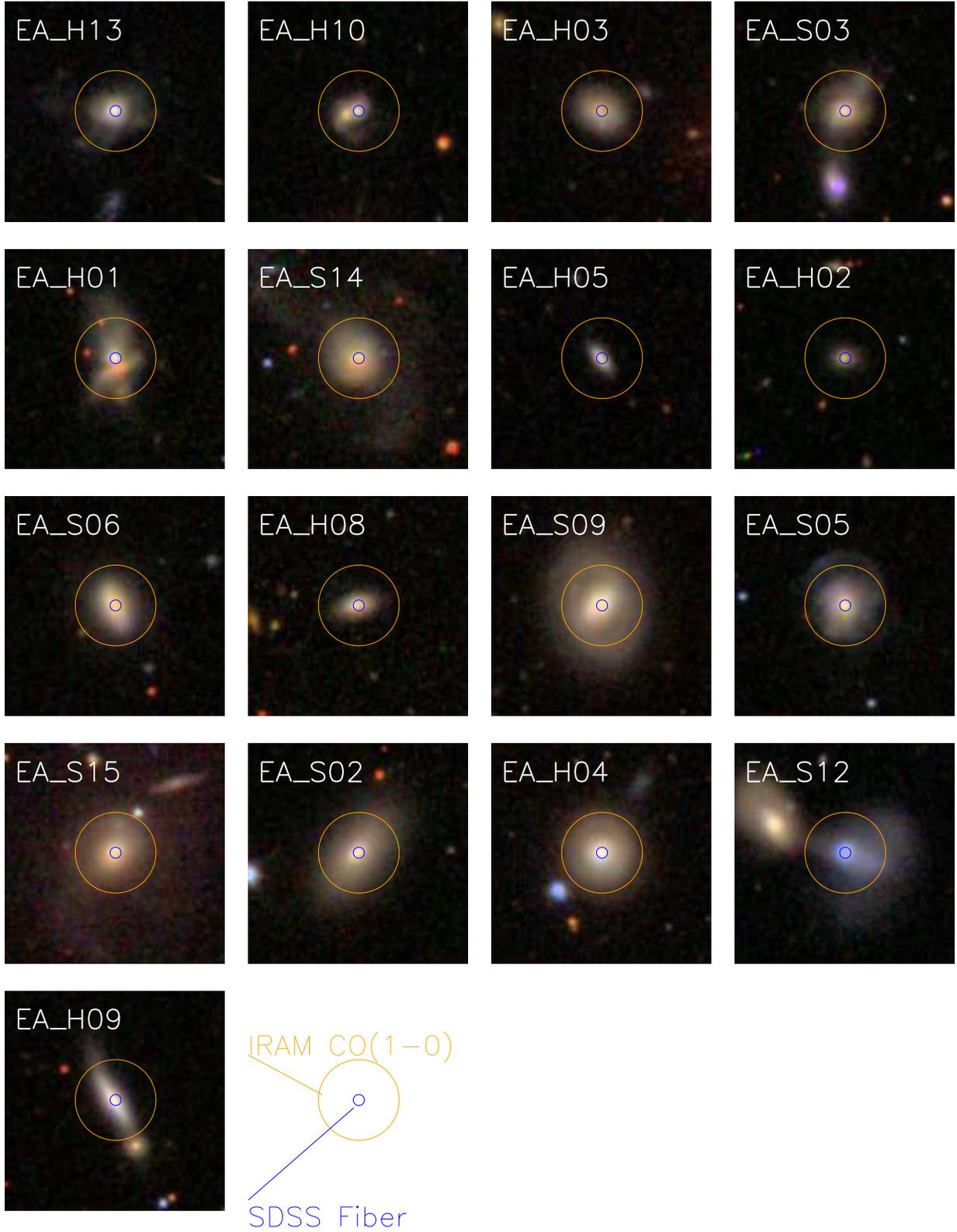}
\caption{60\arcsec \ by 60\arcsec \ SDSS postage stamps of the 17 post-starburst galaxies with CO (1--0) detected  at $>3\sigma$ with the IRAM 30m. The size of the 3\arcsec \ SDSS fiber is overplotted in blue, and the size of the IRAM 30m CO (1--0) 22\arcsec \ beam is overplotted in orange. Galaxies are ordered by decreasing \mhtwo. Given the size estimates in \S\ref{size}, we could be underestimating \lpco by factors of $\sim1.1-2.2\times$, with a median of 1.4, due to aperture effects.}
\label{fig:postagestamps_det}
\end{figure*}

\begin{figure*}
\includegraphics[width=1.0\textwidth]{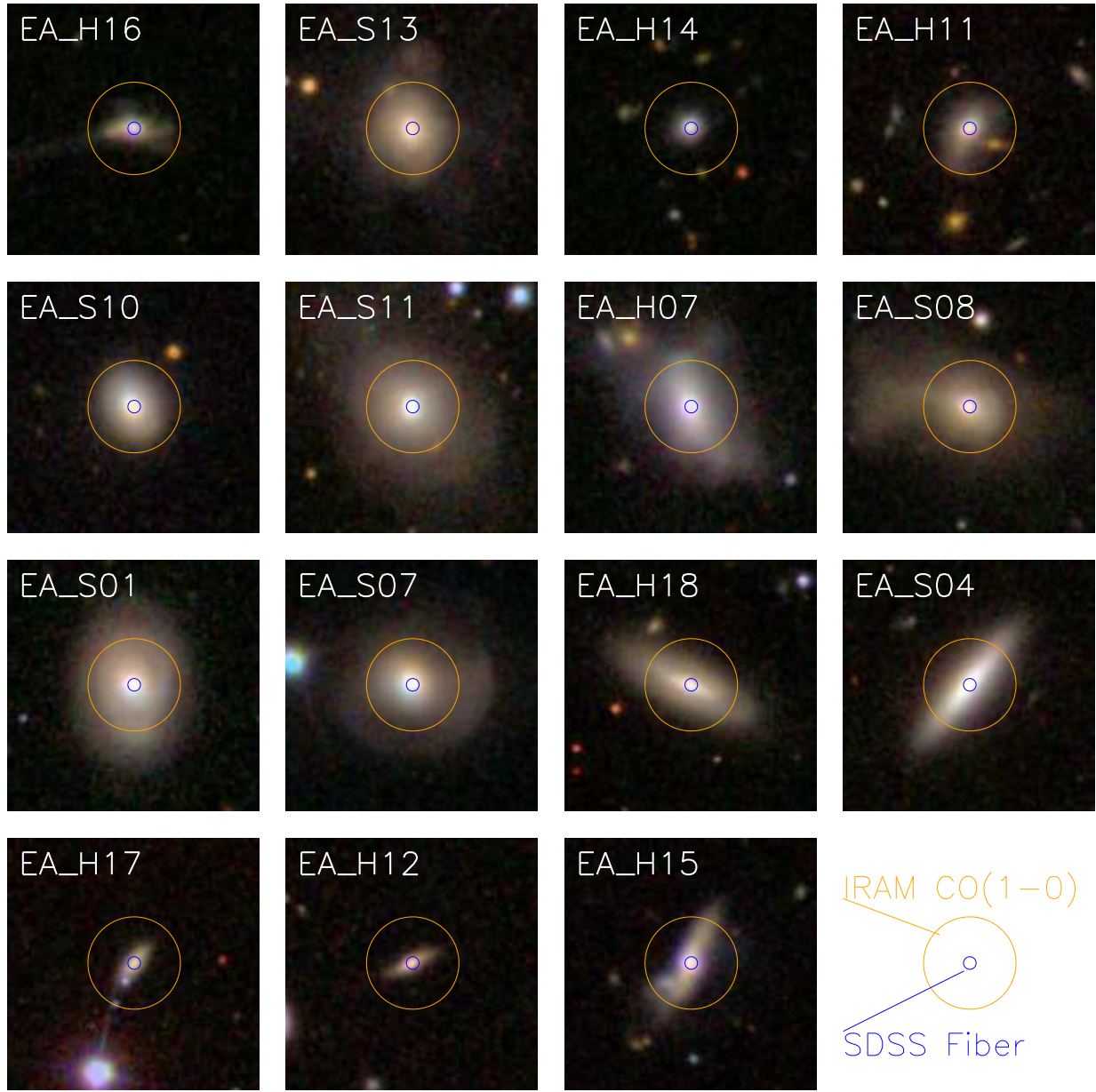}
\caption{Same as Figure \ref{fig:postagestamps_det}, but for the galaxies not detected in CO (1--0).}
\label{fig:postagestamps_non}
\end{figure*}

Molecular gas masses for the post-starburst sample span a broad range, from $3.4\times10^8$ to $6.9\times10^{9}$\msun, with a mean value of $3.0\times10^9$\msun among the detected sample. We measure upper limits for the remaining 15 galaxies, with 3$\sigma$ limits ranging from $4.6\times10^7$ to $5.2\times10^{9}$\msun. Molecular gas masses and upper limits are listed in Table \ref{table:iram10}. Optical postage stamps of the galaxies with and without molecular gas detections are shown in Figures \ref{fig:postagestamps_det} and \ref{fig:postagestamps_non}, respectively.

Next, we compare the molecular gas masses measured here to those from surveys of other galaxy types. 
CO (1--0) measurements have been compiled for the Atlas-3D sample of early type galaxies \citep{Young2011}. 
The COLD GASS \citep{Saintonge2011} sample is a stellar mass-limited sample of galaxies, selected from the SDSS independent of galaxy type.  We divide the COLD GASS sample up by galaxy type assigned by the SDSS based on the optical spectra (galspec \texttt{bptclass}). For now, we only use galaxies classified as star-forming or low signal-to-noise star-forming. 
We assume \alphaco$=4$ to calculate molecular gas mass for the early type and COLD GASS star-forming samples.

We compare the total molecular gas masses of these samples in Figure \ref{fig:molechist}, seeing significant overlap between the star-forming and post-starburst samples. This overlap is surprising, because of the lack of equivalent levels of current star formation in the post-starburst sample. 
The lower mass end, as well as the upper limits, of the post-starburst sample are consistent with the early type sample. 

In addition to comparing \mhtwo, we also compare molecular gas fraction $f_{\rm gas} \equiv M(\rm H_2)/M_\bigstar$ normalized by stellar mass $M_\bigstar$.
We use $M_\bigstar$ calculated from the SDSS spectra, as discussed in \S\ref{sec:sdss}, for both the post-starburst and COLD GASS samples. We calculate stellar masses for the early type galaxies in the same way as Atlas-3D, using K-band measurements \citep{Cappellari2011}.

As with the \mhtwo comparison, we see considerable overlap in $f_{\rm gas}$ between the post-starburst and star-forming samples. These comparisons are shown in the right-hand panels of Figure \ref{fig:molechist}. The molecular gas fractions for the post-starburst sample are primarily above those of the early type sample, while some of the upper limits are more consistent with early types.

\begin{figure}
\includegraphics[width=0.5\textwidth]{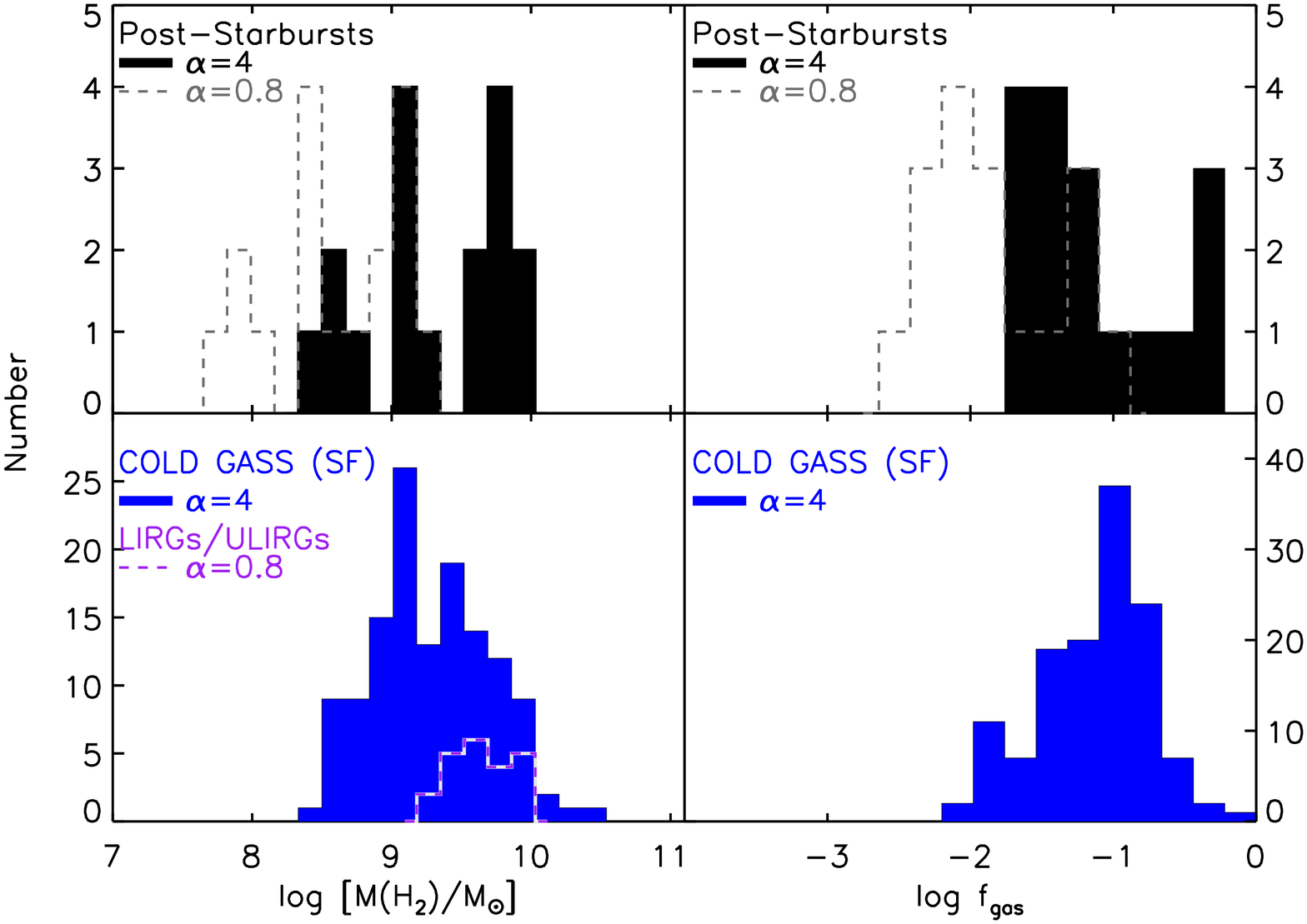}
\caption{Left: Histograms of derived molecular gas masses \mhtwo for a variety of galaxy types:  star-forming \citep[bottom, from COLD GASS;][]{Saintonge2011}, LIRGs and ULIRGs \citep[bottom, from][]{Gao2004}, and our post-starburst sample (top). \alphaco$=4$\unitxco is assumed where data are plotted as solid histograms, and \alphaco$=0.8$\unitxco where histograms are dashed lines. Bin size represents the mean error in the post-starburst sample.
Right: Histograms of molecular gas normalized to stellar mass ($f_{\rm gas}$) for the same samples \citep[except for][]{Gao2004}. Even if a low, ULIRG-like value of \alphaco is used for the post-starburst sample, we still see significant overlap with the star-forming sample, and even some overlap with the LIRG and ULIRG samples.}
\label{fig:molechist_alpha}
\end{figure}

The CO to H$_2$ conversion factor (\alphaco) is a known source of uncertainty in observations of molecular gas \citep[see recent review by][]{Bolatto2013}. Traditionally, a bimodel model has been used, with normal star-forming galaxies assigned a Milky Way-like value of \alphaco$\sim4$, and ULIRGs or starbursting galaxies assigned 
\alphaco$\sim0.8$. This approach was initially motivated by the fact that a high \alphaco applied to ULIRGs produced gas masses higher than the dynamical masses. 
In Figure \ref{fig:molechist_alpha}, we plot \mhtwo and $f_{\rm gas}$ for the post-starbursts and comparison galaxies for different \alphaco assumptions. We also compare to the sample of LIRGs and ULIRGs from \citet{Gao2004}.
Even if a low, ULIRG-like value of \alphaco$=0.8$ is used for the post-starburst sample, we still see significant overlap with the star-forming sample (at \alphaco$=4$), and even some overlap with the LIRG and ULIRG sample (at \alphaco$=0.8$). We expect \alphaco$=0.8$ and \alphaco$=4$ to span the range of possible values of \alphaco in post-starburst galaxies, because recently ended starbursts may reflect ISM physical conditions between  ULIRGs and quiescent disk galaxies. However, the appropriate value of \alphaco for post-starbursts remains largely unconstrained. We discuss the effects of this uncertainty on our results in \S\ref{alphaks}.

\subsection{High Molecular Gas Mass for Given SFR}

\begin{figure*}
\epsscale{1.2}
\includegraphics[width=1\textwidth]{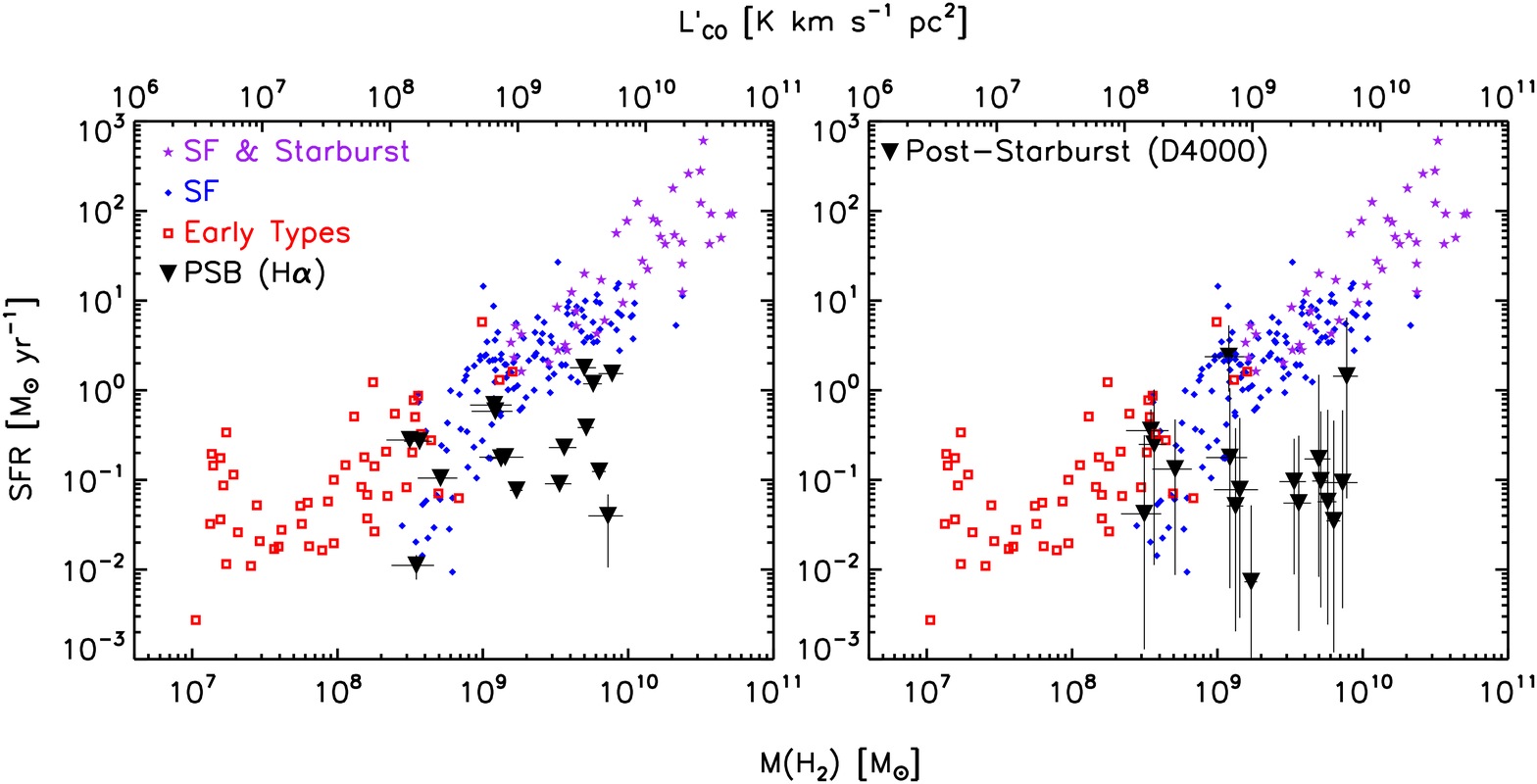}
\caption{Molecular gas mass vs. SFR for post-starburst (PSB) galaxies and comparisons. Whether H$\alpha$ (left) or \dfour (right) SFR upper limits are used for the post-starburst sample, these galaxies fall systematically below the comparison galaxies from the COLD GASS sample classed by the SDSS as star-forming \citep{Saintonge2011, Saintonge2012}, star-forming, LIRG, and ULIRG galaxies from the \citet{Gao2004} sample, and early type galaxies from \citet{Young2011}. Both \dfour and H$\alpha$ are expected to overestimate the SFR in the post-starburst sample, so these galaxies may lie at even lower SFRs. All galaxies have been normalized to the same value of \alphaco$=4$.}
\label{fig:mass_ks}
\end{figure*}

\begin{figure*}
\epsscale{1.2}
\includegraphics[width=1\textwidth]{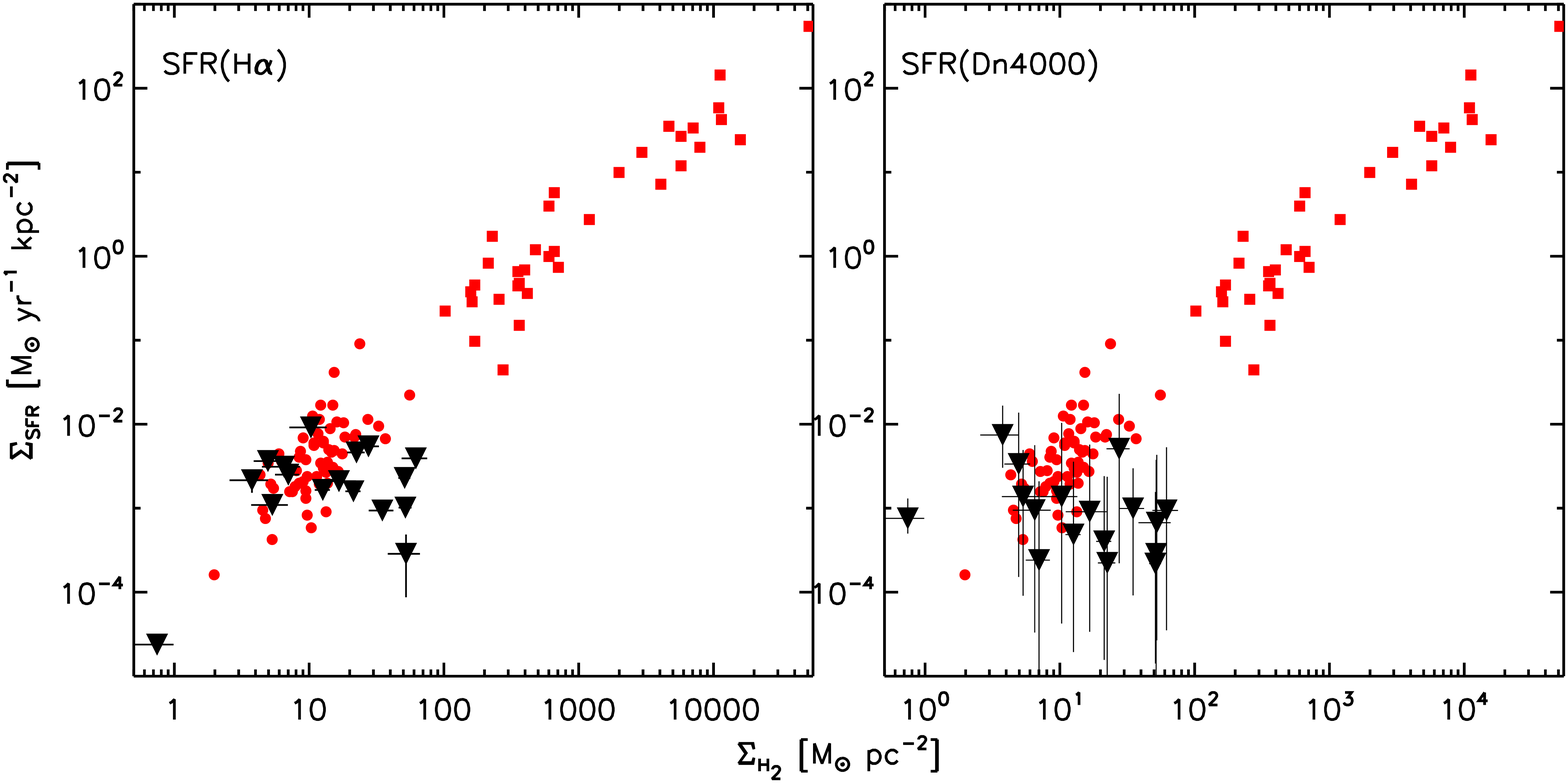}
\caption{Molecular gas surface density vs. SFR surface density from H$\alpha$ (left) and \dfour (right) SFR upper limits.
The post-starburst sample is shown in black, with other local galaxies \citep{Kennicutt1998} shown in red. Red circles are normal star-forming galaxies and red squares indicate local starbursts. Both SFR indicators are upper limits to the true current SFRs for the post-starburst sample. H$\alpha$ has a contribution from the LINER, and \dfour has a contribution from the intermediate age stars produced in the burst. Our post-starburst galaxies are biased low on this relation. For the post-starburst sample, we assume the CO is distributed in the same way as the star formation regions, using the Petrosian 90\% radius $R_{90}$ to calculate the surface densities, $\Sigma_{\rm SFR}=SFR$$/\pi R_{90}^2$ and $\Sigma_{\rm H_2}=$$M(H_2)/\pi R_{90}^2$. Here, \mhtwo includes all the CO detected in the IRAM 30m beam. We use the same value of \alphaco$=4$ for all galaxies. This plot is the most appropriate comparison to the K98 dataset, although we test our assumption that CO and star formation are distributed like the optical light in \S\ref{size}, and that \alphaco$=4$ in \S\ref{alphaks}.}
\label{fig:ksplot_wco_dev}
\end{figure*}

We compare the molecular gas masses measured here to upper limits on SFR derived from H$\alpha$ and \dfour in Figure \ref{fig:mass_ks}. We use several comparison datasets: star-forming galaxies selected from the COLD GASS \citep{Saintonge2011, Saintonge2012}, the star-forming and starburst sample from \citet{Gao2004}, and early type galaxies (with H$\alpha$+PAH SFRs) from \citet{Young2011,Davis2014}. 
The post-starburst sample lies at higher \mhtwo for their SFRs than early-types, star-forming galaxies, and (U)LIRGs. The median H$\alpha$-derived SFR upper-limit for the post-starburst sample is $\sim10\times$ lower than the median SFR for the COLD GASS star-forming sample across the same \mhtwo range. This offset persists for the median of the \dfour-derived SFR upper limits, which is $\sim20\times$ lower than expected given \mhtwo.

Is it possible that our post-starburst selection criteria have generated the observed offset in SFR? Because we selected the post-starburst sample to have low H$\alpha$ equivalent widths, it may include star-forming galaxies whose EW(H$\alpha$) measurement errors have scattered them low. This does not appear to be the case, as star-forming galaxies (classified using {\tt bptclass=1}) that pass our H$\delta$ absorption cut typically have EW(H$\alpha$)$\gg3$\AA, the limit for our post-starburst sample. A Monte-Carlo analysis predicts a 0.004\% contamination rate, and even if the H$\alpha$ equivalent widths had systematic errors 3$\times$ as large as their measurement errors, we still expect $\ll1$ contaminant in the complete SDSS post-starburst sample of 1207 galaxies. Thus, the KS offset does not arise from contamination from normal star-forming galaxies.

The large molecular gas reservoirs in post-starburst galaxies are inconsistent with their SFRs when compared to a broad sample of galaxy types. Thus, the cessation of star formation after the starbust cannot be due to a lack of gas in the nearly half of our sample with detected CO. The question remains: why are these galaxies no longer forming stars at significant rates? One possibility is that the molecular gas is spread out over a larger area, dropping its surface density to be consistent with SFR density on the Kennicutt-Schmidt (KS) relation \citep[hereafter K98]{Kennicutt1998}. We examine the KS relation below.

\subsection{Offset from the Kennicutt-Schmidt Relation}
\label{ks}
While there are clear trends of SFR with molecular gas mass, tighter correlations exist when comparing the surface densities of these quantities for normal star-forming and starburst galaxies.

We determine the molecular gas surface density $\Sigma_{\rm H_2}$ and SFR surface density $\Sigma_{\rm SFR}$ for the post-starburst sample, using the SDSS $r$-band Petrosian 90\% radius to calculate the area as
\begin{equation}
\Sigma_{\rm H_2} = \alpha_{\rm CO} L^\prime_{\rm CO} / (\pi R_{90}^2) ; \: \:  \Sigma_{\rm SFR} = SFR / (\pi R_{90}^2).
\end{equation}
We place the post-starburst sample on a KS plot in Figure \ref{fig:ksplot_wco_dev}. Other local galaxies are shown for comparison, including both normal star-forming galaxies and starbursts from the canonical K98 sample. For now, we apply the same value of \alphaco$=4$ to the post-starburst sample, and the entirety of the K98 sample.
Many of the post-starburst galaxies lie below the relation defined by the other galaxies: the H$\alpha$-derived SFR limits exclude consistency with the relation for all but 4-5 galaxies. The \dfour based SFR limits also lie mostly below the relation. 

The median locus of the 17 post-starburst galaxies lies 4$^{+2}_{-1.5}\times$ lower than the $n=1.4$ power law fit to the K98 galaxies. We perform a Monte Carlo analysis to assess the significance of this result by choosing random sets of 17 galaxies from the K98 disk sample, finding a 5$\sigma$ significant offset for the post-starburst locus. This offset is more extreme than that found by \citet{Davis2014} for their sample of gas-rich early type galaxies. The relationship between the two datasets is unclear.

We see no obvious differences between the properties of our galaxies that are roughly consistent and most discrepant with the KS relation. It is not clear if the post-starbursts are single population of galaxies, or several different families.

The optical size used by K98 to calculate the surface densities is the isophotal radius, where the B-band surface brightness drops to 25 mag arcmin$^{-2}$ and which is comparable to the H$\alpha$ emitting region for normal spiral galaxies (K98, although we test this assumption for our sample in \S\ref{size}). This isophotal radius should be a good estimate of the size if the CO emission is coming from the same region as the optical light from star formation. Here, we use the Petrosian 90\% radius for our post-starbursts, because the isophotal sizes in SDSS are not considered reliable and are not included in the photometric catalogs after DR8. However, the significant offset from the K98 sample remains if we use the $r$-band isophotal radii instead.

There are several other observational uncertainties that could affect our results, which we consider in the Section 4. We discuss the effect of our sample selection criteria in \S\ref{samplebias}. 
Like K98, we assume that the optical size of the galaxies is a good proxy for the spatial extent of both star formation and molecular gas. This assumption may not be valid for post-starburst galaxies. In \S\ref{size}, we test the possibilities that the CO is distributed differently from the optical light and that the star formation is distributed differently from both the optical light and most of the H$_2$, as traced by CO. We also consider how our sampling of the CO region (aperture bias) might affect our results. In \S\ref{alphaks}, we test for the possibility that the measured CO is not tracing the H$_2$ as we expect, resulting in a different \alphaco.

\section{Discussion}

\subsection{Possible Sample Selection Biases}
\label{samplebias}
Given the way we selected our CO targets, the sample observed here might not represent the gas properties of the overall post-starburst sample.
To study any biases that may occur within our sample, we test whether the galaxies with CO (1--0) detections lie at the extremes of our selection criteria. 

The two parts of our sample (labeled ``H'' and ``S'') were selected from the parent sample of SDSS post-starbursts using different criteria (more details in Smercina et al., in prep). The ``H'' sample was selected based on post-starburst galaxies bright in the WISE 12$\mu$m band. One might expect these galaxies to have more gas if the 12$\mu$m band is a proxy for hot dust content and their dust traces their gas. However, we see no mean offset in 12$\mu$m luminosity between the galaxies detected and not detected in CO. 

Both samples were selected to have a variety of times elapsed since the starburst (post-burst age). If the molecular gas is depleted over time, younger post-starbursts may be easier to detect in CO. 
While there is a shift towards detections with younger post-starburst ages, it is not statistically significant. The relation between molecular gas content and age since the burst is not straightforward, and is heavily dependent on the pre-burst gas mass of the galaxy and on the mechanics of the burst itself. 

Because no statistically significant boost in CO detections occurs with either younger post-burst age or higher 12$\mu$m luminosity, the molecular gas properties of our sample here are not significantly biased by the selection criteria. Therefore, our sample is likely to have a molecular gas detection rate representative of the overall post-starburst population.

\subsection{Effect of Spatial Distributions of Gas and Stars}
\label{size}

In \S\ref{ks}, we made the assumptions that both the CO and current star formation were well-traced by the optical light, in order to calculate their surface densities. This is a good assumption for star-forming galaxies \citep{Regan2001,Leroy2008,Schruba2011}, but may not apply to our sample, especially if the reason for the end of the starburst is a disruption of the gas. Additionally, the spatial extent of any residual star formation is unknown and may not overlap with the optical light, which is dominated  ($\sim60-90$\%) by the A stars formed in the recent burst. Without resolved observations, we are limited in how accurately we can know the distributions of gas and current star formation. If the CO emission or any currently star forming regions have sizes different than the optical size, it might be possible to resolve the observed offset from the KS relation. First, we test these assumptions for the post-starburst sample by estimating the CO emitting size with a model for the CO emission. We continue here to assume that CO traces H$_2$ well, with a conversion factor of \alphaco$=4$. Second, we use our estimate of the CO emitting size to compare the scaled amount of CO near the center of the galaxy to the SDSS fiber-based SFRs.

By comparing the CO emission in two differently sized beams, we can roughly constrain the CO emitting size by assuming a Gaussian model for the shape of the emitting region.
We estimate the source size for each galaxy by combining the CO (2--1) line measurements from IRAM 30m and SMT 10m, using the method from \citet{Lavezzi1999}. $I_{CO}$, the integrated line intensity, is related to the surface brightness, so it should scale with the convolved size of the beam and the source as
\begin{equation}
\frac{I_{SMT}}{I_{IRAM}} = \frac{\theta_s^2+\theta_{b,IRAM}^2}{\theta_s^2+\theta_{b,SMT}^2}
\label{eq:thetas0}
\end{equation}
for the same line, assuming a Gaussian distribution of the CO emission, where $\theta_b$ are the different beam sizes for IRAM 30m and SMT, $\theta_{b,SMT}=1.2 \lambda/D \approx 33$ arcsec, and $\theta_{b,IRAM}=1.166 \lambda/D \approx 11$ arcsec, with the different
coefficients due to the taper of each telescope. 
 The source size $\theta_s$ is then given by,
\begin{equation}
\theta_s = \sqrt{\frac{I_{IRAM} \: \theta_{b,IRAM}^2 - I_{SMT} \:  \theta_{b,SMT}^2}{I_{SMT}-I_{IRAM}}}
\label{eq:thetas}
\end{equation}
for a Gaussian source

\begin{figure}
\includegraphics[width=0.5\textwidth]{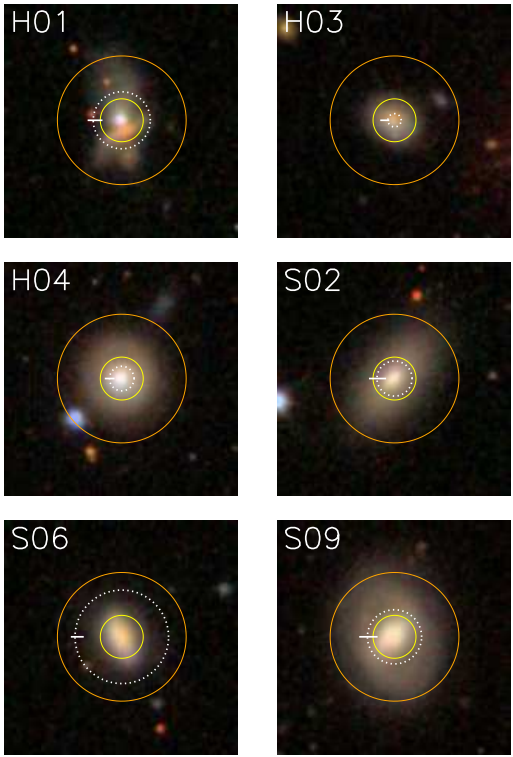}
\caption{Estimated Gaussian half-light sizes of CO emitting regions (white dotted circles), with horizontal lines representing the Monte Carlo estimated error. FWHM beam sizes are shown for comparison: the inner yellow circle represents the IRAM 30m CO (2--1) $\approx$11\arcsec \ beamsize, and the outer orange circle is the SMT CO (2--1) $\approx$33\arcsec \ beamsize. Optical images are from SDSS, and are 60\arcsec \ $\times$ 60\arcsec \ . These CO sizes are $\sim2\times$ the optical half-light sizes (not shown), due to the concentrated optical profile from the stars produced during the burst.}
\label{fig:ps_size}
\end{figure}

We estimate source sizes for the 6 galaxies with IRAM 30m CO (1--0), CO (2--1) and SMT CO (2--1) detections. Because the measurement errors propagate non-linearly, we use a Monte Carlo technique to estimate the formal error on the size estimates, excluding any systematic error from the Gaussian model assumption. Sizes are listed in Table \ref{table:all21}, and plotted over SDSS postage stamp images in Figure \ref{fig:ps_size}.

Previous studies of star-forming galaxies have found comparable exponential scale lengths of the CO emission and optical emission using resolved data \citep{Regan2001,Leroy2008,Schruba2011}. 
However, the CO to optical sizes are on average $\sim2\times$ larger for our post-starburst sample, using either the optical half-light radii from an exponential fit or the Petrosian half-light radii.  Because of the centrally peaked distribution of A stars in the post-starburst sample, this difference is due to the concentrated optical light, rather than extended CO emission. Concentrations measured for post-starburst galaxies \citep{Yang2008} are high, consistent with the half light radii being smaller than for galaxies with an exponential profile.

To test the effect of assuming a Gaussian model, we perform a similar calculation for each galaxy, but for a uniform disk emitting region instead of a Gaussian. The half-light sizes for each source model are consistent within the measurement errors except for EAS06, where the uniform disk model predicts a size smaller than the Gaussian prediction (20\arcsec \ vs. 25\arcsec \ ). 

The systematic errors associated with this method may be significant, especially as several of the estimated sizes are larger than the CO (2--1) IRAM 30-m beamsize. Thus, we use these sizes only to roughly estimate aperture bias in the CO observations, and to test whether use of this alternate size measure can eliminate the offset in the observed KS relation. The CO size estimates play no role in our main conclusions.

\begin{figure*}
\epsscale{1.2}
\includegraphics[width=1\textwidth]{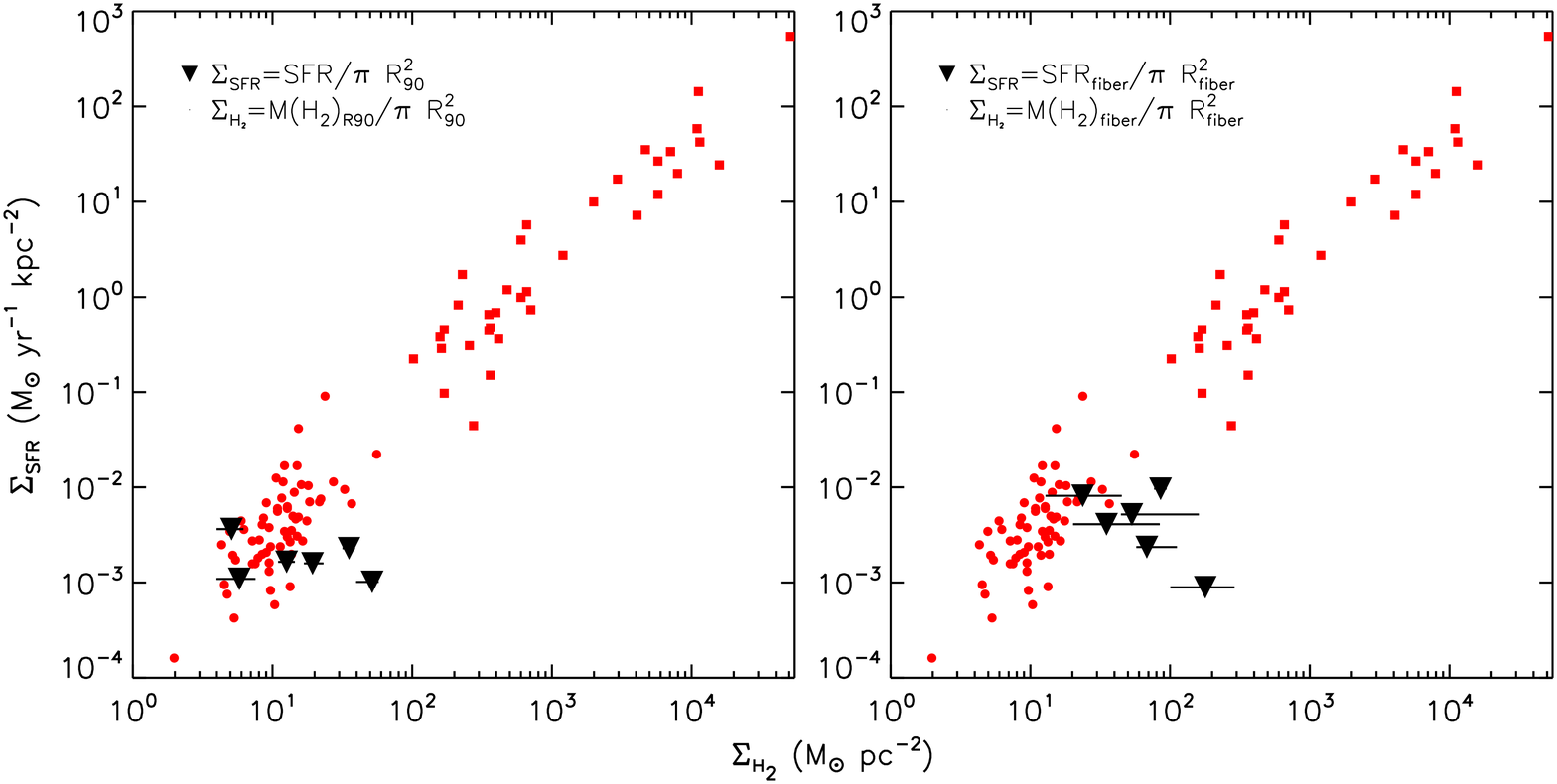}
\caption{Alternative KS plots (red comparison points same as in Figure \ref{fig:ksplot_wco_dev}), accounting for possible spatial differences between molecular gas, star formation, and optical light. Left: $\Sigma_{\rm SFR}$ (from H$\alpha$) and $\Sigma_{H_2}$ are both calculated within the optical radius $R_{90}$ for the post-starburst sample. We use the CO size estimate from \S\ref{size} and assume a Gaussian distribution to rescale \mhtwo to that within $R_{90}$ instead of within the larger IRAM 30m 22\arcsec \ beam as in Figure \ref{fig:ksplot_wco_dev}. Despite this rescaling, which allows for the possibility that the optical light and CO are distributed differently, we still observe an offset of the post-starbust galaxies from the K98 sample. Right: $\Sigma_{\rm SFR}$ and $\Sigma_{\rm H_2}$ are both calculated within the 3\arcsec \ SDSS fiber. We rescale \mhtwo to within this radius. SFR is calculated using only the H$\alpha$ flux from the fiber, without rescaling. This is a test of whether the star formation is distributed differently than either the optical light or CO, and the continued offset shows that this possibility is unlikely unless the Gaussian assumption for the CO distribution is poor, or the star formation is even more concentrated within the fiber size.}
\label{fig:radius_comp}
\end{figure*}

We use the CO emitting size estimates to model the effect of aperture bias in our CO (1--0) observations taken with the IRAM 30-m $\approx$22\arcsec \ beam. Given that the CO-emitting region is not a point source, the CO line luminosity calculated in \S3.1 is likely an underestimate by a factor of $(\theta_s^2+\theta_b^2)/\theta_b^2$, where $\theta_s$ and $\theta_b$ are the source and beam sizes, respectively. If we assume the mean CO source size of 9.2 kpc from our estimates above, then the corrections needed for the CO line luminosity, thus also \mhtwo are $\sim$1.1-2.2$\times$. The mean correction is an increase of 1.4$\times$ from the original calculation, indicating that the molecular gas masses reported here may be conservative, with even higher amounts of molecular gas present.

Using the models of the CO emitting region that we construct from the size estimates and Gaussian assumption, we test the possibility that some of the observed CO does not participate in star formation. In \S\ref{ks}, and in Figure \ref{fig:ksplot_wco_dev}, we assumed that all of the observed CO lay within the optical size $R_{90}$, which is always smaller than the IRAM 30m CO (1--0) beamsize. Our CO size estimates, however, indicate that CO may extend beyond this aperture. We now estimate $\Sigma_{\rm H_2}$ within $R_{90}$ by rescaling \mhtwo using the Gaussian model to include only the gas mass within $R_{90}$ (Figure \ref{fig:radius_comp}a). While still assuming that the current star formation is distributed like the optical light, this method allows us to compare the surface densities within the same aperture.  In \S\ref{detect}, we assumed that the source sizes were much smaller than the beam sizes to calculate $L^\prime_{CO (1-0)}$, but we now calculate $\Omega_{s*b}$ explicitly in Equation 1. Despite the decrease in $M(H_2)$ (and $\Sigma_{H_2}$), the post-starburst sample is still significantly ($>4\sigma$) offset from the K98 galaxies. Allowing for the optical light and CO to be distributed differently does not change our results.

Next, we study the effect of allowing the spatial extent of any current star formation to differ from both the CO and optical light distributions.
\citet{Swinbank2012} use resolved IFU observations of post-starburst galaxies, and find the nebular emission lines [O II] and [O III] to be spatially offset from the A star population in some cases, although these lines are particularly contaminated by LINER emission.
If the current star formation is even more concentrated than the optical light, this could drive the observed offset of the post-starbursts on the KS relation. If this is the case, our assumptions about the aperture correction in the SFRs in \S\ref{sec:sdsssfr} are not valid, and the SFRs are even more of an over-estimate. We consider now the case where the star formation is restricted to the optical fiber.

Using the CO sizes estimates and Gaussian model, we rescale $M(H_2)$ to that within the 3\arcsec \  SDSS optical fiber, again with a Monte-Carlo method to estimate the errors. We then estimate $\Sigma_{\rm H_2}$ inside the fiber aperture by scaling the CO luminosity, assuming a Gaussian profile with the CO size estimate.  To calculate $\Sigma_{\rm SFR}$ , we use only the H$\alpha$ flux from the fiber, not the aperture-corrected flux used elsewhere throughout this paper. We plot the resulting surface densities within the fiber apertures in Figure \ref{fig:radius_comp}b.
 The post-starburst galaxies are still significantly offset from the K98 comparison galaxies. Thus, even if we allow for the possibility that some of the CO in the outer regions of the galaxy does not participate in star formation, or that the current star formation is more compact than the optical light, the offset from the K98 galaxies remains.

Even if the CO in the post-starburst sample and the K98 sample is distributed in the same way with respect to the optical light, aperture bias in the datasets could generate an apparent offset. We do not see evidence of severe aperture bias in the post-starburst sample.  Although we do observe higher molecular gas masses at higher redshifts, this is due to a combination of our decreased sensitivity and the higher stellar mass SDSS-selection at higher redshifts. We see no statistically significant trends with redshift of $f_{\rm gas}$, or in the offset from the $n=1.4$ power law KS relation from K98. If aperture bias only affected the post-starburst sample, it would result in an under-estimation of the offset.

An offset could be generated, however, if the K98 sample was not measured out to the same physical radii as in the post-starburst sample. The CO isophotal sizes in the K98 star-forming sample are at most 60\% of the optical isophotal sizes, and often much smaller. Galaxies with CO measured out to at least half the isophotal radii are the only ones included in the K98 sample. Assuming a worst-case, uniform disk distribution, the actual CO flux could be up to 70\% higher than measured, which is not enough to resolve the observed offset of $\sim400$\%.

The offset observed in Figure \ref{fig:mass_ks}, that the post-starburst sample lies at lower SFR for a given \mhtwo than other galaxies, and the similar robust offset in the KS relation (Figures \ref{fig:ksplot_wco_dev} and \ref{fig:radius_comp}), suggest that 1) the CO does not trace the H$_2$ as in our comparison galaxies or 2) the H$_2$ is not turning into stars in the same manner as our comparison galaxies (i.e., the star formation efficiency (SFE) is lower or the IMF is bottom-heavy). We explore 1) in Section \ref{alphaks} and 2) in Section \ref{sec:evo}. For now we note that suppressing the SFE by allowing CO to extend beyond any current star formation region does not resolve the KS offset. Ultimately, interferometric CO observations will be required to test the spatial distribution of CO.

\subsection{Effect of \alphaco Choice}
\label{alphaks}

As discussed in \S3, the CO to H$_2$ conversion factor (\alphaco) is a known source of uncertainty in measuring molecular gas masses from CO observations. Here, we assume that H$_2$ traces the current star formation as expected for other galaxies, and explore variations in how CO traces the H$_2$.

\begin{figure}
\epsscale{1.2}
\includegraphics[width=0.5\textwidth]{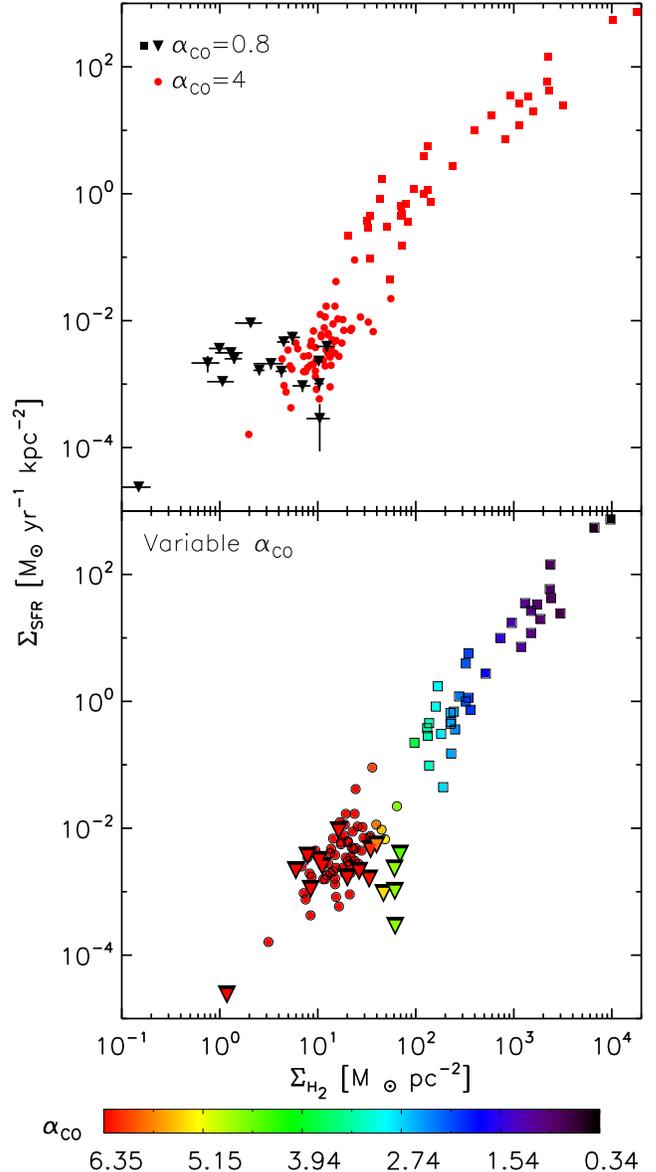}
\caption{Top: Same as Figure \ref{fig:ksplot_wco_dev}a (see description in \S\ref{ks}), but assuming \alphaco$=4$ for the K98 star-forming galaxies, and \alphaco$=0.8$ for the K98 starbursts and our post-starburst sample. Here, the offset previously observed can be reconciled by assuming a ULIRG-type value of \alphaco$=0.8$ for the post-starburst sample. Bottom: Same as Figure \ref{fig:ksplot_wco_dev}, but now assuming the variable \alphaco for the K98 sample and post-starburst sample using the formula from \citet{Narayanan2012}. As before, triangles, circles, and squares represent the post-starburst sample, K98 star-forming galaxies, and K98 starbursts, respectively. We still observe a systematic shift low on the Kennicutt-Schmidt relation, which suggests that the variable \alphaco model does not resolve the differences between the post-starbursts and other galaxies.}
\label{fig:ks-varalpha1}
\end{figure}

So far, in plotting the KS relation we have assumed a single value of \alphaco$=4$ for all samples. 
Now assuming a bimodel \alphaco model instead, we apply a ULIRG-like value of \alphaco$=0.8$ to the K98 starbursts and to our post-starburst sample, and leave the K98 star-forming galaxies with \alphaco$=4$. We obtain the results in Figure \ref{fig:ks-varalpha1}a. This low value of \alphaco applied to the post-starburst sample can remove their observed offset from the modified KS relation.

We can understand why starbursting galaxies may require a lower value of \alphaco using the following toy model. \alphaco is proportional to the column density of molecular gas $N(H_2)$ over the CO line intensity, as
\begin{equation}
\alpha_{CO}\propto \frac{N(H_2)}{I_{CO}} \propto \frac{N(H_2)}{T\times\sigma}
\end{equation}
\citep{Narayanan2011a}.
In a merger, the column density is increased. However, the line intensity goes as the temperature $T$ times the velocity dispersion $\sigma$, which both increase during the merger. In total, these factors result in a lower value of \alphaco. After the merger, the gas kinetic temperature may decline to match conditions in early-type galaxies, but it is not clear what this simple model predicts for the post-starburst sample.

We cannot assume the post-starbursts will have a similar value of \alphaco as ULIRGs or starbursts simply because they are the likely progenitors. We expect the physical state of the gas to have changed significantly in the 0.3-1 Gyr since the burst, as the dynamical timescales for ULIRGs are of order $10^6-10^{7.5}$ yr \citep{Genzel2010}.  
However, if the state of the gas after the merger changes in such a way as to keep the gas heated, but as to lower the column density of molecular gas, \alphaco may have a low value. Additionally, if an AGN heated the bulk of the molecular gas, the CO brightness temperature could increase for a given H$_2$ mass, lowering \alphaco.

If much of the CO emission comes from outside of GMCs, the CO (1--0) linewidth could be strongly affected by the gravitational potential in the galaxy, instead of just by its own turbulence \citep{Downes1998}. The fact that starbursts have more diffuse gas, and less gas bound in molecular clouds, has been used as justification for their low values of \alphaco. We can estimate the influence of the stellar potential on \alphaco using the prescription from \citet{Bolatto2013}. They suggest that \alphaco scales down from the Galactic value as $\alpha_{CO}/\alpha_{CO,MW} = (\Sigma_{\text{total}}/100\: $M$_\sun$pc$^{-2})^{-0.5}$, where $\Sigma_{\text{total}}$ is the combined surface density of stars and gas. Adding the stellar mass to the molecular gas mass, and calculating the surface density within the optical radius $R_{90}$, our post-starbursts have total surface densities of 130-460 M$_\sun$pc$^{-2}$, predicting \alphaco$=1.9-3.5$, scaled down from the Galactic value of \alphaco$=4$. This difference between the assumed \alphaco for comparison star-forming galaxies and the post-starburst sample is not enough to resolve their observed offset. However, this relation is subject to scatter, and there may be significant variation in the influence of diffuse gas on \alphaco \citep{Liszt2012}.

The  use of a bimodel \alphaco is not necessarily physical \citep[although see, e.g., ][]{Tacconi2008, Daddi2009}, because a variety of ISM conditions should exist, resulting in a continuum of \alphaco values.
In particular, because post-starburst galaxies may be at an intermediate stage between being dynamically hot and more relaxed, the appropriate \alphaco for these systems may lie between the Galactic average and the typical value used for ULIRGs.  The \citet{Narayanan2012} formulation suggests that \alphaco varies smoothly with galaxy physical properties, and may be parametrized in terms of the gas phase metallicity and the CO surface brightness. Our galaxies do not have abnormal metallicities. \citet{Goto2007} study the metallicities of post-starburst galaxies and do not find them to be anomalous.  The mass-metallicity relation \citep{Tremonti2004} also predicts metallicities that are similar to the comparison galaxies, so we vary \alphaco only with CO brightness. The result of applying this variable \alphaco model to both the post-starburst and K98 samples can be seen in Figure \ref{fig:ks-varalpha1}b. The variable \alphaco model does not remove the offset.

We are unable to rule out a low value of \alphaco$=0.8$ as a potential explanation for the offset from the KS relation. We can resolve this question observationally using higher J$_{up}$ lines of CO to constrain the temperature and density of the gas, as well as with denser gas tracers such as HCN to probe denser regions of the gas, bypassing the uncertainties arising from any CO outside of GMCs.

\subsection{Implications for Galaxy Evolution}
\label{sec:evo}

Our findings that post-starburst galaxies can have large gas reservoirs, and that they are offset low
from the KS relation, help to discriminate among the physical processes proposed to end the burst. 
Clearly, scenarios that require the molecular gas to be absent, such as the complete 1) expulsion or removal of the gas \citep[e.g., in outflows or some environmental mechanisms;][]{Feruglio2010,Rupke2011,Boselli2006}, 2) consumption of the gas in the burst, and 3) prevention of gas infall into the galaxy and of the subsequent formation of new GMCs \citep[``starvation";][]{Larson1980}, are now excluded, at least in the $~$half of our sample with detected molecular gas.
Alternatively, the molecular gas within the galaxies could be 4) heated \citep{Nesvadba2010}, 5) kinematically prevented from collapsing into GMCs \citep[``morphological quenching";][]{Martig2009}, or 6) dispersed. Here, we comment on the implications of our results for these latter scenarios, and on what data are needed to complete the picture.

The offset observed in both Figures \ref{fig:mass_ks} and \ref{fig:ksplot_wco_dev} could be caused by either a breakdown in the relation between CO and H$_2$ (a different value of \alphaco), or between H$_2$ and the SFR (a different star formation efficiency), such that either is different than for the comparison galaxies.
The burst-ending mechanisms of molecular gas heating (scenario 4) or morphological quenching (scenario 5) might alter the state of the gas relative to normal star-forming galaxies, acting
to lower \alphaco (see Eqn. 5) or to reduce the SFE for a given $M(H_2)$. In the case where \alphaco {\it is} low, dispersal of the gas (scenario 6) would drive the post-starbursts down along the KS relation. 

Analyzing the gas state is outside the scope of this paper, as we cannot constrain the temperature or density of the CO emitting region using only the CO (1-0) and (2-1) lines \citep[e.g.,][]{Carilli2013}. We need higher J$_{up}$ lines and denser gas tracers to do so. Determining if the galaxy is undergoing morphological quenching requires resolved kinematics of the molecular gas.
If the offset from the KS relation is not due to incorrect assumptions about \alphaco (see \S\ref{alphaks}), the SFR (\S\ref{sec:sdsssfr}), or the relative spatial distribution of $H_2$, optical light, and star formation (\S\ref{size}, assuming CO traces $H_2$), the intriguing possibility of a lower star formation efficiency in post-starbursts remains.

Another potential explanation for the offset is that our galaxies are still forming stars, but with a bottom-heavy IMF dominated by low mass stars that are not detectable by our SFR indicators. Regardless of the IMF during
the burst, any subsequent star formation in our sample might be expected to track that of early types, for which bottom-heavy IMFs have been suggested \citep[e.g.,][]{vanDokkum2010}. We test this hypothesis by estimating the change in the H$\alpha$-derived SFRs with IMF slope. For consistency, so far we have used the same conversion factor from $L(H\alpha)$ to SFR as K98, who employed a Salpeter IMF ($x = 2.35$ slope). However, \citet{vanDokkum2010} favor a steeper, more bottom heavy slope of $x = 3$. We use Starburst 99 v7.0.0 \citep{Leitherer2014} models with a variety of IMF slopes over the range $0.1-1 M_\sun$ . For a bottom heavy IMF slope of $x = 3$, we would underestimate the SFR significantly (by $\sim20\times$), more than enough to explain the offset of our galaxies from the KS relation.

\section{Conclusions}
We study molecular gas in a sample of 32 nearby ($0.01<z<0.12$) post-starburst (aka ``E+A'') galaxies, whose optical spectra indicate a recent starburst that ended within the last $\sim$Gyr.
We target the CO lines (1--0) and (2--1) with the IRAM 30m and SMT 10m telescopes, constraining the molecular gas mass remaining in these galaxies after the starburst. 
Our conclusions are as follows:
\begin{itemize}
\item Molecular gas is detected in 17 (53\%) galaxies with CO (1-0) observations from the IRAM 30m. We obtain molecular gas masses of $M(H_2)=10^{8.6}$-$10^{9.8} M_\sun$ (\alphaco/4\unitxco), and molecular gas to stellar mass fractions of $f_{\rm gas}\sim10^{-2}$-$10^{-0.5}$ (\alphaco/4\unitxco), roughly comparable to those of star-forming galaxies and generally larger than for early types, for a range of likely CO-to-H$_2$ conversion factors (\alphaco). The upper limits on $M(H_2)$ for the 15 non-detected galaxies range from $10^{7.7} M_\sun$ to $10^{9.7} M_\sun$, with the median more consistent with early-type galaxies than with star-forming galaxies.

\item We compare $M(H_2)$ to the star formation rate (SFR), using H$\alpha$ and \dfour to calculate upper limits on the current SFR in this sample. When compared to other star-forming, starbursting, and early type galaxies, the post-starbursts have $\sim10-20\times$ lower SFRs for a given \mhtwo.

\item The post-starburst sample falls $\sim4\times$ below other local galaxies on the Kennicutt-Schmidt relation \citep{Kennicutt1998} of SFR surface density vs. \mhtwo surface density. The median locus of the post-starburst galaxies is offset from the relation defined by normal star-forming galaxies (K98) at 5$\sigma$ significance. After considering sample selection effects, aperture bias, varying spatial extents of current star formation, optical light and H$_2$, CO not tracing H$_2$ (a different \alphaco), and the effect of IMF assumptions, we conclude the observed offset is likely due to suppressed SFE, a low value of \alphaco consistent with ULIRGs, and/or a bottom-heavy IMF.

\end{itemize}
Our results show that the end of the starburst in these galaxies cannot be attributed to the complete consumption, expulsion, or starvation of the molecular gas reservoirs. Resolved interferometric CO maps of these galaxies, higher J$_{up}$ lines of CO for density-temperature constraints, denser gas tracers such as HCN, and resolved star formation maps are necessary to more thoroughly study the current state of gas in post-starbursts, and to more accurately compare to the residual star formation.
Understanding this possibly common phase in galaxy evolution will help reveal the physics of star formation in galaxies as well as their evolution through mergers.

\acknowledgments
We thank Adam Smercina for helpful discussions regarding his ongoing analysis of the dust emission from the post-starburst sample. Based on observations carried out with the IRAM-30m Telescope and the ARO SMT 12m. IRAM is supported by INSU/CNRS (France), MPG (Germany) and IGN (Spain). The SMT is operated by the Arizona Radio Observatory (ARO), Steward Observatory, University of Arizona. We thank Dennis Zaritsky for his suggestion to explore a bottom-heavy IMF. We thank the operators and staff of the ARO. KDF is supported by NSF grant DGE-1143953. AIZ thanks the John Guggenheim Foundation and the Center for Cosmology and Particle Physics at NYU. AIZ acknowledges support from NASA ADP grant 3227400 and NSF grant 3002520.  DN acknowledges financial support from the NSF via awards AST-1009452 and AST-1442650.
Funding for the Sloan Digital Sky Survey (SDSS) has been provided by the Alfred P. Sloan Foundation, the Participating Institutions, the National Aeronautics and Space Administration, the National Science Foundation, the U.S. Department of Energy, the Japanese Monbukagakusho, and the Max Planck Society. The SDSS Web site is http://www.sdss.org/.

\vspace{2cm}
Facilities: \facility{IRAM-30m, SMT}

\bibliographystyle{apj}
\bibliography{earefs.bib}

\clearpage

\begin{turnpage}
\begin{table}

\caption{Post-Starburst Targets}
\label{table:targets}
\begin{tabular}{r r r r r r r r r r r r r}
\hline
\hline
 \multicolumn{1}{c}{Target} &
 \multicolumn{1}{c}{R.A.} &
 \multicolumn{1}{c}{decl.} &
 \multicolumn{1}{c}{z} &
 \multicolumn{1}{c}{PlateID \footnotemark[1]} &
 \multicolumn{1}{c}{MJD \footnotemark[1]} &
 \multicolumn{1}{c}{FiberID \footnotemark[1]} &
 \multicolumn{1}{c}{log($M_{\rm *}/M_\sun$) \footnotemark[2]} &
 \multicolumn{1}{c}{\dfour \footnotemark[3]} &
 \multicolumn{1}{c}{$R_{90}$ \footnotemark[4]} &
 \multicolumn{1}{c}{$SFR_{lim}$(H$\alpha$)  \footnotemark[5]} &
 \multicolumn{1}{c}{$SFR_{lim}$(\dfour)  \footnotemark[6]} &
 \multicolumn{1}{c}{ap. corr. \footnotemark[7]} \\

 \multicolumn{1}{c}{} &
 \multicolumn{1}{c}{(deg)} &
 \multicolumn{1}{c}{(deg)} &
 \multicolumn{1}{c}{} &
 \multicolumn{1}{c}{} &
 \multicolumn{1}{c}{} &
 \multicolumn{1}{c}{} &
 \multicolumn{1}{c}{} &
 \multicolumn{1}{c}{} &
 \multicolumn{1}{c}{(arcsec)} &
 \multicolumn{1}{c}{\unitsfr} &
 \multicolumn{1}{c}{\unitsfr} &
 \multicolumn{1}{c}{} \\

\hline
EAH01 &      128.640457 &       17.346207 &          0.0478 &         2276 &        53712 &      444&           10.45 &          1.4122 &            9.35 &          0.0586 &          0.0149 &          6.5416 \\
EAH02 &      141.580383 &       18.678055 &          0.0541 &         2360 &        53728 &      167&            9.96 &          1.3798 &            5.26 &          0.0256 &          0.0271 &          3.5339 \\
EAH03 &      222.066864 &       17.551651 &          0.0449 &         2777 &        54554 &      258&           10.34 &          1.4227 &            7.06 &          0.0196 &          0.0055 &          6.3128 \\
EAH04 &      318.502258 &        0.535107 &          0.0269 &          986 &        52443 &      468&           10.18 &          1.2445 &            8.98 &          0.0674 &          0.0621 &          4.0061 \\
EAH05 &      184.260117 &       39.077038 &          0.0653 &         2001 &        53493 &      473&           10.00 &          1.4148 &            3.44 &          0.0625 &          0.0150 &          3.6750 \\
EAH06 &      116.456268 &       31.378378 &          0.0441 &          755 &        52235 &       42&           10.53 &          1.4477 &            5.52 &          0.3737 &          0.0221 &          2.8889 \\
EAH07 &      167.824844 &       11.554388 &          0.0380 &         1604 &        53078 &      161&           10.65 &          1.3226 &           12.50 &          0.2206 &          0.0292 &          9.8156 \\
EAH08 &      147.077820 &        2.501155 &          0.0604 &          480 &        51989 &      580&           10.41 &          1.3765 &            4.48 &          0.0407 &          0.0176 &          4.4127 \\
EAH09 &      227.229538 &       37.558273 &          0.0291 &         1352 &        52819 &      610&           10.21 &          1.3668 &            5.34 &          0.0572 &          0.0086 &          4.8709 \\
EAH10 &      158.427979 &       21.127987 &          0.1053 &         2376 &        53770 &      454&           10.24 &          1.4019 &            3.44 &          0.0419 &          0.0985 &          0.9470 \\
EAH11 &      166.419617 &        5.998405 &          0.0542 &         1003 &        52641 &       87&           10.61 &          1.4173 &            4.67 &          0.1690 &          0.0271 &          1.8951 \\
EAH12 &      223.772690 &       13.281012 &          0.0826 &         2750 &        54242 &       18&           10.55 &          1.4002 &            2.83 &          0.1834 &          0.0217 &          1.5728 \\
EAH13 &      155.503281 &       22.163177 &          0.1129 &         2365 &        53739 &      624&           11.00 &          1.2897 &            4.61 &          0.6212 &          0.5817 &          2.4663 \\
EAH14 &      178.276855 &       64.299026 &          0.0622 &          598 &        52316 &      170&           10.04 &          1.2878 &            3.15 &          0.1005 &          0.0272 &          2.1201 \\
EAH15 &      163.085205 &        5.828218 &          0.0411 &         1001 &        52670 &       48&           10.40 &          1.4178 &            4.72 &          0.0596 &          0.0258 &          5.6546 \\
EAH16 &      141.740372 &       42.526840 &          0.1113 &          870 &        52325 &      208&           10.74 &          1.1959 &            4.52 &          0.5076 &          5.2118 &          1.5974 \\
EAH17 &      191.215393 &       -1.759901 &          0.0481 &          336 &        51999 &      469&           10.05 &          1.4225 &            4.67 &          0.0479 &          0.0634 &          1.6746 \\
EAS01 &       11.246839 &       -8.889684 &          0.0196 &          656 &        52148 &      404&           10.24 &          1.4919 &           13.08 &          0.0130 &          0.0255 &          5.0585 \\
EAS02 &       49.228809 &       -0.041979 &          0.0231 &          413 &        51929 &      238&           10.08 &          1.4068 &           11.83 &          0.0253 &          0.0318 &          4.1478 \\
EAS03 &      117.809624 &       34.418201 &          0.0628 &          756 &        52577 &      424&           10.86 &          1.4849 &            7.46 &          0.1710 &          0.0083 &          6.8937 \\
EAS04 &      126.755821 &       21.706779 &          0.0153 &         1927 &        53321 &      584&            9.99 &          1.4244 &            8.84 &          0.0140 &          0.0091 &         45.2321 \\
EAS05 &      146.112335 &        4.499120 &          0.0467 &          570 &        52266 &      537&           10.57 &          1.3109 &            8.42 &          0.1078 &          0.0328 &          5.4079 \\
EAS06 &      159.488983 &       46.244514 &          0.0227 &          962 &        52620 &      212&           10.14 &          1.4450 &            7.10 &          0.0575 &          0.0055 &          1.3282 \\
EAS07 &      169.781738 &       58.053974 &          0.0325 &          951 &        52398 &      128&           10.54 &          1.4897 &            7.23 &          0.0447 &          0.1355 &          2.7343 \\
EAS08 &      189.900208 &       12.438888 &          0.0408 &         1616 &        53169 &       71&           10.67 &          1.3406 &           13.55 &          0.0550 &          0.1720 &          6.2527 \\
EAS09 &      191.611816 &       50.792061 &          0.0270 &         1279 &        52736 &      362&           10.56 &          1.5223 &           10.68 &          0.0432 &          0.0127 &          4.0306 \\
EAS10 &      196.357605 &       53.591759 &          0.0381 &         1039 &        52707 &       42&           10.53 &          1.5089 &            6.44 &          0.0189 &          0.0397 &          2.5848 \\
EAS11 &      242.585358 &       41.854881 &          0.0395 &         1170 &        52756 &      189&           10.74 &          1.5299 &           10.55 &          0.0765 &          0.0583 &          6.6460 \\
EAS12 &      243.375778 &       51.059879 &          0.0336 &          623 &        52051 &      209&           10.01 &          1.1354 &           18.20 &          0.0314 &          0.2572 &          1.3772 \\
EAS13 &      246.760666 &       43.476093 &          0.0462 &          815 &        52374 &      586&           10.95 &          1.4548 &           11.41 &          0.0860 &          0.0101 &          7.2333 \\
EAS14 &      316.286133 &       -5.399832 &          0.0826 &          637 &        52174 &      584&           11.31 &          1.5036 &            9.68 &          0.4120 &          0.0396 &          4.3280 \\
EAS15 &      343.778320 &        0.977756 &          0.0533 &          379 &        51789 &      579&           10.83 &          1.3498 &            9.70 &          0.1697 &          0.5866 &          4.0352 \\

\hline
\multicolumn{10}{l}{%
  \begin{minipage}{12cm}%
    Notes: \footnotetext[1]{SDSS Spectra Identification from DR7 \citep{Abazajian2009}. }
\footnotetext[2]{MPA-JHU Stellar Masses, method described in \citet{Brinchmann2004}.}
\footnotetext[3]{\dfour from SDSS spectra.}
\footnotetext[4]{Petrosian 90\% size in $r$ band, from SDSS photometry.}
\footnotetext[5]{Limit on SFR from H$\alpha$ measurements, before aperture correction. Conversion from \citet{Kennicutt1994}, line fluxes from MPA-JHU catalogs \citep{Aihara2011}. See \S\ref{sec:sdss}.}
\footnotetext[6]{Limit on SFR from \dfour measurements, from MPA-JHU catalogs \citep{Brinchmann2004}, before aperture correction. See \S\ref{sec:sdsssfr}.}
\footnotetext[7]{Aperture correction to convert fiber-based SFRs to global SFRs, from MPA-JHU catalogs \citep{Brinchmann2004}. See \S\ref{sec:sdsssfr}.}
  \end{minipage}%
}\\
\end{tabular}
\end{table}
\end{turnpage}

\clearpage

\begin{table}
\centering
\caption{IRAM 30m CO (1-0) observations}
\label{table:iram10}
\begin{tabular}{r r r r r r}
\hline
\hline
 \multicolumn{1}{c}{Target \footnotemark[1]} &
 \multicolumn{1}{c}{$t_{obs}$} &
 \multicolumn{1}{c}{$I_{CO}$ \footnotemark[2] } &
 \multicolumn{1}{c}{\lpco \footnotemark[3]} &
 \multicolumn{1}{c}{\mhtwo \footnotemark[4]} &
 \multicolumn{1}{c}{FWHM \footnotemark[5]} \\

 \multicolumn{1}{c}{ } &
 \multicolumn{1}{c}{(hours)} &
 \multicolumn{1}{c}{(K km s$^{-1}$)} &
 \multicolumn{1}{c}{($10^7$ \unitlpco)} &
 \multicolumn{1}{c}{($10^7$ \msun)} &
 \multicolumn{1}{c}{(km s$^{-1}$)} \\

\hline
{\bf EAH01} &  0.30 &  2.58$\pm$ 0.33 &           128.6$\pm$           16.4 &           514.5$\pm$           65.7 &           309.8$\pm$           18.4 \\
{\bf EAH02} &  0.31 &  1.32$\pm$ 0.27 &            84.3$\pm$           17.4 &           337.0$\pm$           69.7 &           172.9$\pm$           29.8 \\
{\bf EAH03} &  0.20 &  3.60$\pm$ 0.39 &           158.0$\pm$           17.2 &           632.1$\pm$           68.9 &           158.7$\pm$            8.6 \\
{\bf EAH04} &  0.51 &  0.59$\pm$ 0.13 &             9.2$\pm$            2.0 &            36.7$\pm$            7.9 &           106.7$\pm$           20.3 \\
{\bf EAH05} &  0.60 &  0.97$\pm$ 0.21 &            90.7$\pm$           19.5 &           362.7$\pm$           77.9 &           278.7$\pm$           44.8 \\
EAH06 &  0.71 & $<$ 0.59 & $<$          24.84 & $<$          99.34 &...\\
EAH07 &  1.02 & $<$ 0.33 & $<$          10.31 & $<$          41.25 &...\\
{\bf EAH08} &  0.83 &  0.45$\pm$ 0.15 &            35.6$\pm$           12.0 &           142.5$\pm$           47.8 &           233.7$\pm$           56.9 \\
{\bf EAH09} &  0.61 &  0.43$\pm$ 0.13 &             7.8$\pm$            2.4 &            31.4$\pm$            9.7 &           119.5$\pm$           28.5 \\
{\bf EAH10} &  0.43 &  0.73$\pm$ 0.20 &           181.5$\pm$           48.6 &           726.1$\pm$          194.3 &           275.1$\pm$           58.8 \\
EAH11 &  0.40 & $<$ 0.60 & $<$          38.53 & $<$         154.12 &...\\
EAH12 &  0.70 & $<$ 0.44 & $<$          65.87 & $<$         263.47 &...\\
{\bf EAH13} &  0.94 &  0.68$\pm$ 0.13 &           193.8$\pm$           37.6 &           775.2$\pm$          150.3 &           190.1$\pm$           27.8 \\
EAH14 &  0.80 & $<$ 0.53 & $<$          44.96 & $<$         179.85 &...\\
EAH15 &  0.60 & $<$ 0.75 & $<$          27.57 & $<$         110.27 &...\\
EAH16 &  0.30 & $<$ 0.56 & $<$         155.85 & $<$         623.40 &...\\
EAH17 &  0.43 & $<$ 0.51 & $<$          25.67 & $<$         102.66 &...\\
EAS01 &  0.33 & $<$ 0.76 & $<$           6.23 & $<$          24.93 &...\\
{\bf EAS02} &  0.40 &  1.11$\pm$ 0.34 &            12.8$\pm$            3.9 &            51.2$\pm$           15.5 &           162.7$\pm$           44.0 \\
{\bf EAS03} &  0.47 &  1.66$\pm$ 0.24 &           143.6$\pm$           21.1 &           574.4$\pm$           84.5 &           270.5$\pm$           30.3 \\
EAS04 &  1.56 & $<$ 0.28 & $<$           1.38 & $<$           5.51 &...\\
{\bf EAS05} &  0.74 &  0.64$\pm$ 0.20 &            30.5$\pm$            9.5 &           121.8$\pm$           38.2 &           346.9$\pm$           82.0 \\
{\bf EAS06} &  0.31 &  3.83$\pm$ 0.39 &            42.6$\pm$            4.3 &           170.6$\pm$           17.2 &           112.9$\pm$            4.0 \\
EAS07 &  0.61 & $<$ 0.47 & $<$          10.82 & $<$          43.27 &...\\
EAS08 &  1.14 & $<$ 0.28 & $<$          10.06 & $<$          40.24 &...\\
{\bf EAS09} &  0.54 &  2.12$\pm$ 0.27 &            33.3$\pm$            4.3 &           133.3$\pm$           17.2 &           267.8$\pm$           20.5 \\
EAS10 &  0.80 & $<$ 0.49 & $<$          15.53 & $<$          62.12 &...\\
EAS11 &  0.40 & $<$ 0.51 & $<$          17.20 & $<$          68.80 &...\\
{\bf EAS12} &  1.10 &  0.36$\pm$ 0.12 &             8.7$\pm$            2.8 &            34.9$\pm$           11.3 &           141.7$\pm$           29.1 \\
EAS13 &  0.39 & $<$ 0.70 & $<$          32.75 & $<$         131.01 &...\\
{\bf EAS14} &  1.61 &  0.83$\pm$ 0.17 &           124.6$\pm$           25.3 &           498.5$\pm$          101.1 &           428.7$\pm$           73.4 \\
{\bf EAS15} &  0.74 &  0.48$\pm$ 0.15 &            29.9$\pm$            9.5 &           119.8$\pm$           38.0 &           166.2$\pm$           57.2 \\

\hline
\multicolumn{6}{l}{%
  \begin{minipage}{12cm}%
    Notes: \footnotetext[1]{Lines in {\bf bold} represent $>3\sigma$ detections in IRAM 30m CO (1--0) observations.}
\footnotetext[2]{Upper limits are shown at the $3\sigma$ level.}
\footnotetext[3]{Calculated using $L^\prime_{\rm CO} =   23.5 \: \Omega_{b} \:  D_L^2 \: I_{CO}\:  (1+z)^{-3}$.}
\footnotetext[4]{Masses calculated assuming \alphaco$=4$\unitxco, $M(\rm H_2) = \alpha_{\rm CO} L^\prime_{\rm CO}$.}
\footnotetext[5]{FWHM from Gaussian fit to data.}
  \end{minipage}%
}\\
\end{tabular}
\end{table}
\clearpage

\begin{table}[h!]
\centering
\caption{IRAM 30m and SMT CO (2-1) observations}
\label{table:all21}
\begin{tabular}{r r r r r r r r}
\hline
\hline
 \multicolumn{1}{c}{Target} &
 \multicolumn{1}{c}{$t_{obs}$ \footnotemark[1]} &
 \multicolumn{1}{c}{$I_{IRAM}^{2-1}$ \footnotemark[2]} &
 \multicolumn{1}{c}{IRAM 30m FWHM \footnotemark[3]} &
 \multicolumn{1}{c}{$t_{obs}$ \footnotemark[4]} &
 \multicolumn{1}{c}{$I_{SMT}^{2-1}$ \footnotemark[2]} &
 \multicolumn{1}{c}{SMT FWHM  \footnotemark[3]} &
 \multicolumn{1}{c}{$\theta_s$ \footnotemark[5]} \\

 \multicolumn{1}{c}{} &
 \multicolumn{1}{c}{(hours)} &
 \multicolumn{1}{c}{(K km s$^{-1}$)} &
 \multicolumn{1}{c}{(km s$^{-1}$)} &
 \multicolumn{1}{c}{(hours)} &
 \multicolumn{1}{c}{(K km s$^{-1}$)} &
 \multicolumn{1}{c}{(km s$^{-1}$)} &
 \multicolumn{1}{c}{(arcsec)} \\

\hline
EAH01 &  0.30 & 4.67$\pm$0.57 & 337.7$\pm$18.4 &  3.10 & 1.12$\pm$0.27 & 275.7$\pm$31.2 & $14.6^{+2.9}_{-4.7}$\\
EAH02 &  0.26 & 1.39$\pm$0.46 & 81.6$\pm$18.2 & ... & ... & ... & ...\\
EAH03 &  0.20 & 5.60$\pm$0.61 & 144.0$\pm$6.3 &  2.85 & 0.64$\pm$0.15 & 149.5$\pm$25.2 & $3.4^{+3.9}_{-0.7}$\\
EAH04 &  0.41 & 1.51$\pm$0.23 & 125.3$\pm$17.0 &  6.60 & 0.20$\pm$0.03 & 126.5$\pm$16.9 & $6.3^{+2.4}_{-2.6}$\\
EAH05 &  0.54 & 1.21$\pm$0.30 & 266.4$\pm$54.0 &  6.00 & $< $0.72 & ... & ...\\
EAH06 &  0.71 & $< $1.53 & ... & ... & ... & ... & ...\\
EAH07 &  1.03 & $< $0.48 & ... & ... & ... & ... & ...\\
EAH08 &  0.76 & $< $1.06 & ... &  3.80 & $< $0.18 & ... & ...\\
EAH09 &  0.61 & 0.78$\pm$0.18 & 127.0$\pm$24.2 &  9.50 & $< $0.20 & ... & ...\\
EAH10 &  0.43 & $< $1.83 & ... & ... & ... & ... & ...\\
EAH11 &  0.40 & $< $0.66 & ... & ... & ... & ... & ...\\
EAH12 &  0.70 & $< $1.15 & ... & ... & ... & ... & ...\\
EAH13 &  0.94 & $< $1.17 & ... & ... & ... & ... & ...\\
EAH14 &  0.80 & $< $0.98 & ... & ... & ... & ... & ...\\
EAH15 &  0.60 & $< $1.14 & ... & ... & ... & ... & ...\\
EAH16 &  0.32 & $< $2.05 & ... & ... & ... & ... & ...\\
EAH17 &  0.43 & $< $1.69 & ... & ... & ... & ... & ...\\
EAS01 &  0.34 & $< $0.81 & ... & ... & ... & ... & ...\\
EAS02 &  0.40 & 2.31$\pm$0.70 & 202.2$\pm$62.5 &  4.60 & 0.38$\pm$0.09 & 131.2$\pm$16.9 & $9.0^{+4.1}_{-4.7}$\\
EAS03 &  0.47 & 1.84$\pm$0.43 & 71.0$\pm$18.0 &  6.80 & $< $0.28 & ... & ...\\
EAS04 &  1.61 & $< $0.47 & ... & ... & ... & ... & ...\\
EAS05 &  0.75 & $< $0.81 & ... &  4.60 & $< $0.77 & ... & ...\\
EAS06 &  0.31 & 5.19$\pm$0.54 & 117.6$\pm$3.6 &  1.50 & 2.10$\pm$0.21 & 111.7$\pm$8.0 & $24.0^{+2.1}_{-4.5}$\\
EAS07 &  0.61 & $< $0.74 & ... & ... & ... & ... & ...\\
EAS08 &  1.00 & $< $0.61 & ... & ... & ... & ... & ...\\
EAS09 &  0.55 & 3.39$\pm$0.39 & 276.1$\pm$13.8 &  2.60 & 0.79$\pm$0.26 & 234.6$\pm$46.3 & $13.9^{+4.2}_{-5.2}$\\
EAS10 &  0.80 & $< $0.82 & ... & ... & ... & ... & ...\\
EAS11 &  0.39 & $< $0.52 & ... & ... & ... & ... & ...\\
EAS12 &  1.10 & 0.62$\pm$0.14 & ... &  5.55 & $< $0.27 & ... & ...\\
EAS13 &  0.44 & $< $0.81 & ... & ... & ... & ... & ...\\
EAS14 &  1.61 & 1.15$\pm$0.30 & 598.1$\pm$199.7 & ... & ... & ... & ...\\
EAS15 &  0.70 & 1.04$\pm$0.22 & 299.0$\pm$58.1 &  4.50 & $< $0.16 & ... & ...\\

\hline

\multicolumn{8}{l}{%
  \begin{minipage}{12cm}%
    Notes: \footnotetext[1]{Time on source (hours) IRAM 30m.}
\footnotetext[2]{Upper limits are shown at the $3\sigma$ level.}
\footnotetext[3]{FWHM from Gaussian fit to data.}
\footnotetext[4]{Time on source (hours) SMT.}
\footnotetext[5]{Approximate Gaussian source size (FWHM) of CO emitting region, see text (\S\ref{size}) for details}

  \end{minipage}%
}\\
\end{tabular}
\end{table}

\newpage
%
\begin{figure*}

\includegraphics[width=0.32\textwidth]{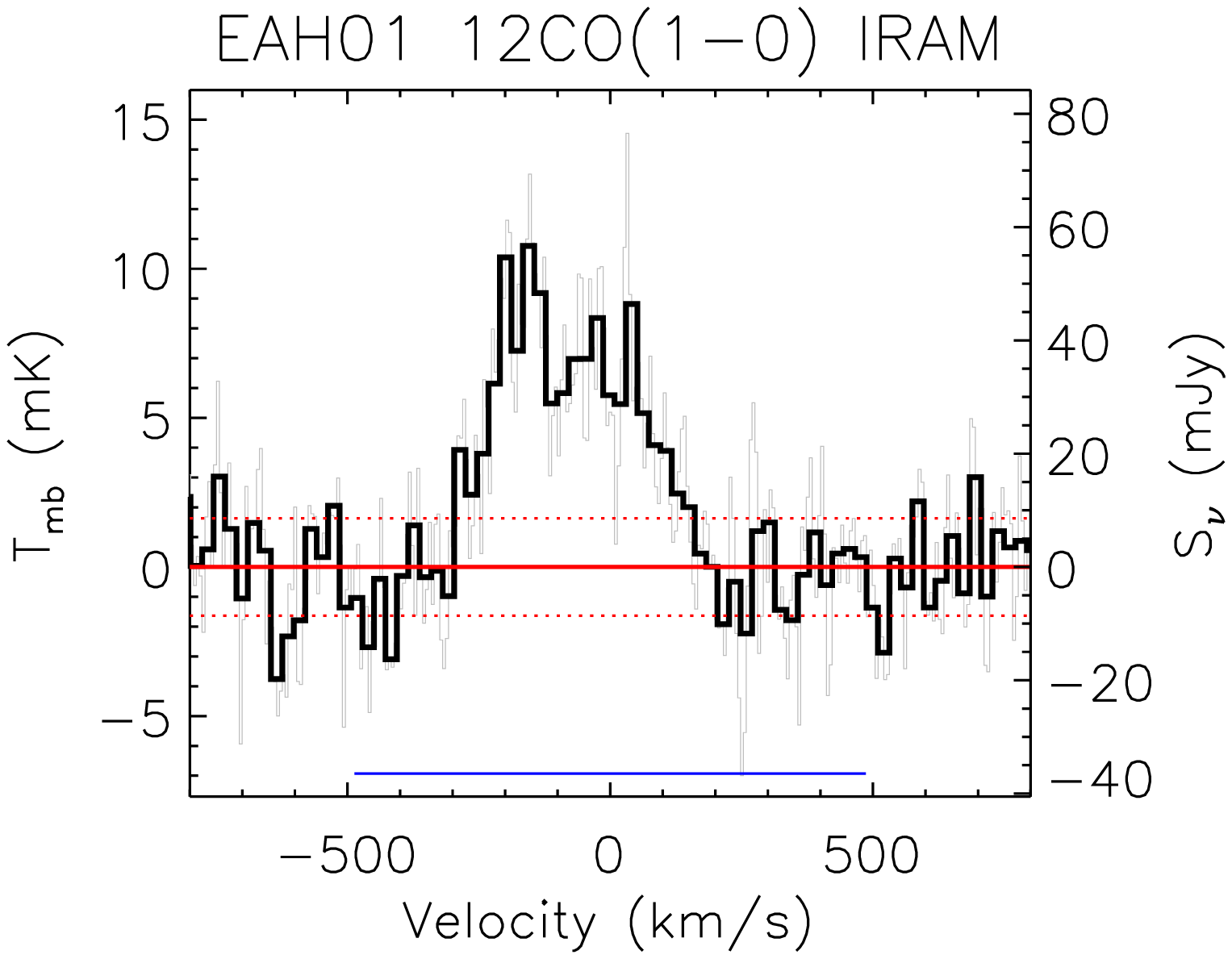}
\includegraphics[width=0.32\textwidth]{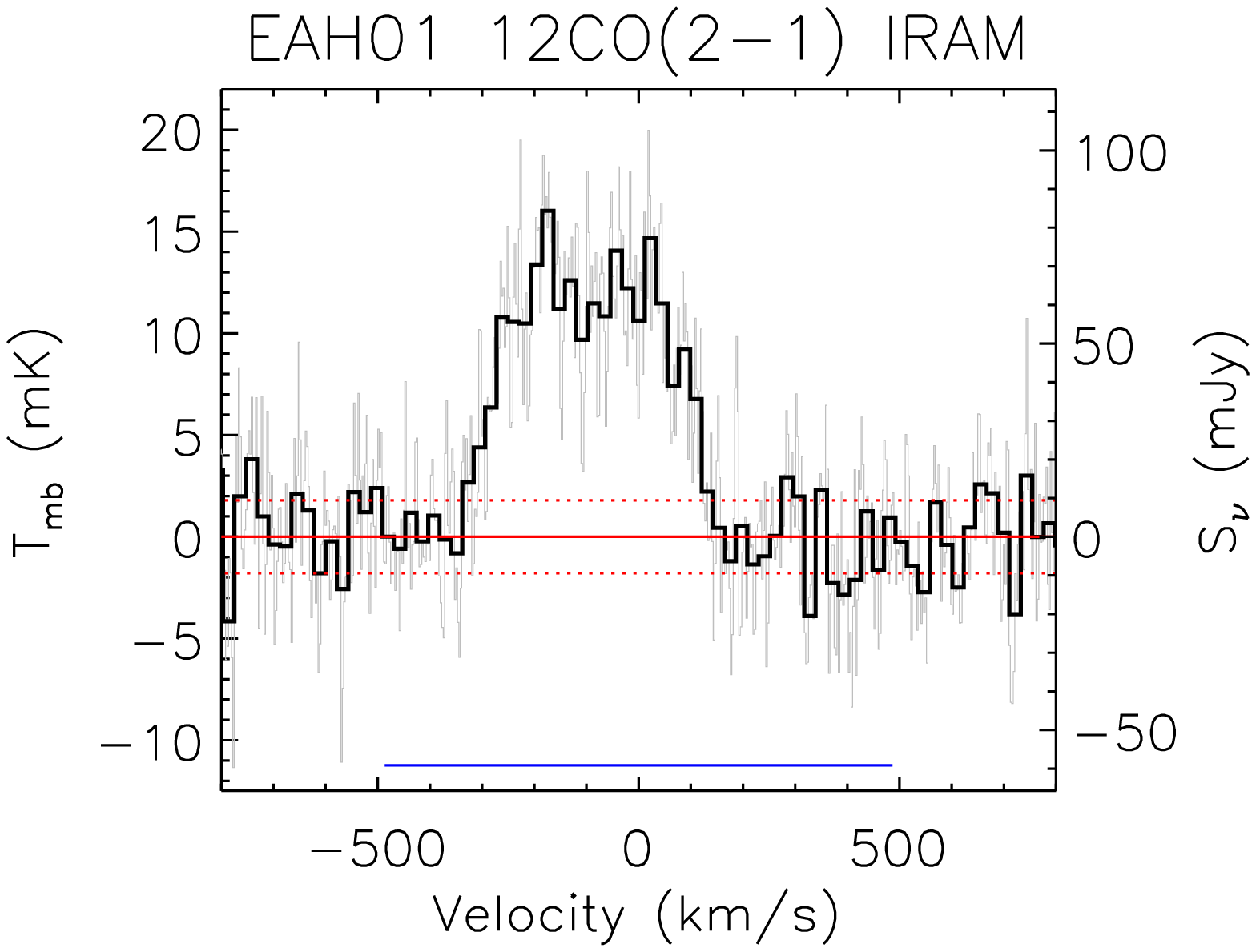}
\includegraphics[width=0.32\textwidth]{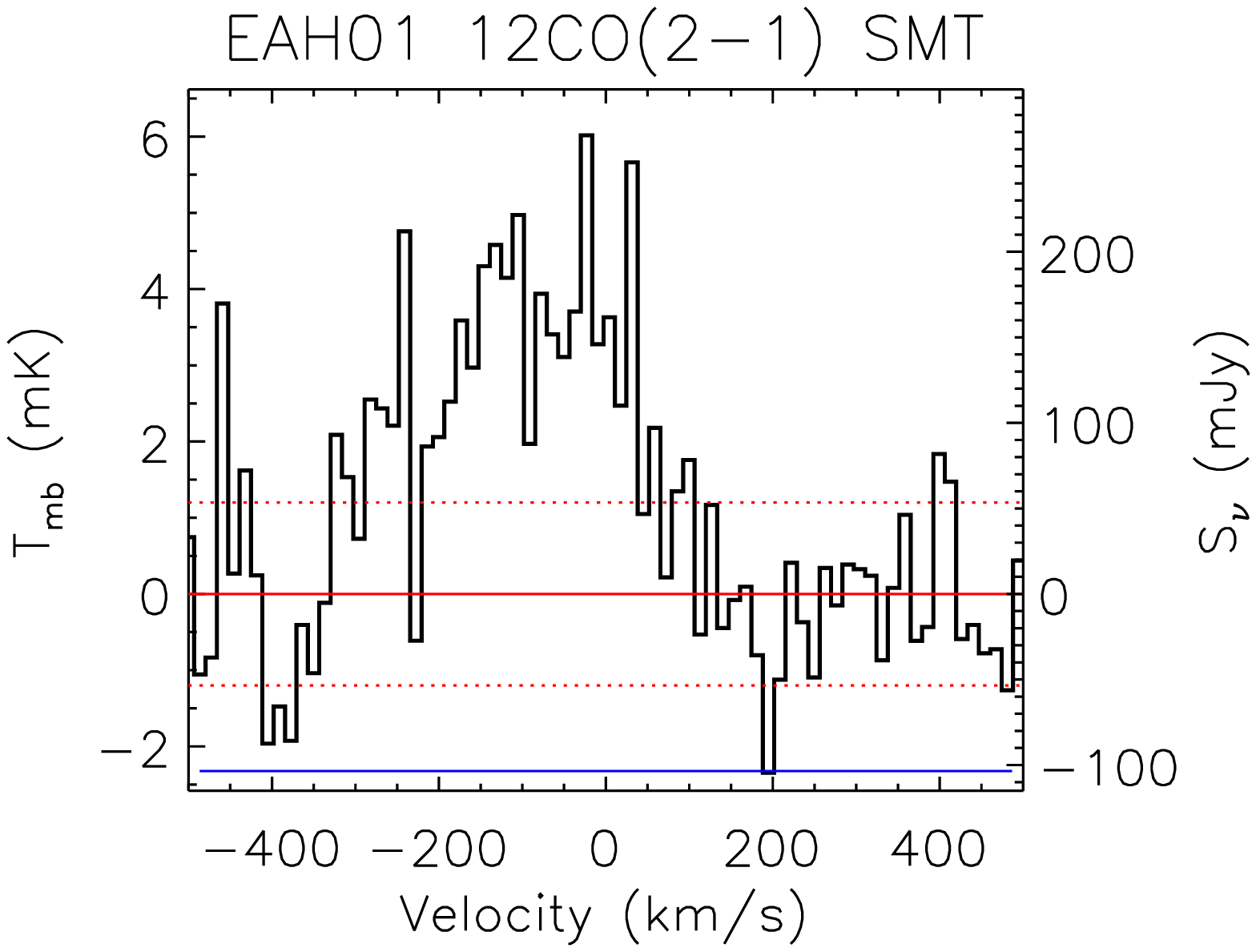}

\includegraphics[width=0.32\textwidth]{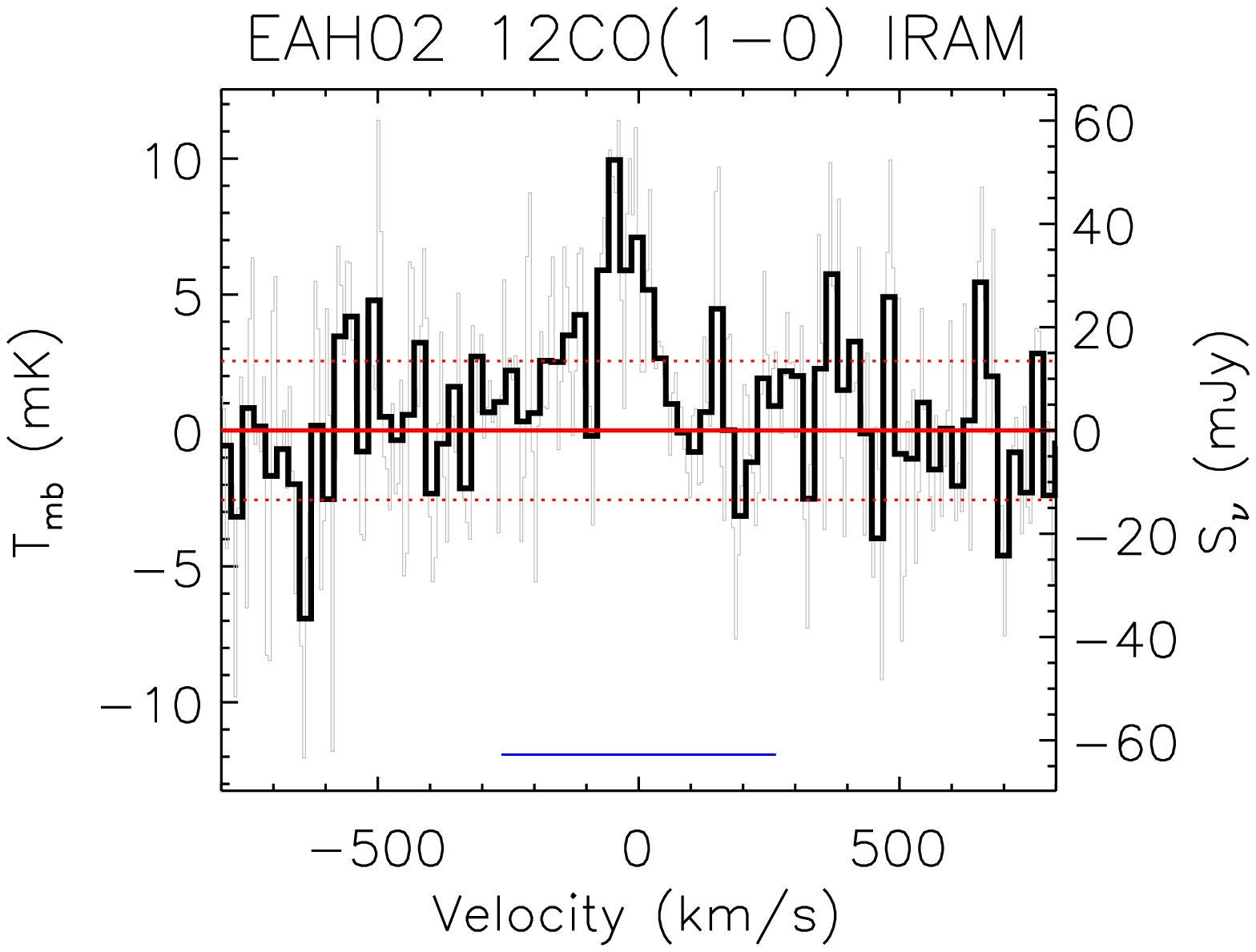}
\includegraphics[width=0.32\textwidth]{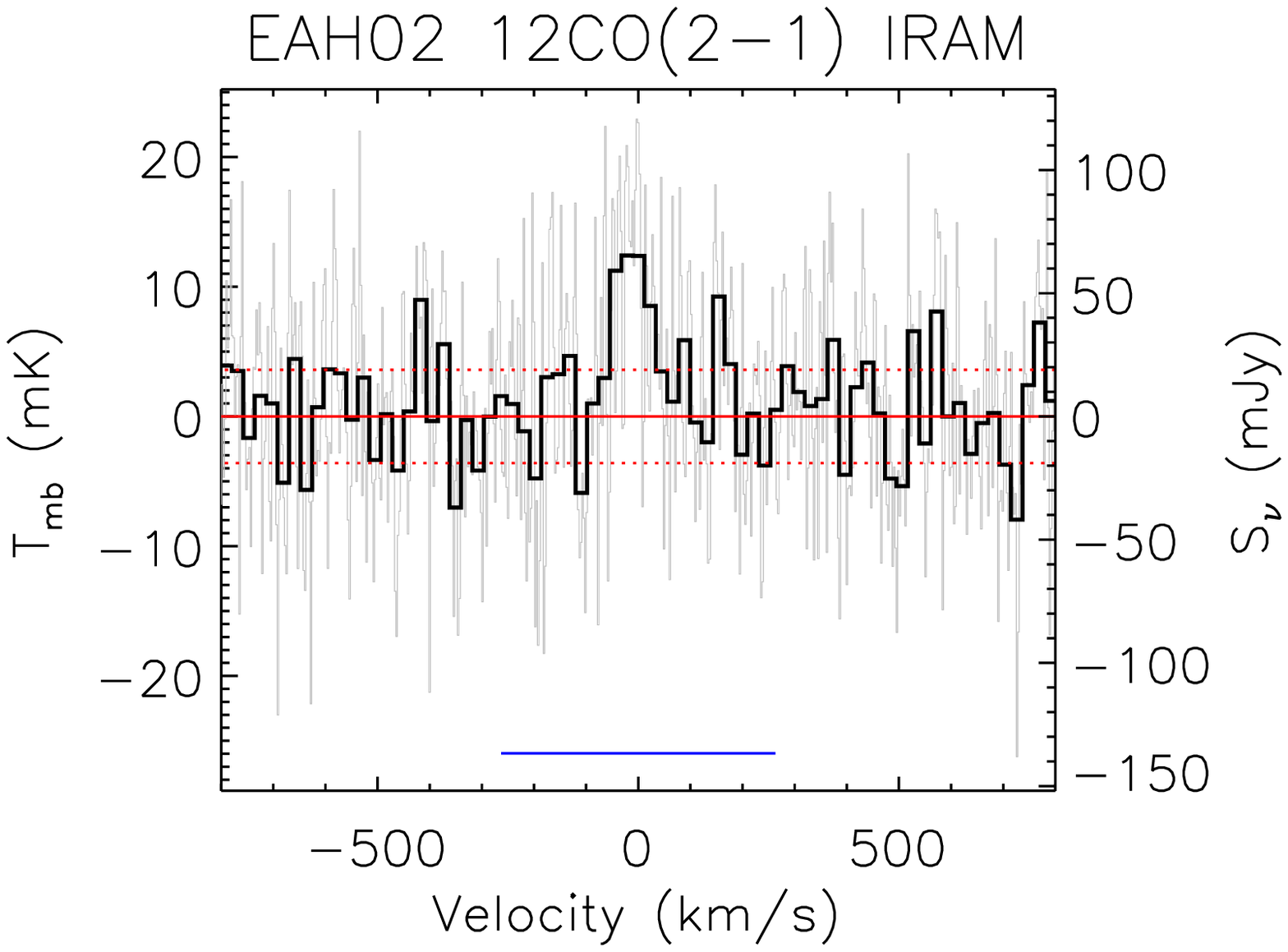}

\includegraphics[width=0.32\textwidth]{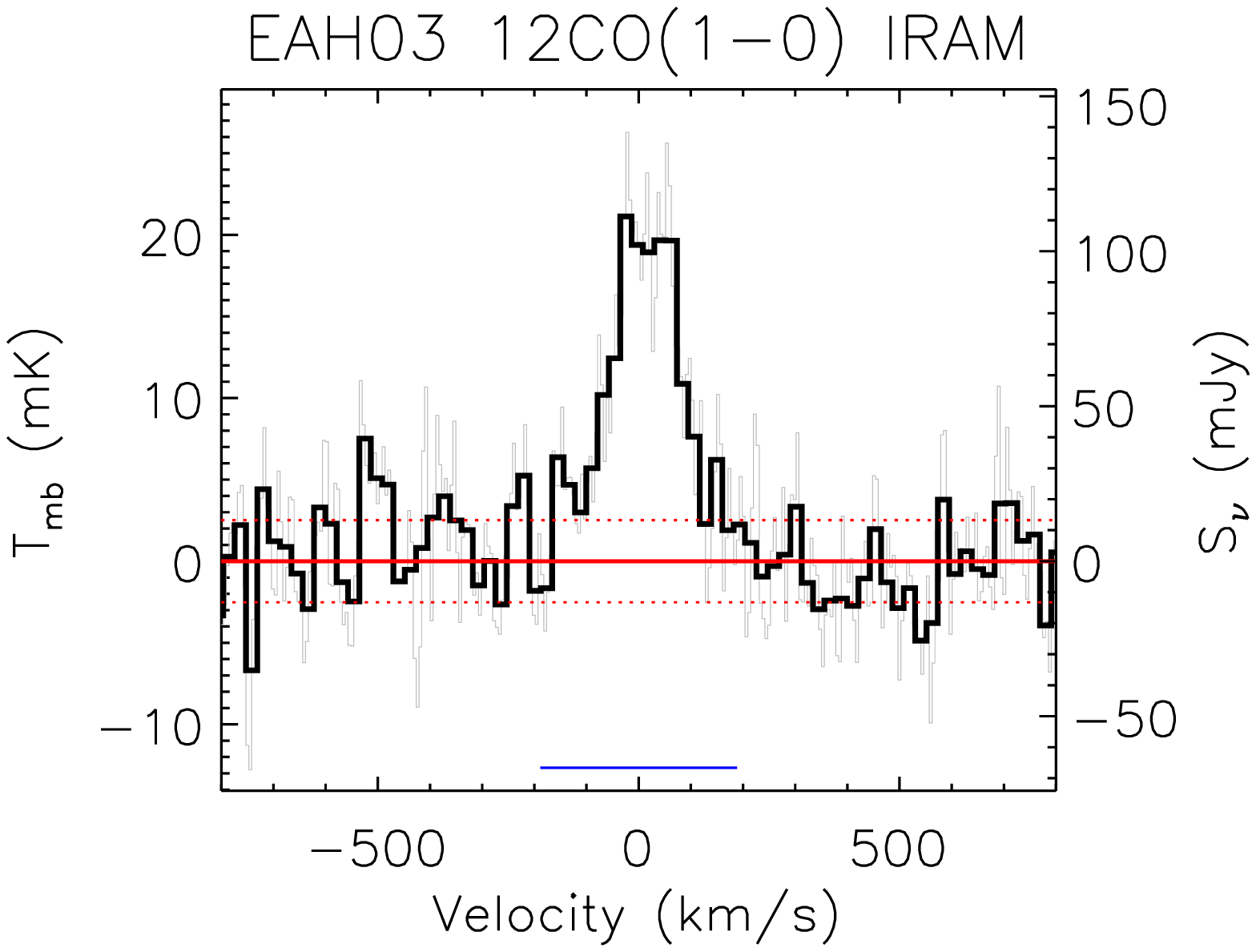}
\includegraphics[width=0.32\textwidth]{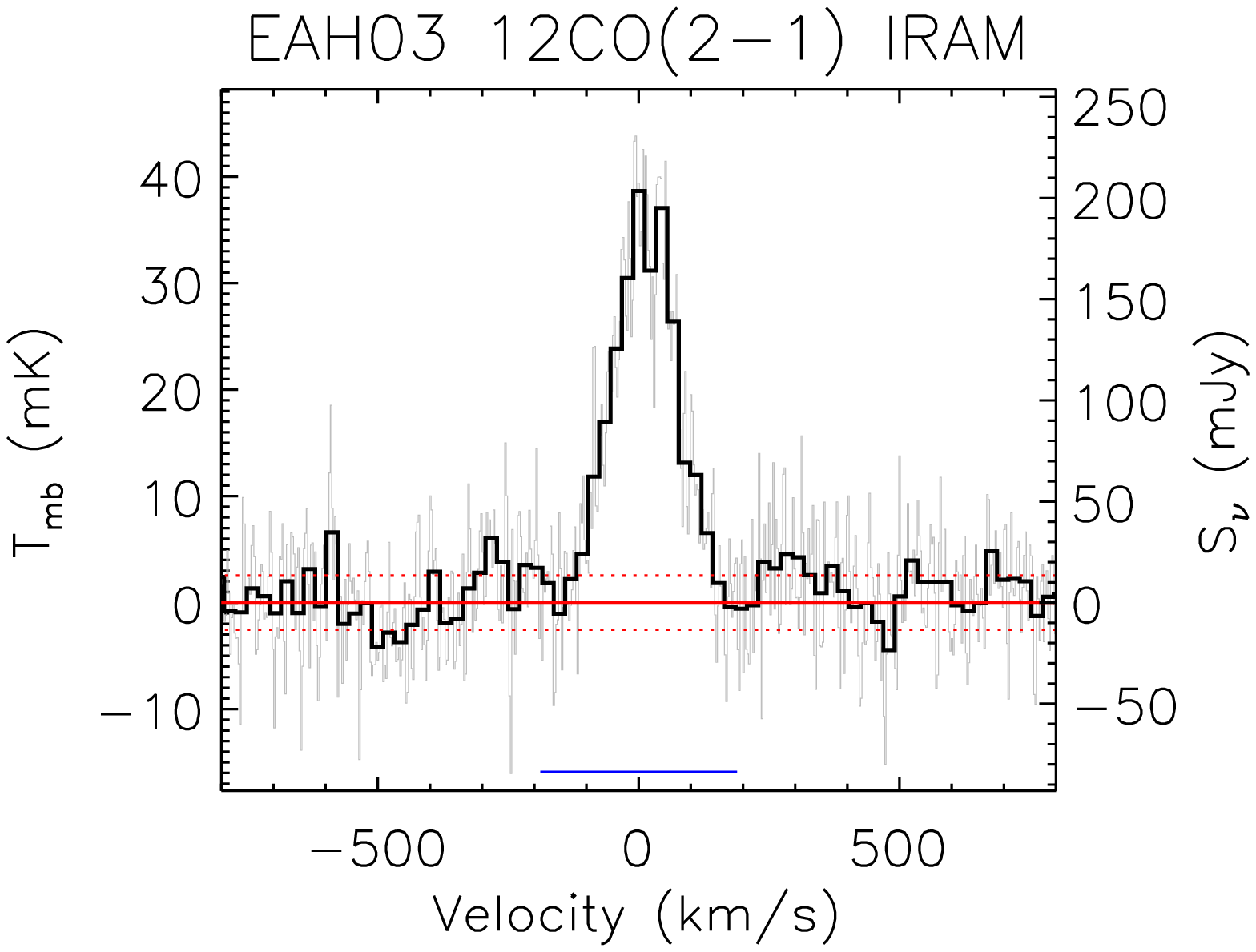}
\includegraphics[width=0.32\textwidth]{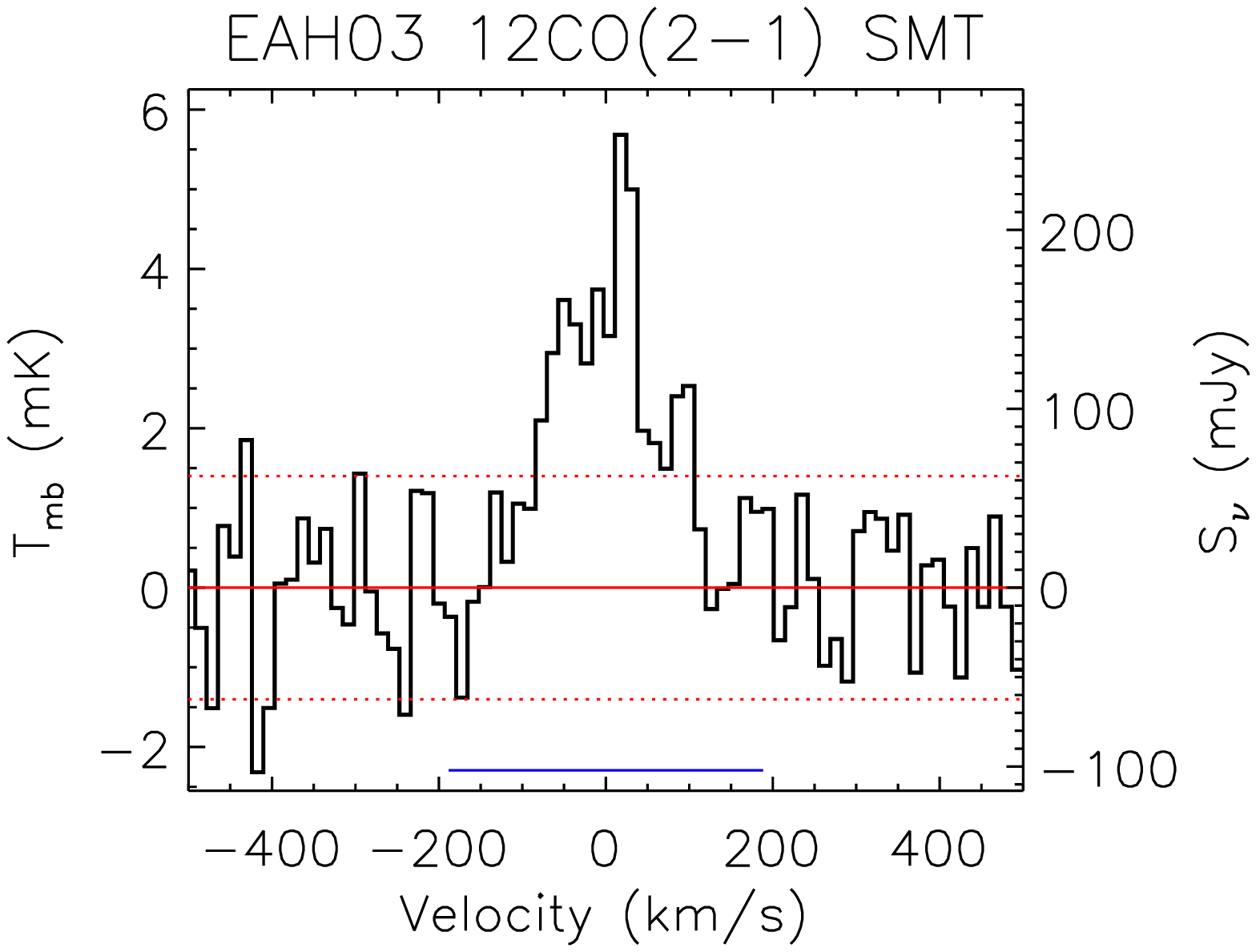}

\includegraphics[width=0.32\textwidth]{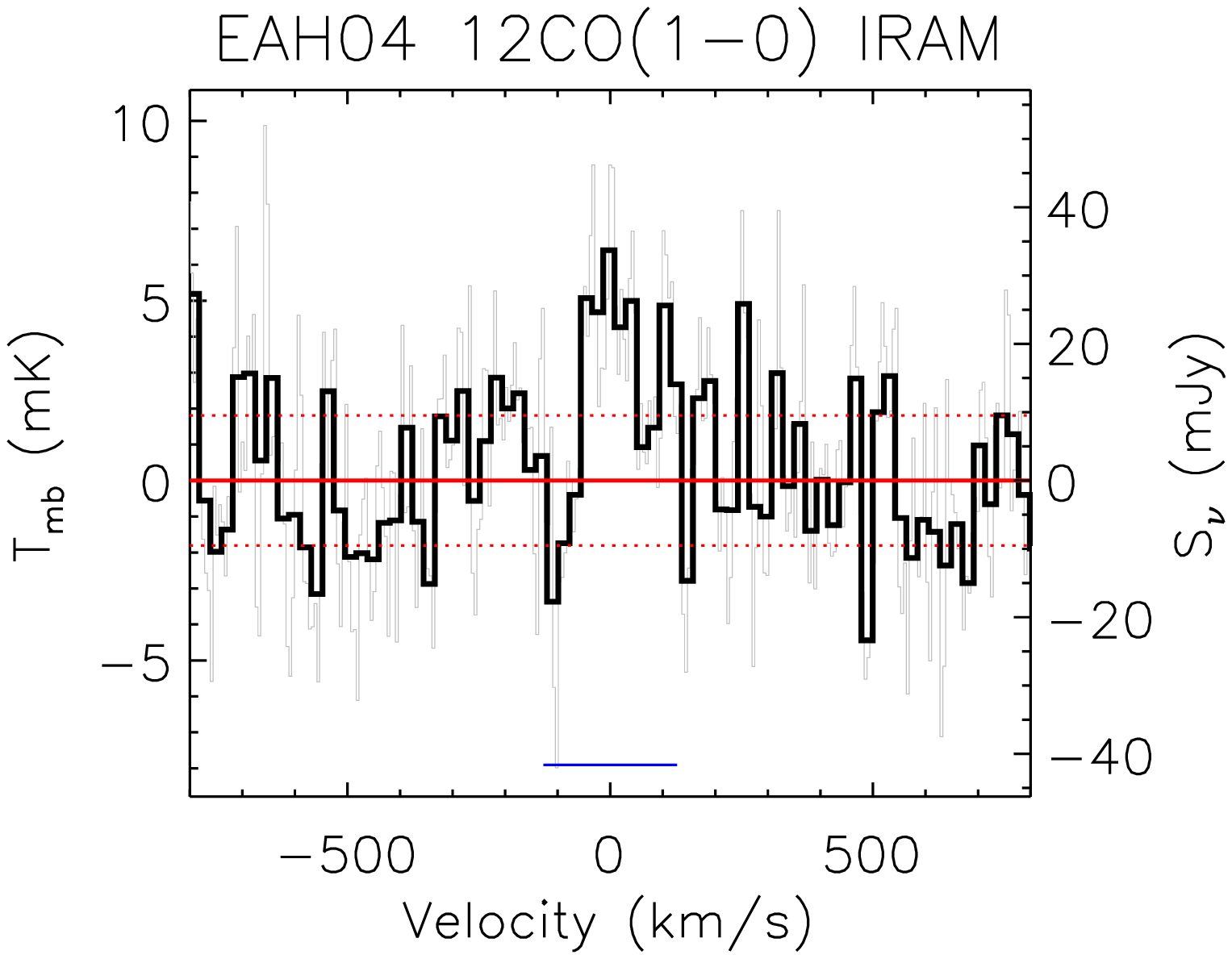}
\includegraphics[width=0.32\textwidth]{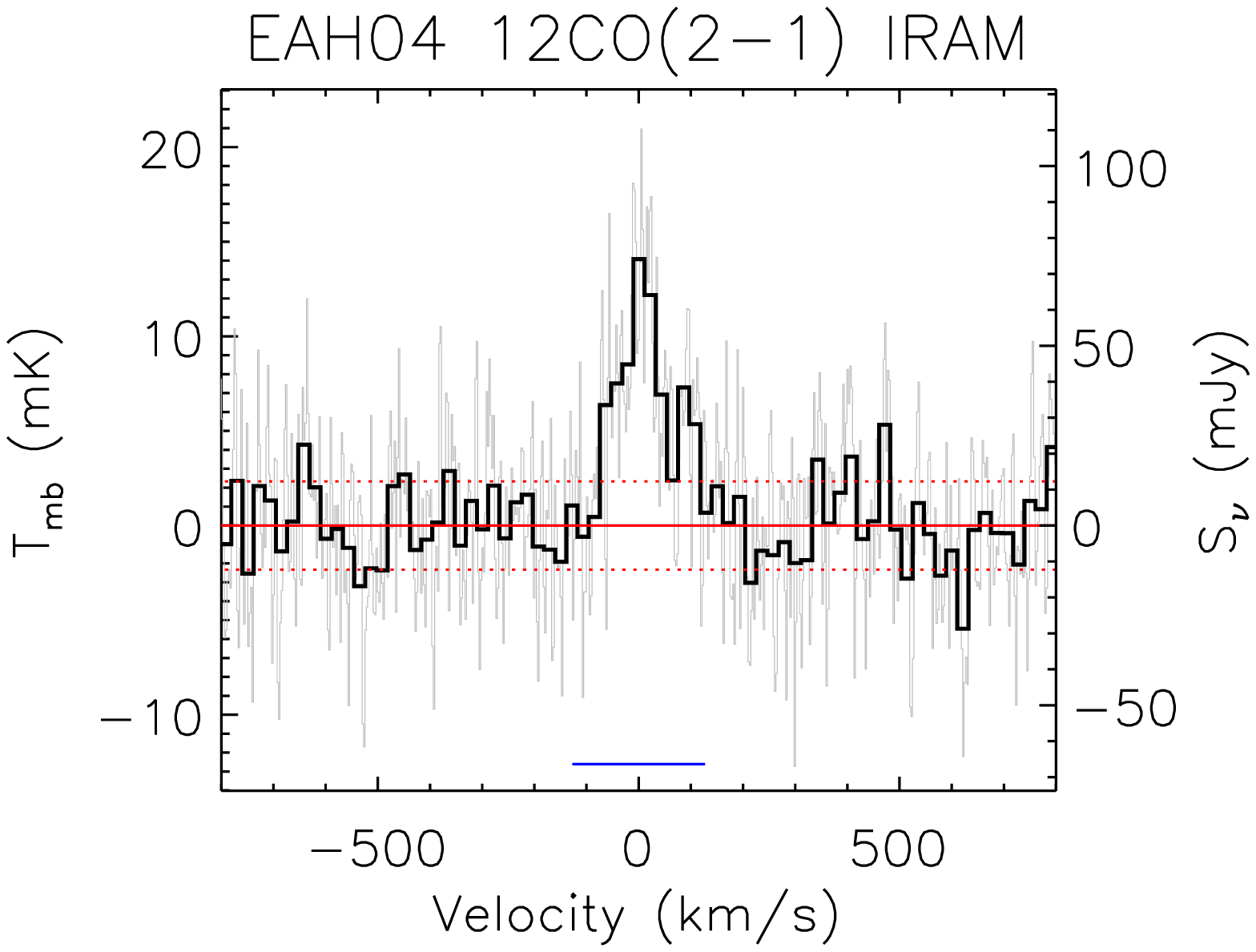}
\includegraphics[width=0.32\textwidth]{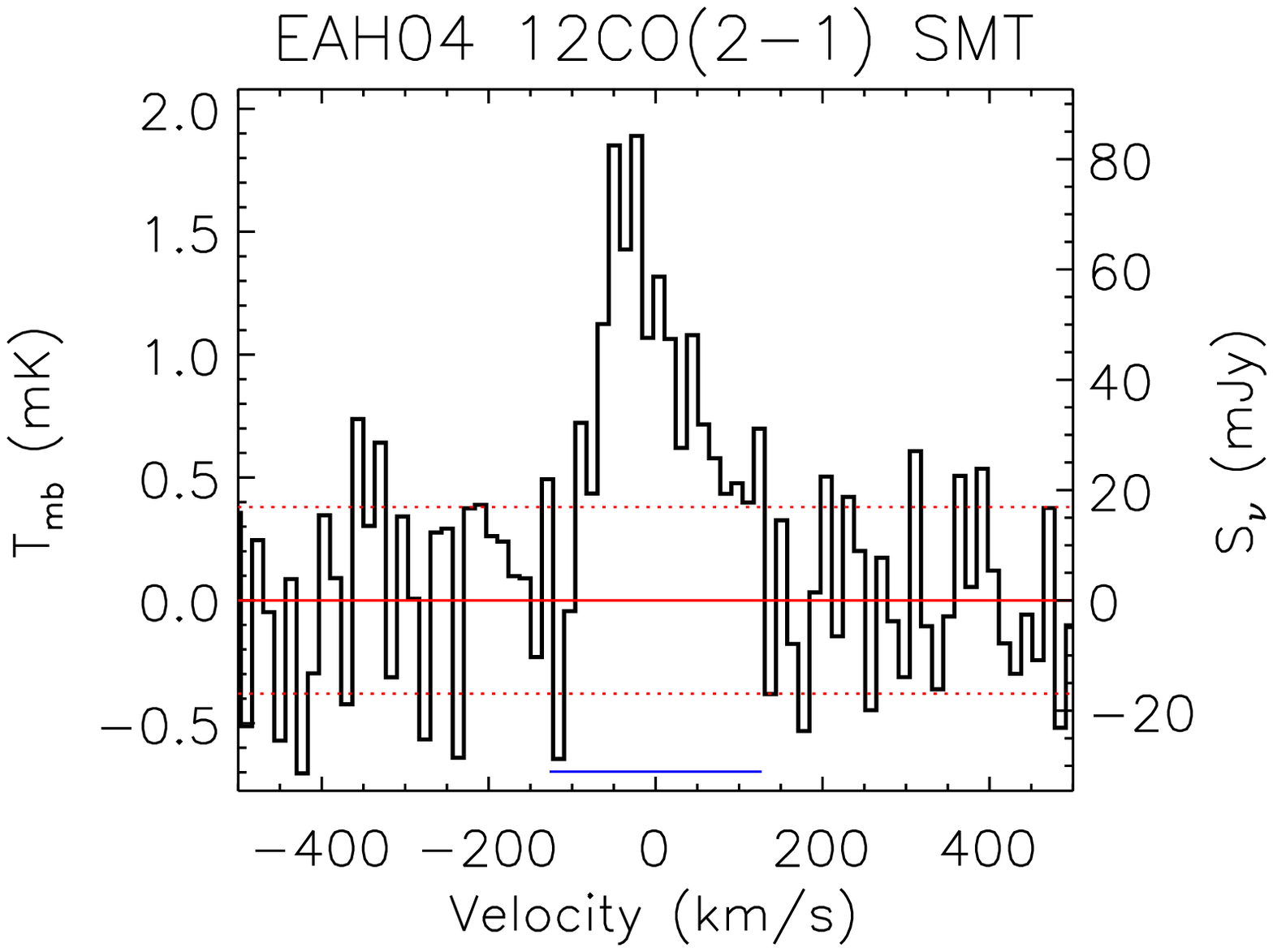}

\includegraphics[width=0.32\textwidth]{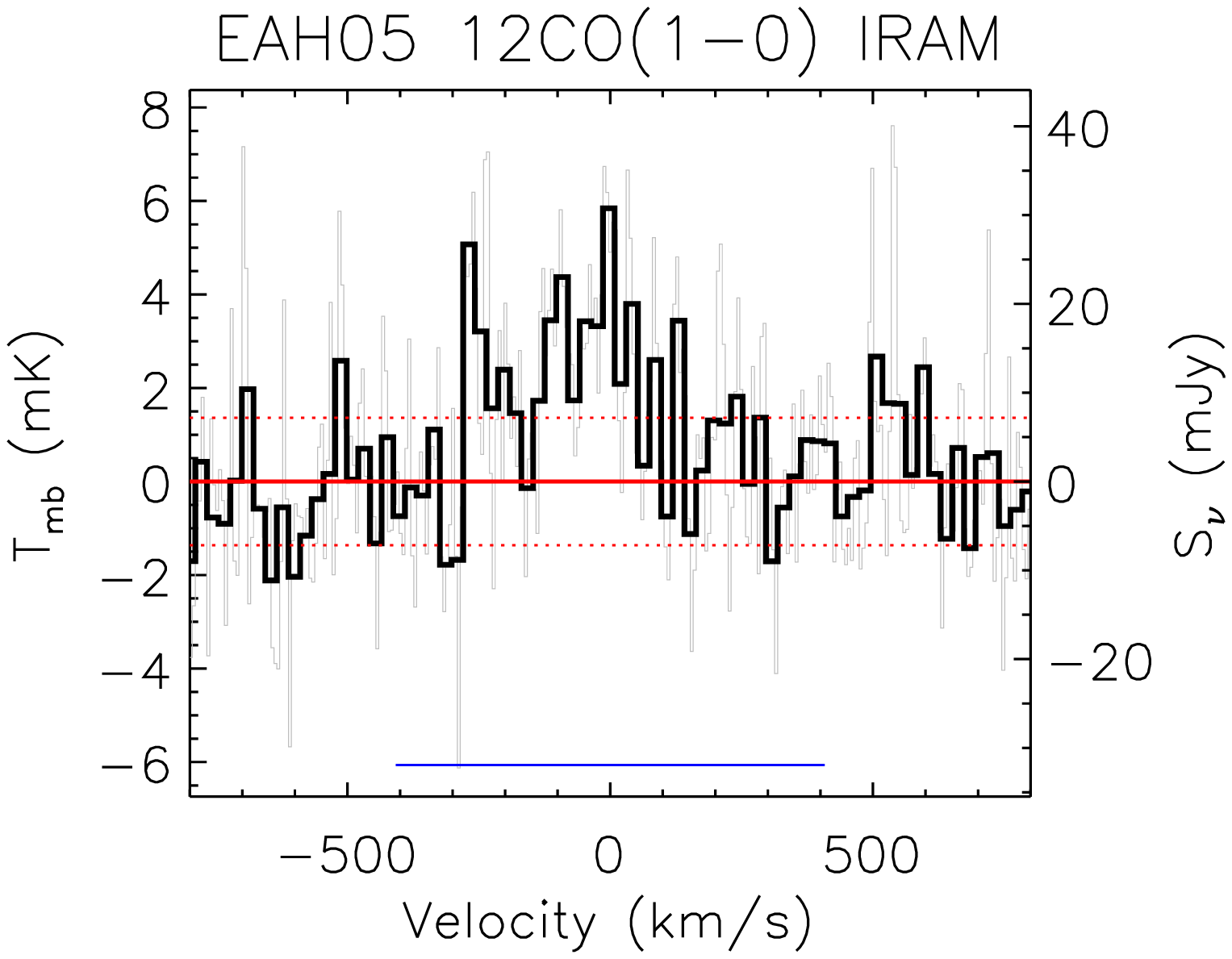}
\includegraphics[width=0.32\textwidth]{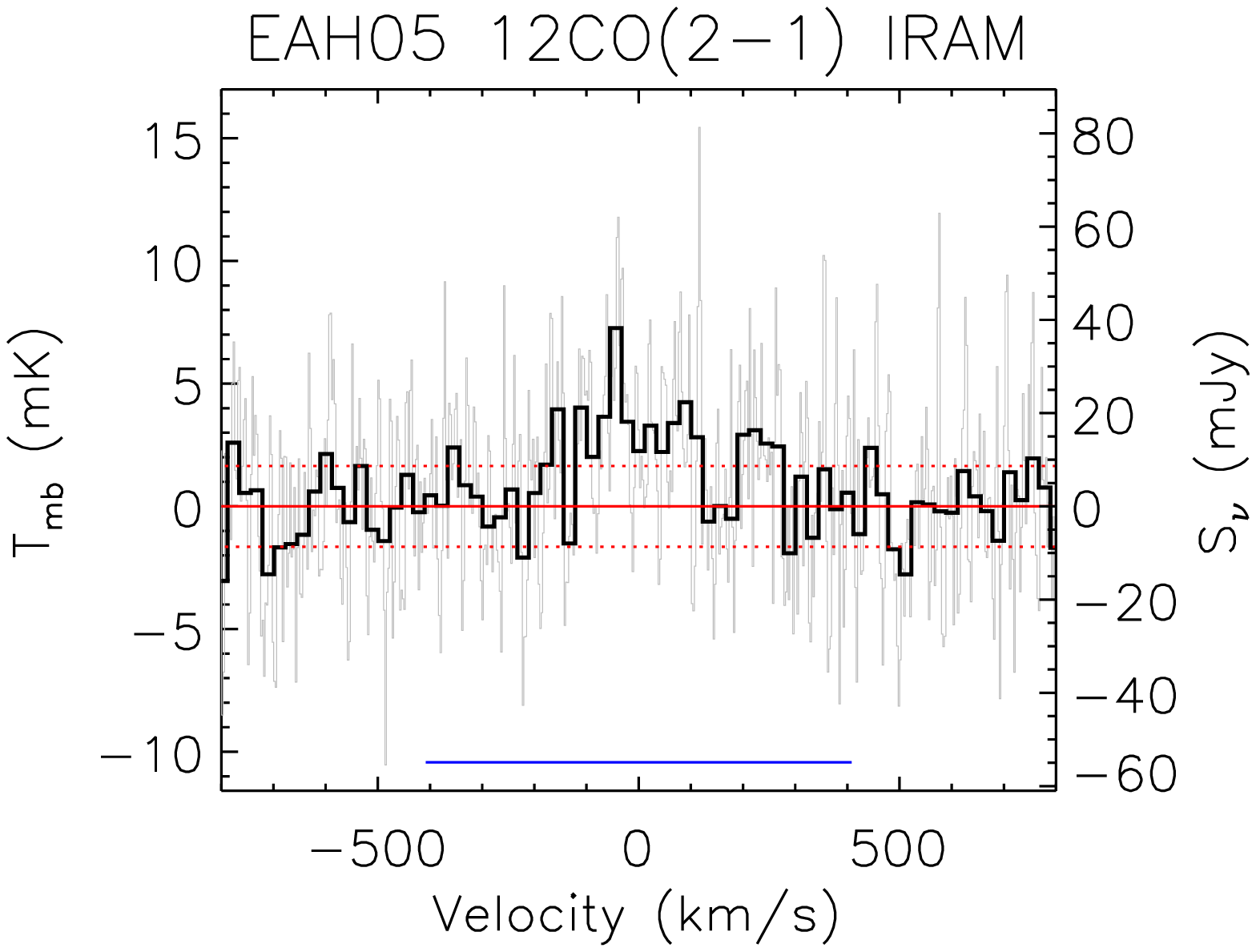}
\includegraphics[width=0.32\textwidth]{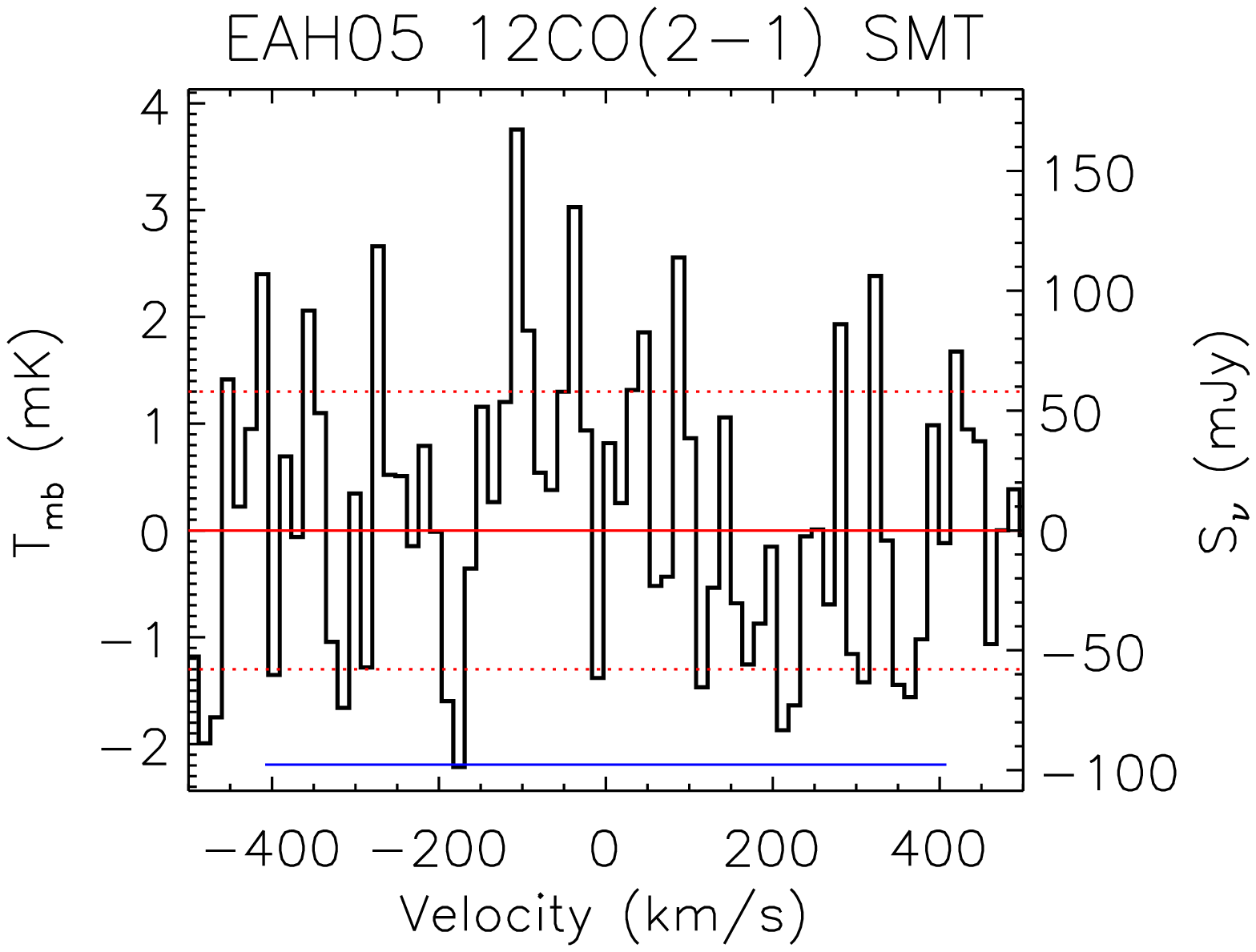}

\caption{CO(1-0) and CO(2-1) spectra from IRAM and SMT for galaxies with IRAM-30m CO (1--0) detections ($>3\sigma$) in our post-starburst sample. Spectra are shown in units of both main beam temperature T$_{mb}$ [mK] and S$_\nu$ [Jy]. Grey lines show the unbinned IRAM data for 5 km/s channels, and black lines show the data binned to 20 km/s. Dashed red lines represent the rms of the binned data. SMT data are shown in 13 km/s bins. Blue horizontal lines at bottom represent the integration intervals, as described in the text.}
\label{fig:CO1}
\end{figure*}

\begin{figure*}
\ContinuedFloat

\includegraphics[width=0.32\textwidth]{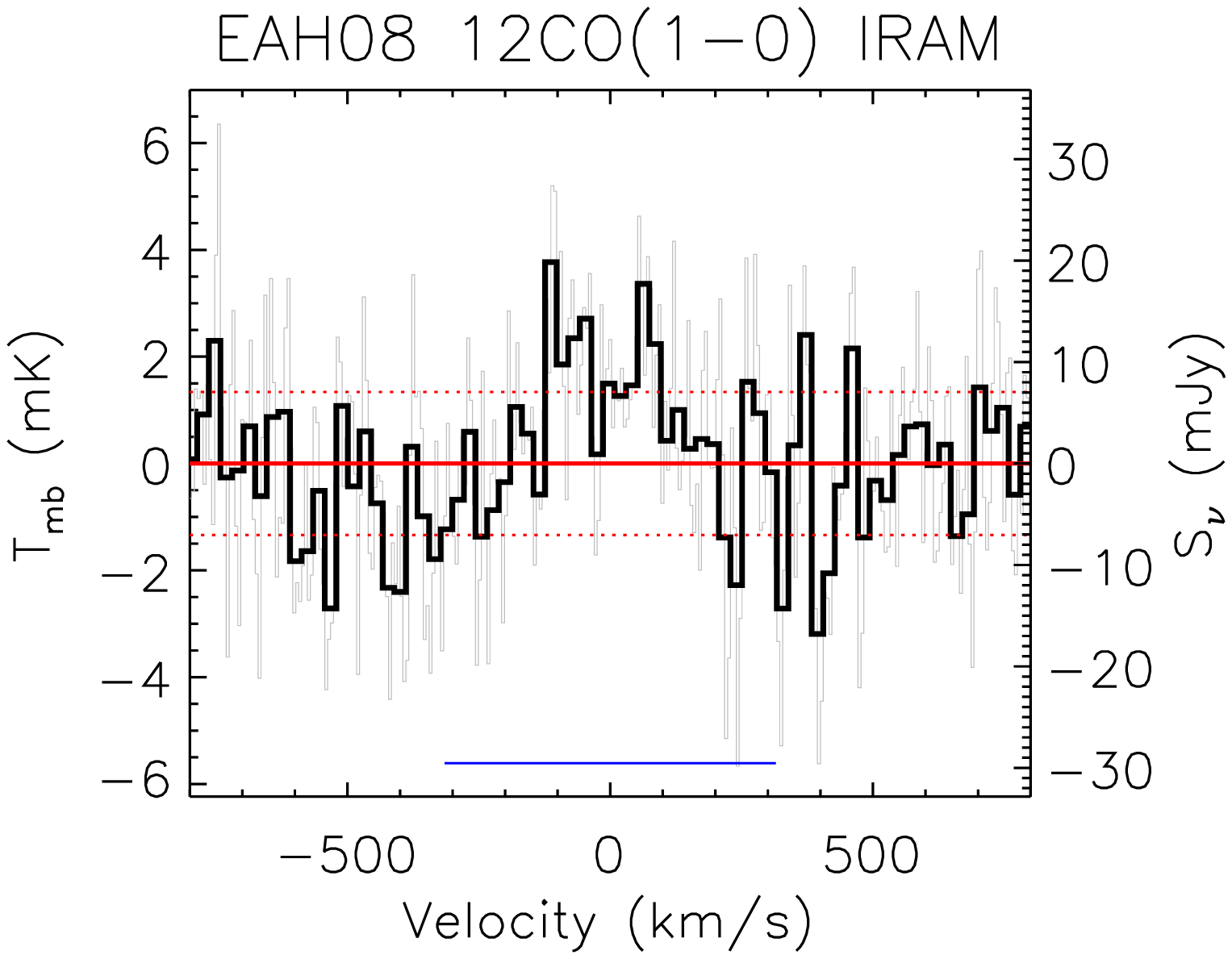}
\includegraphics[width=0.32\textwidth]{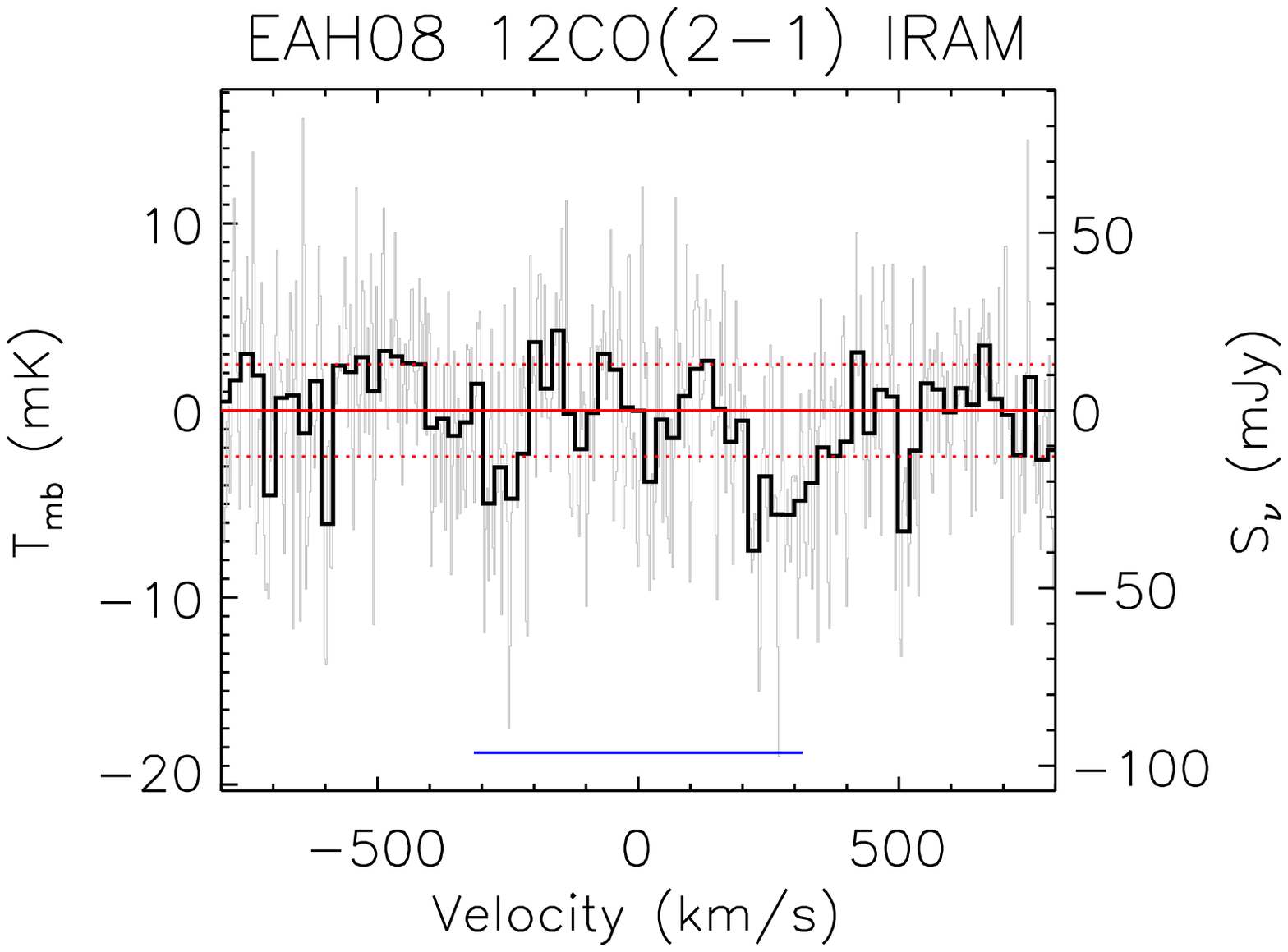}
\includegraphics[width=0.32\textwidth]{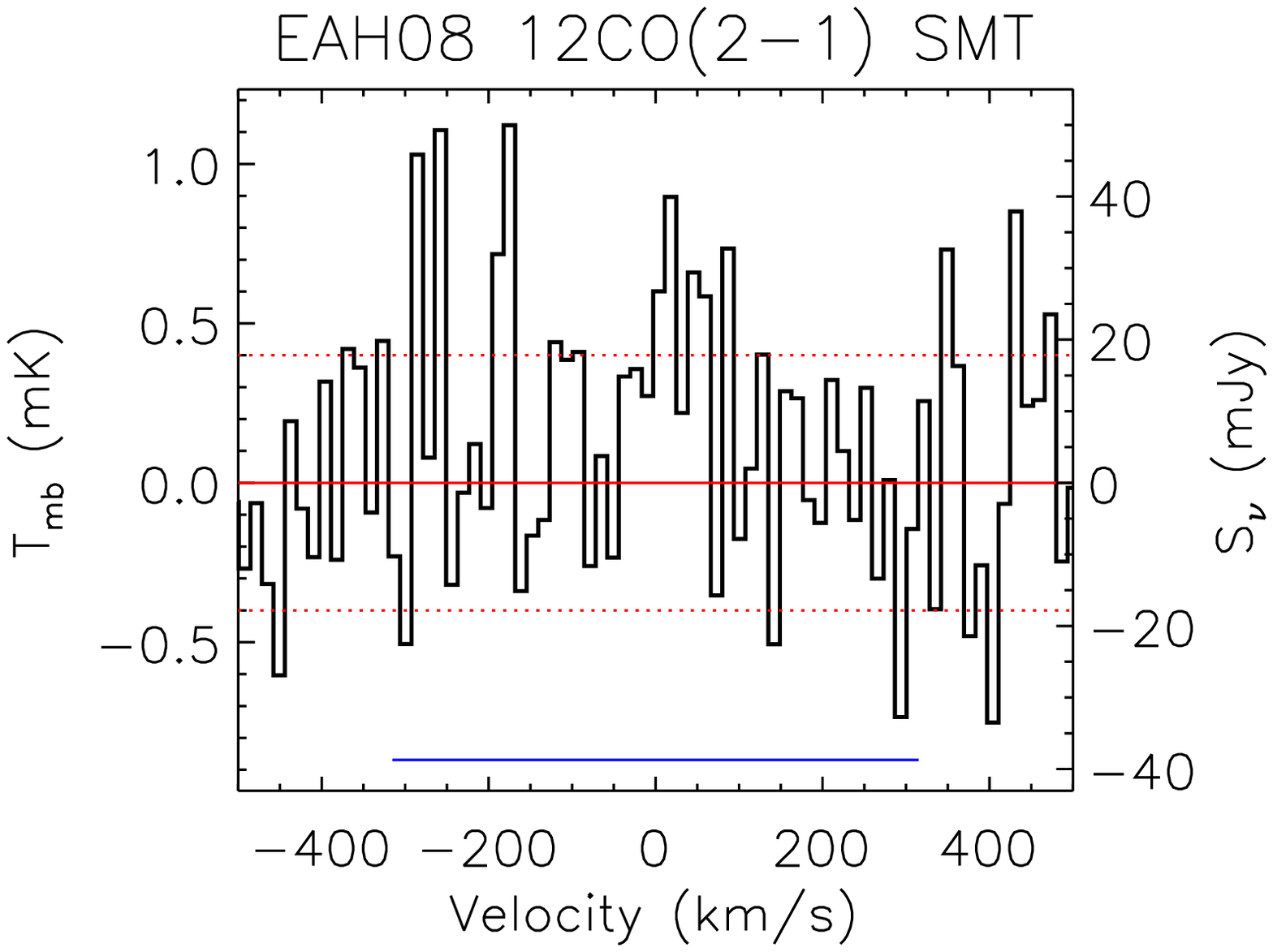}

\includegraphics[width=0.32\textwidth]{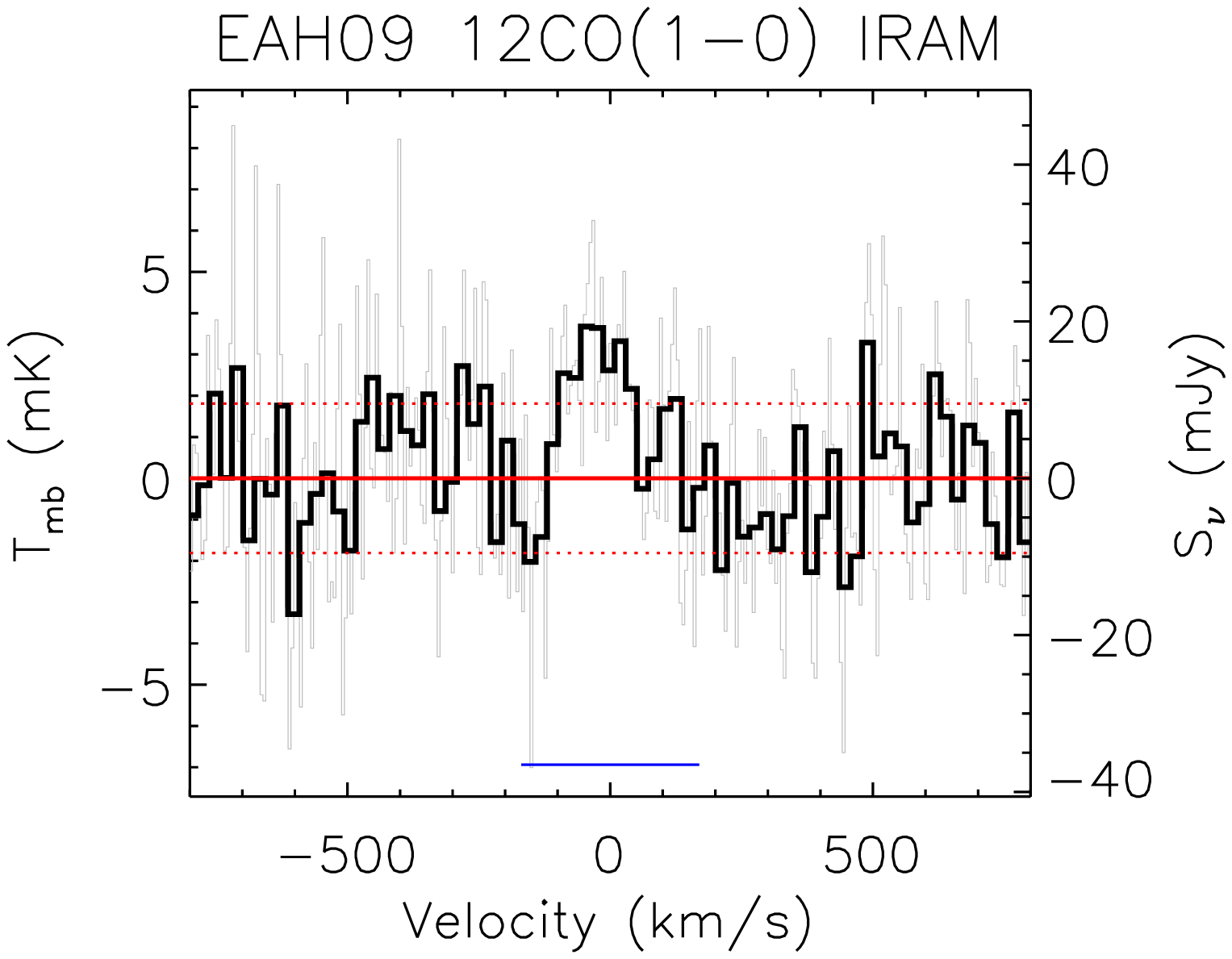}
\includegraphics[width=0.32\textwidth]{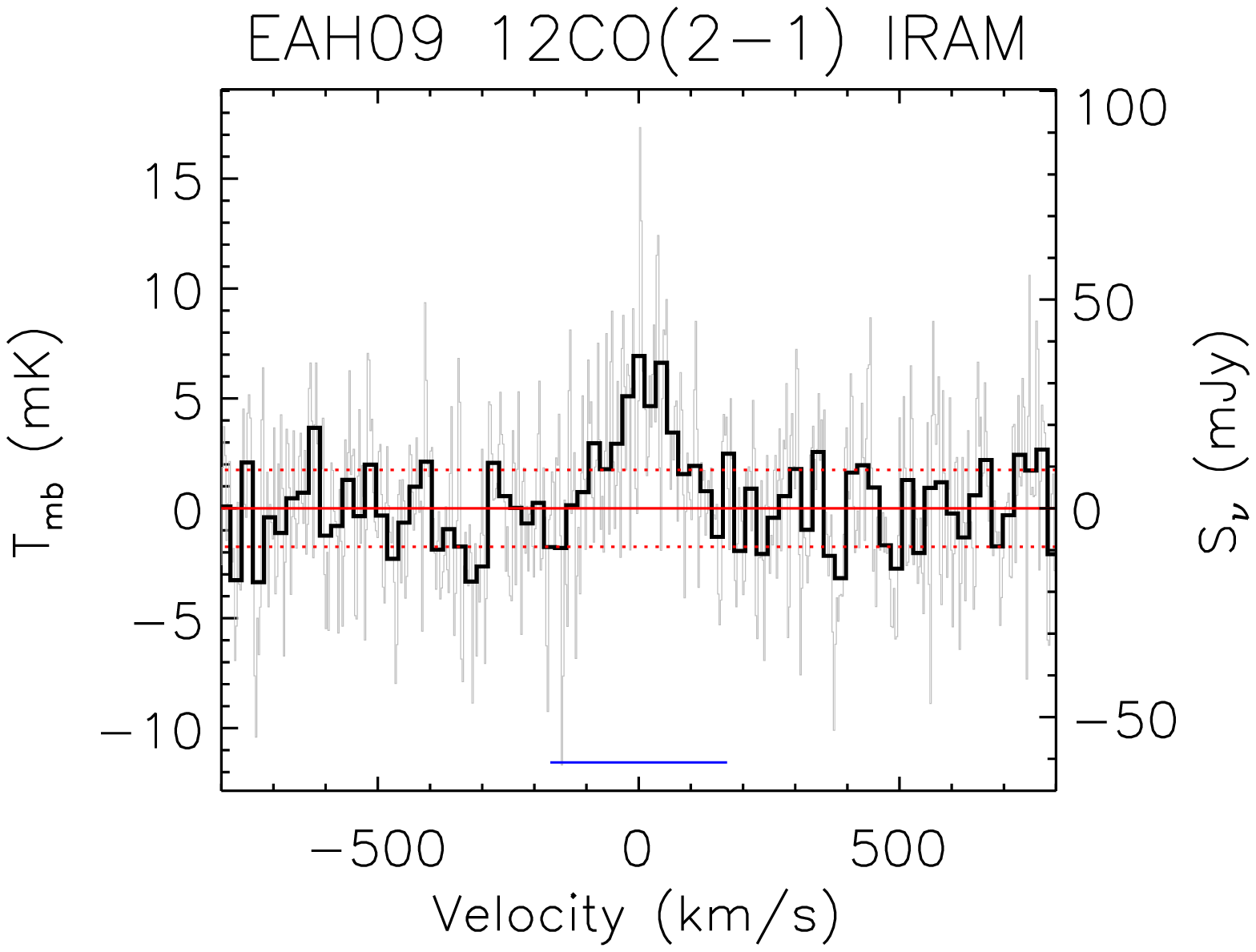}
\includegraphics[width=0.32\textwidth]{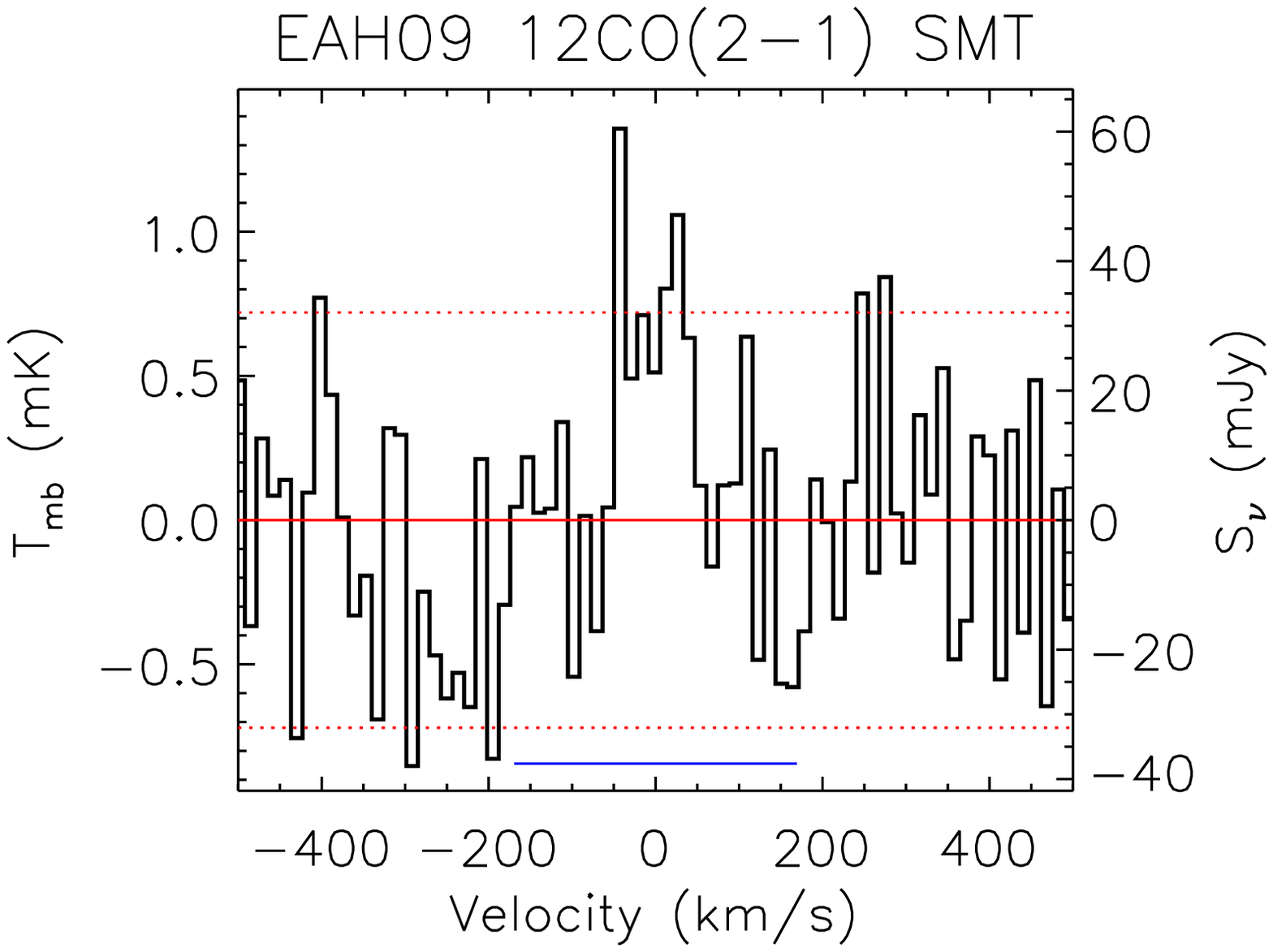}

\includegraphics[width=0.32\textwidth]{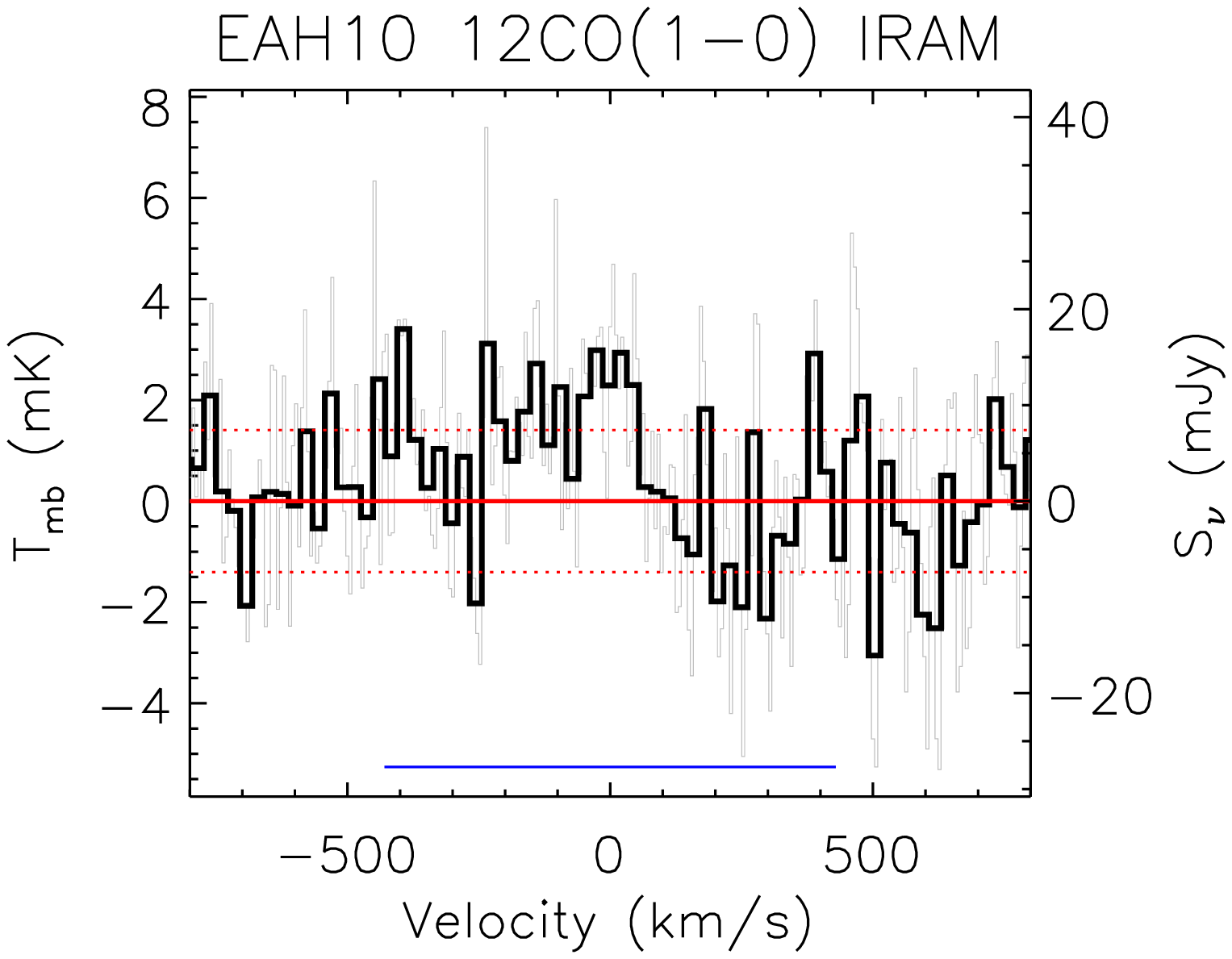}
\includegraphics[width=0.32\textwidth]{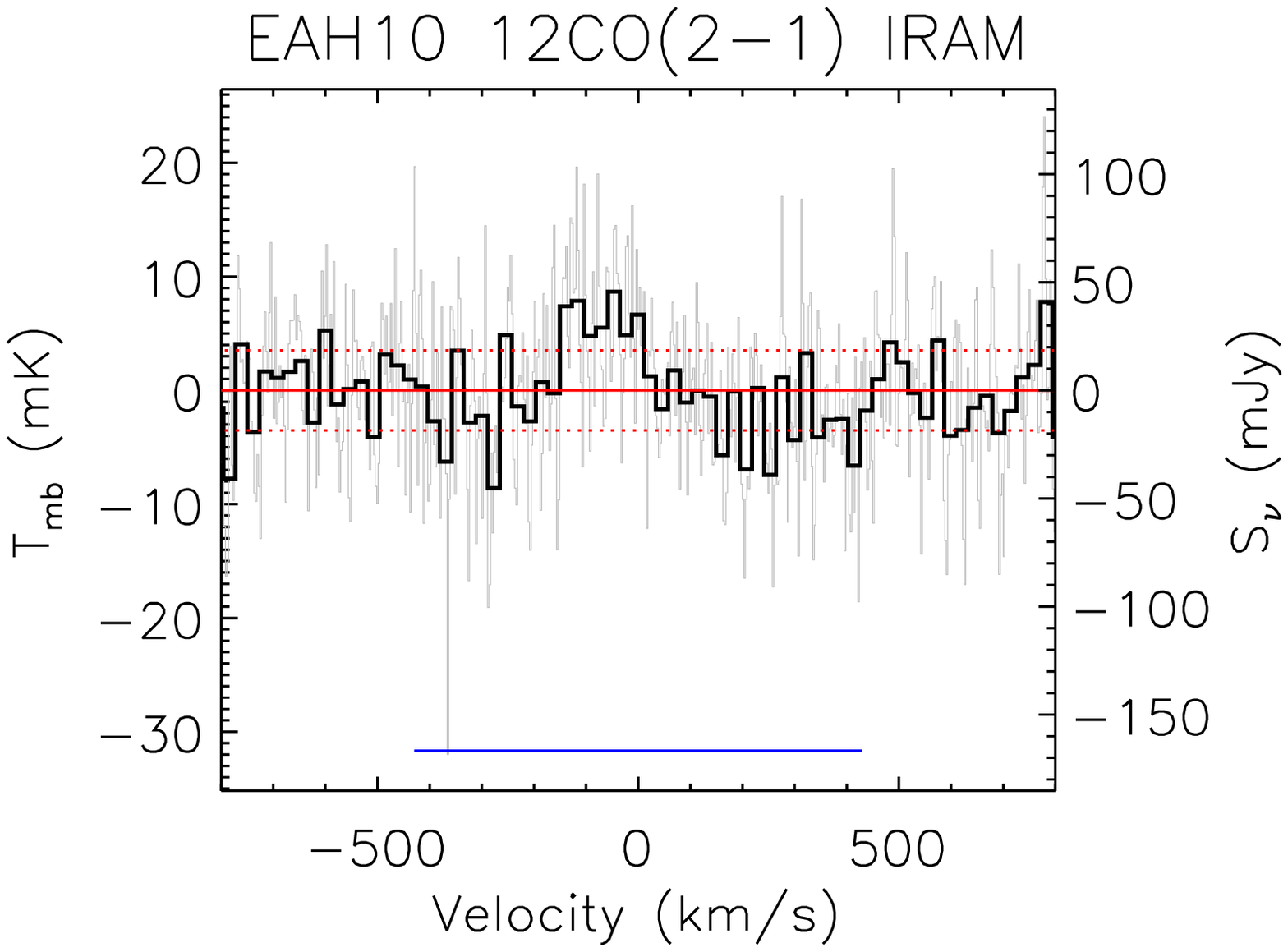}

\includegraphics[width=0.32\textwidth]{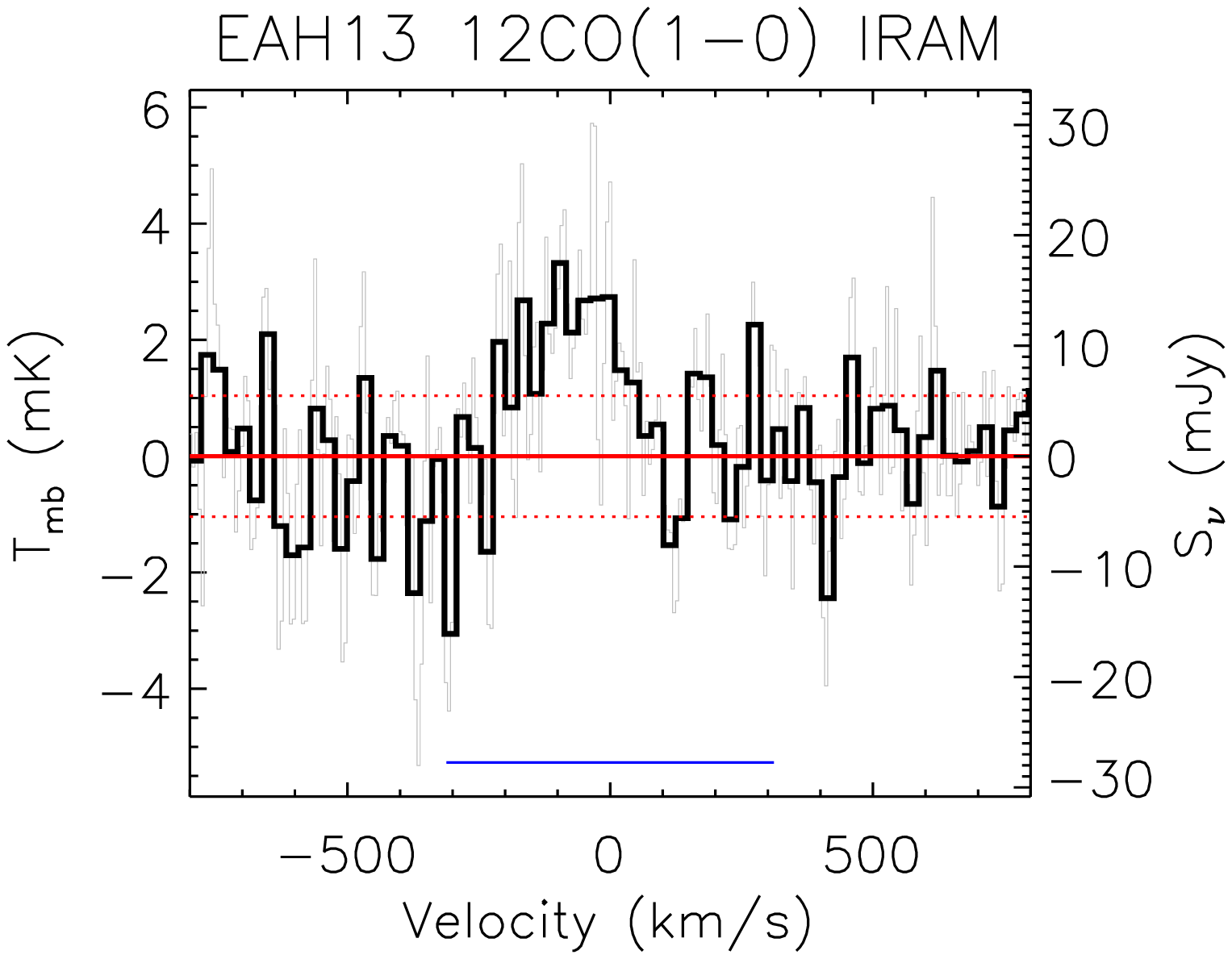}
\includegraphics[width=0.32\textwidth]{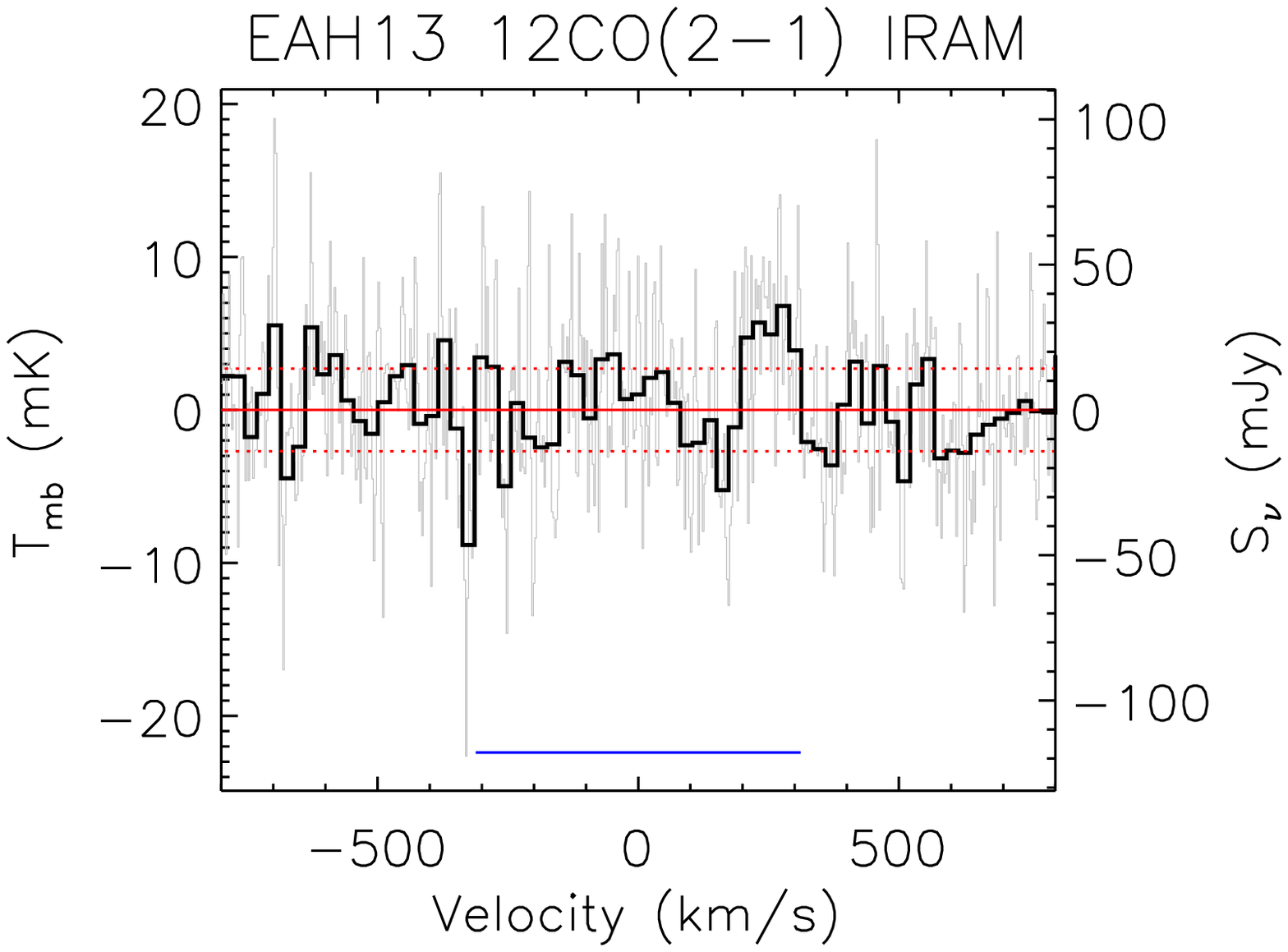}

\includegraphics[width=0.32\textwidth]{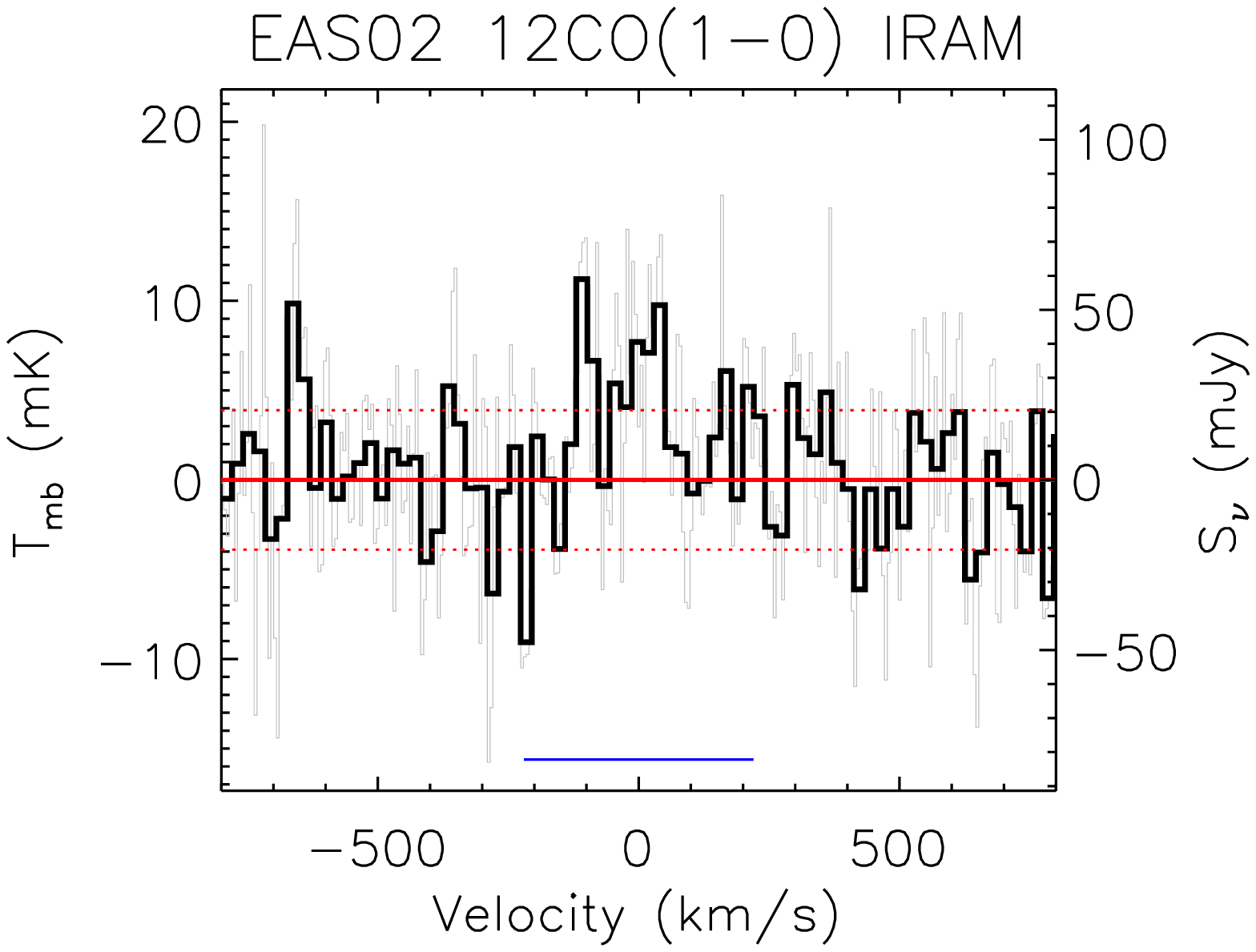}
\includegraphics[width=0.32\textwidth]{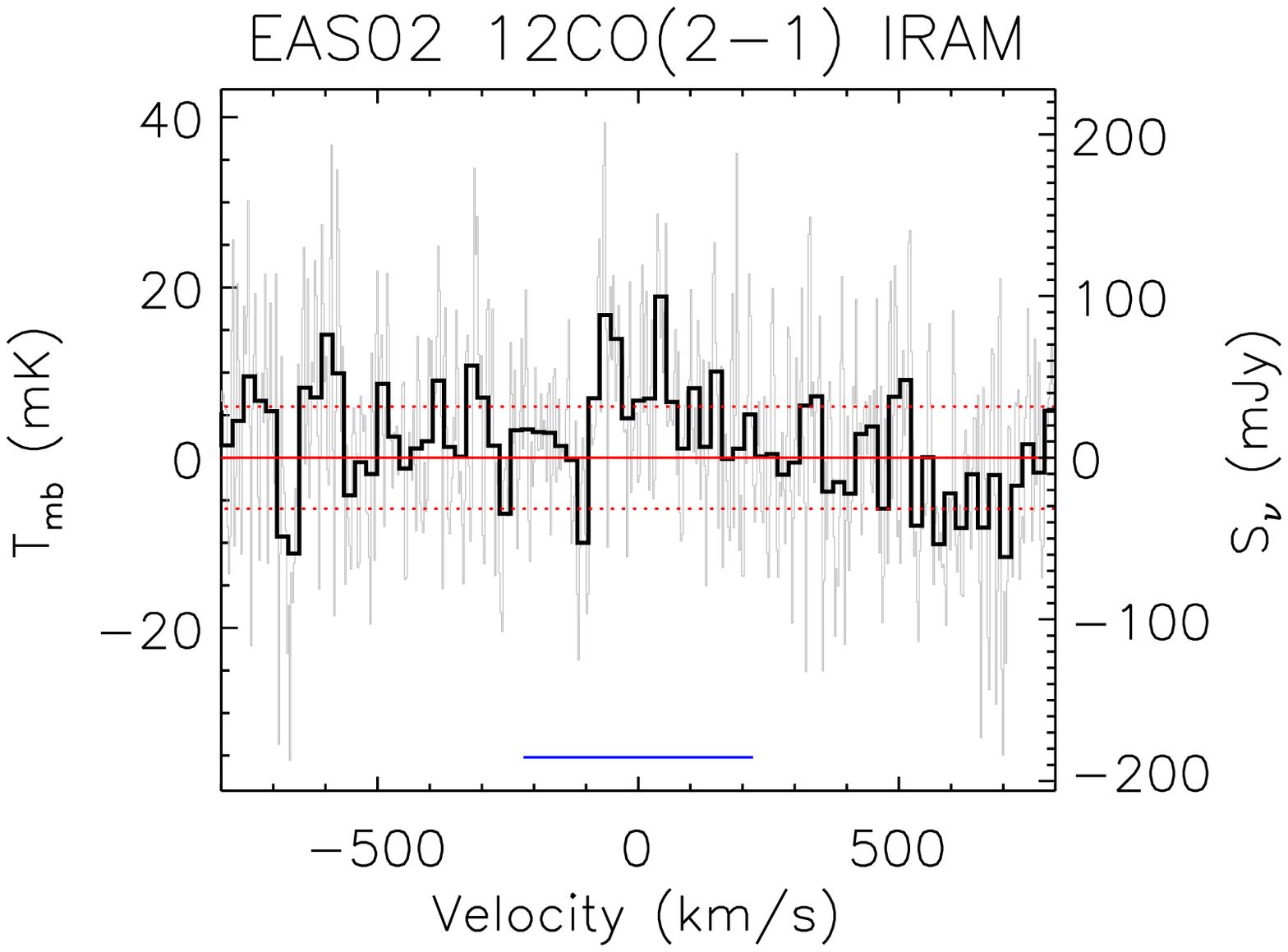}
\includegraphics[width=0.32\textwidth]{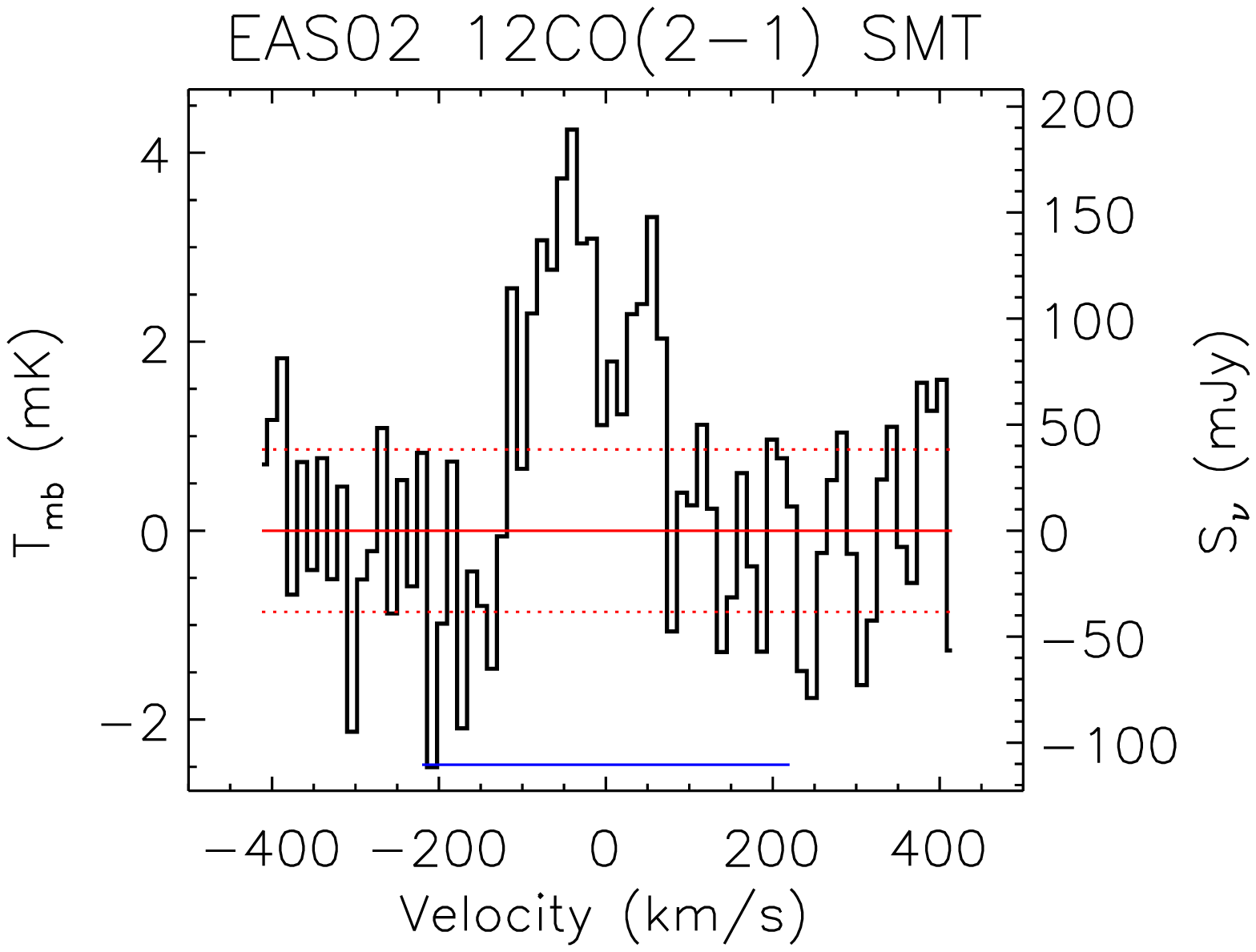}

\caption[]{continued}
\end{figure*}

\begin{figure*}
\ContinuedFloat

\includegraphics[width=0.32\textwidth]{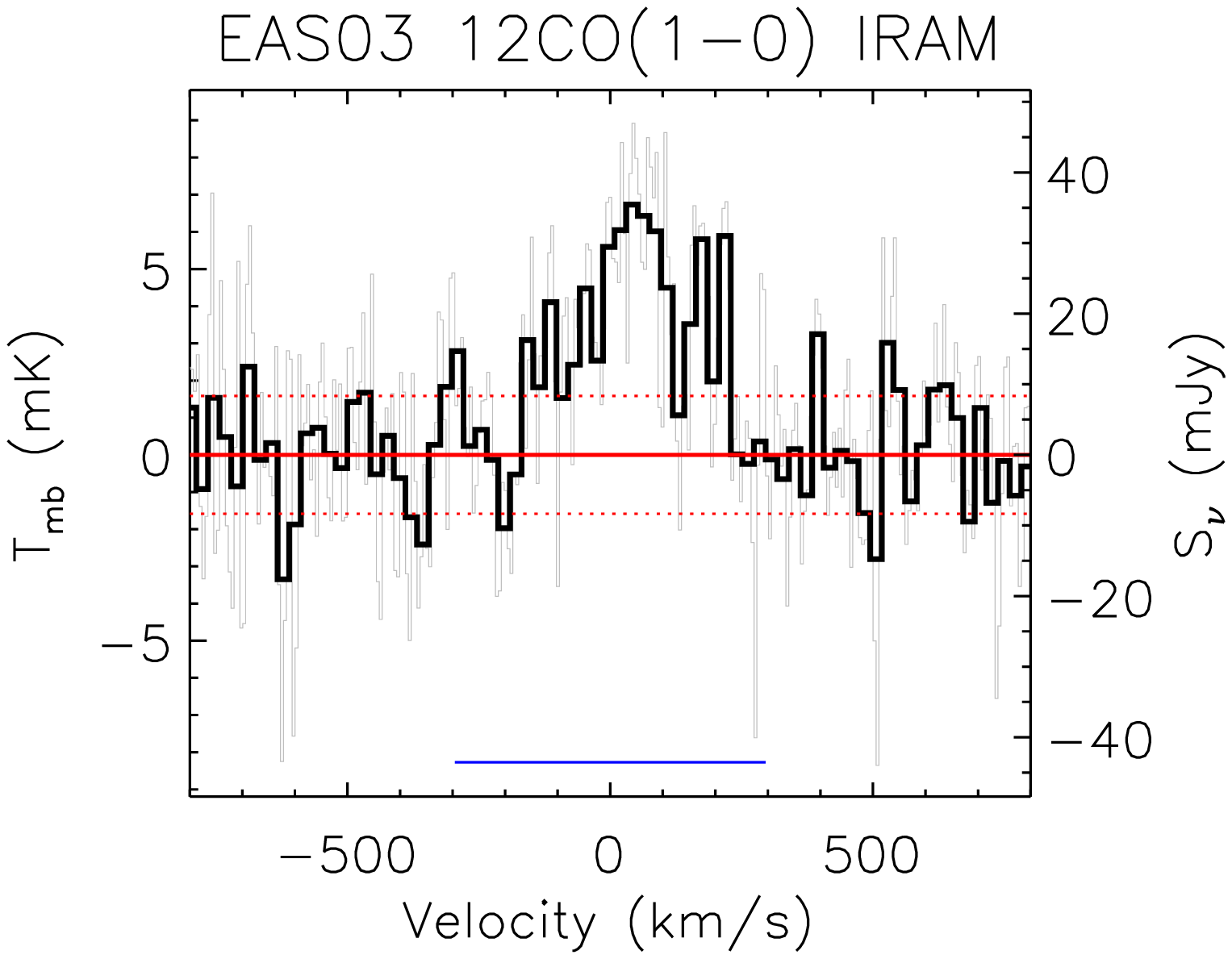}
\includegraphics[width=0.32\textwidth]{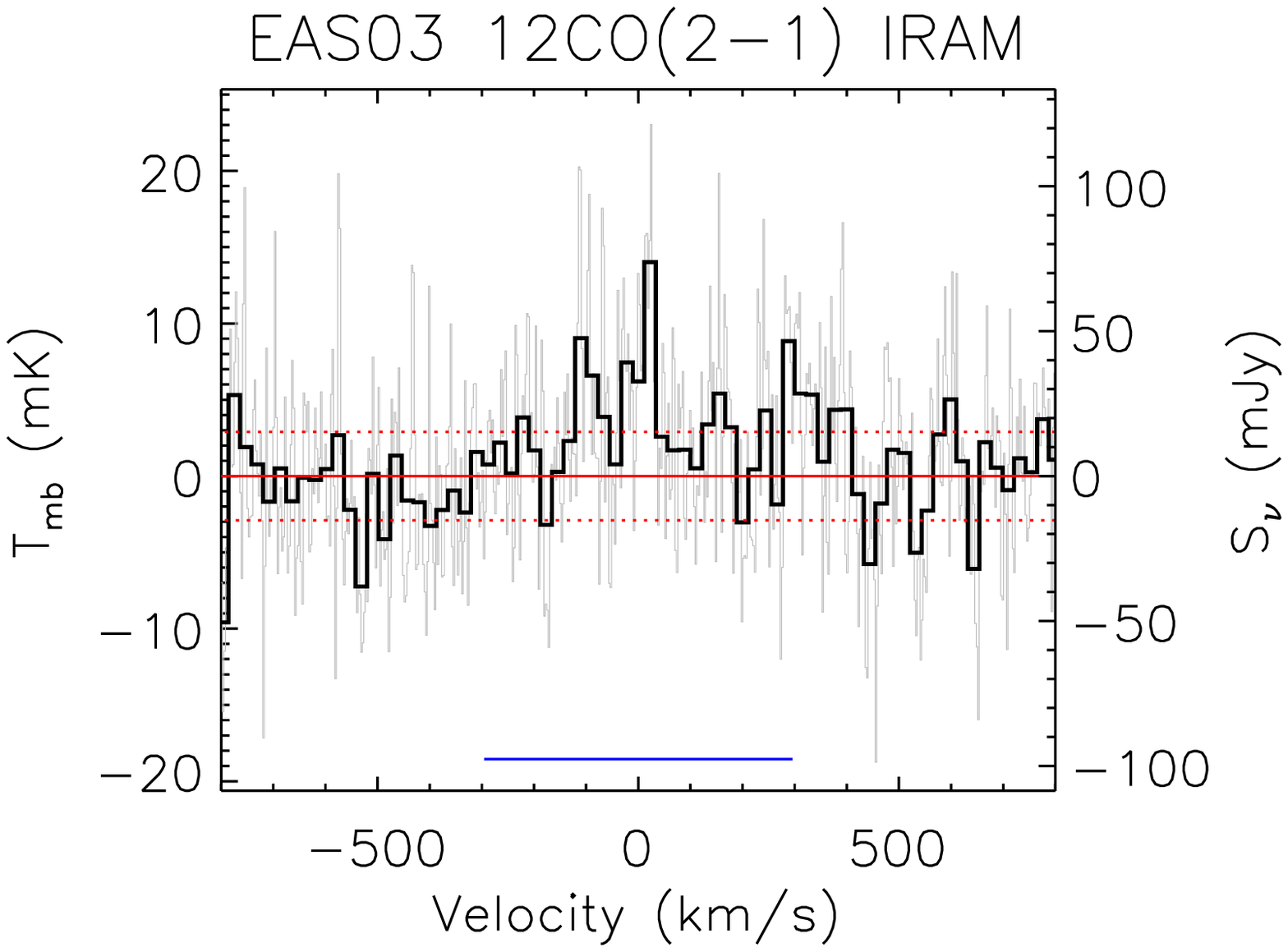}
\includegraphics[width=0.32\textwidth]{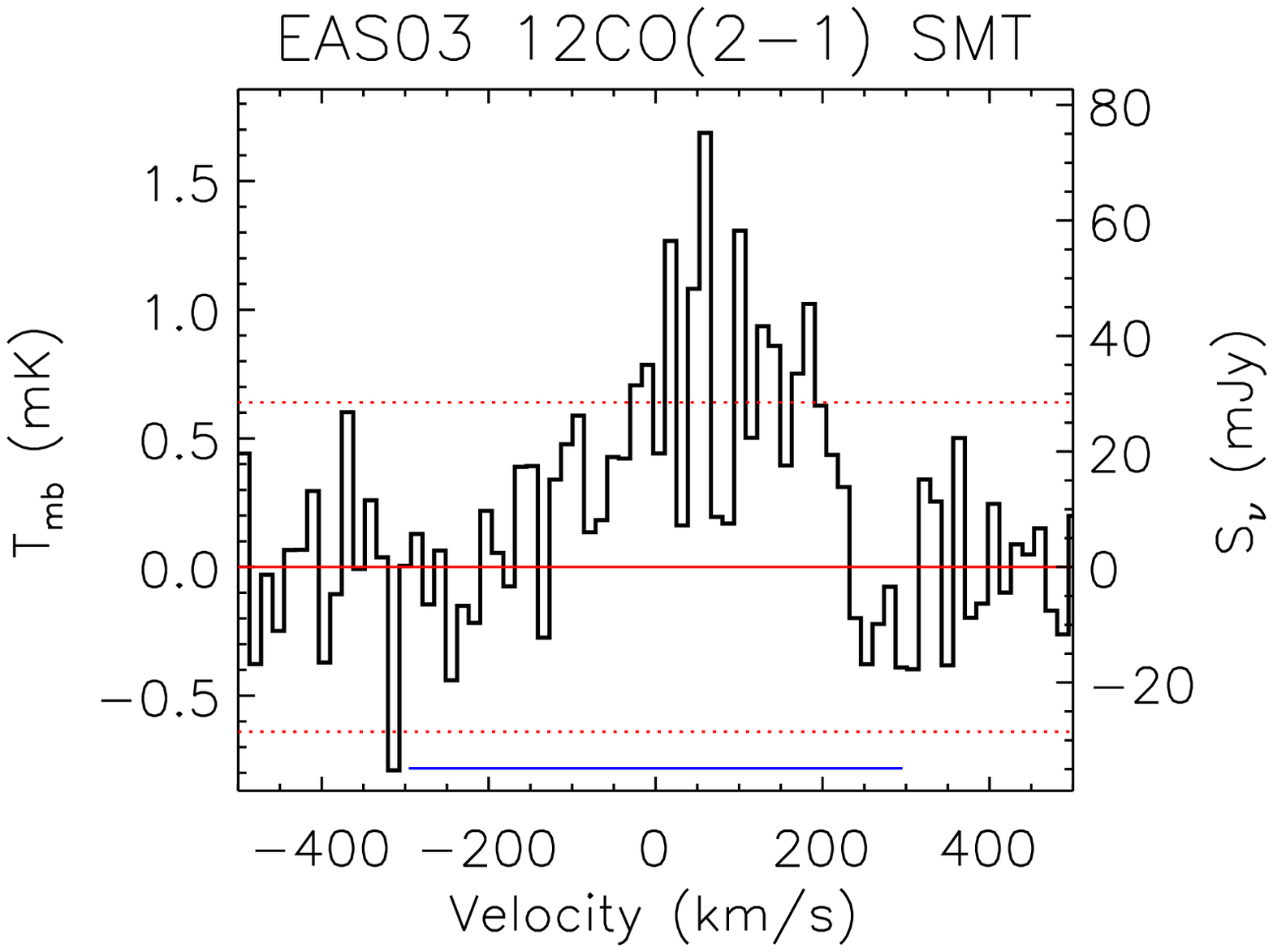}

\includegraphics[width=0.32\textwidth]{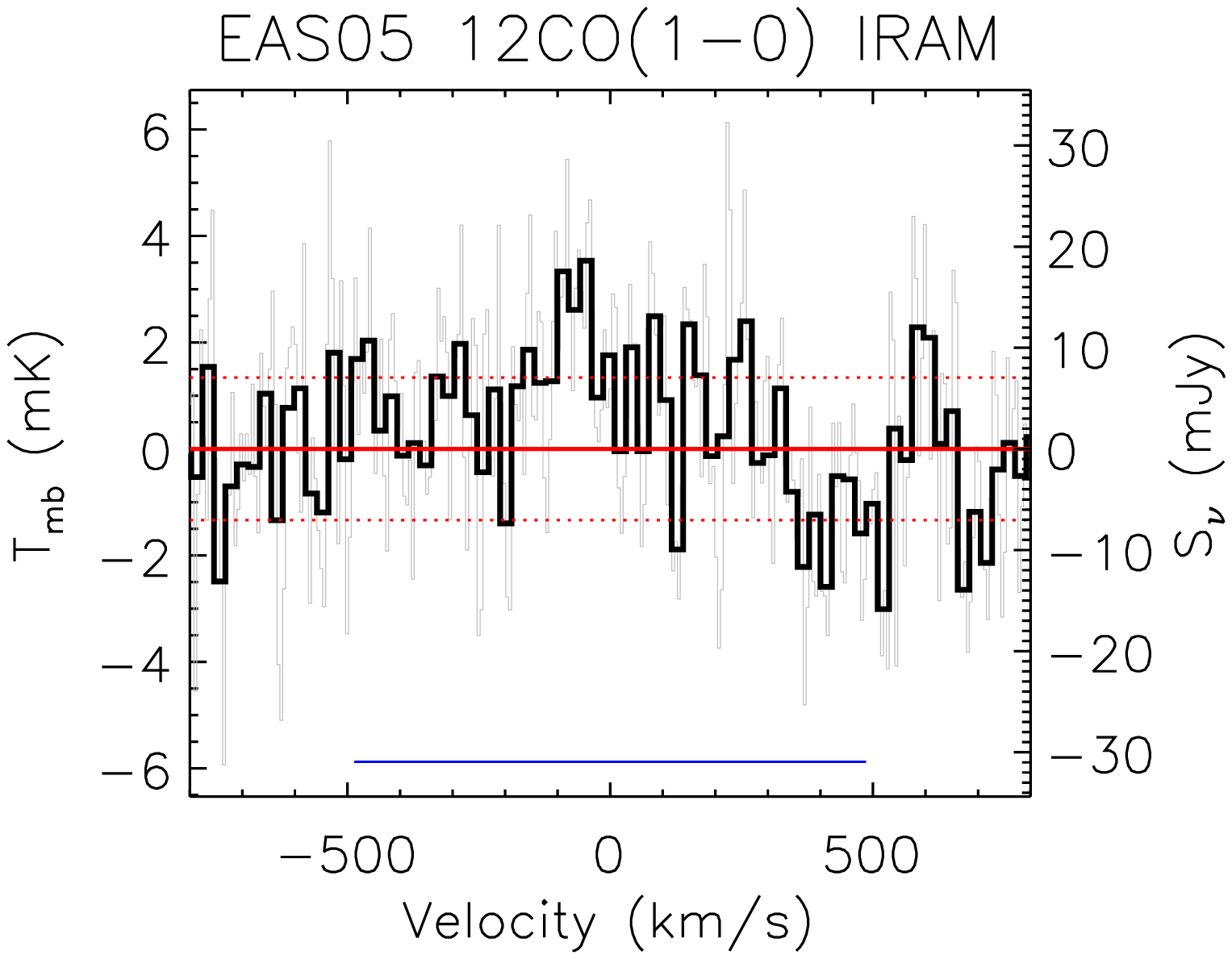}
\includegraphics[width=0.32\textwidth]{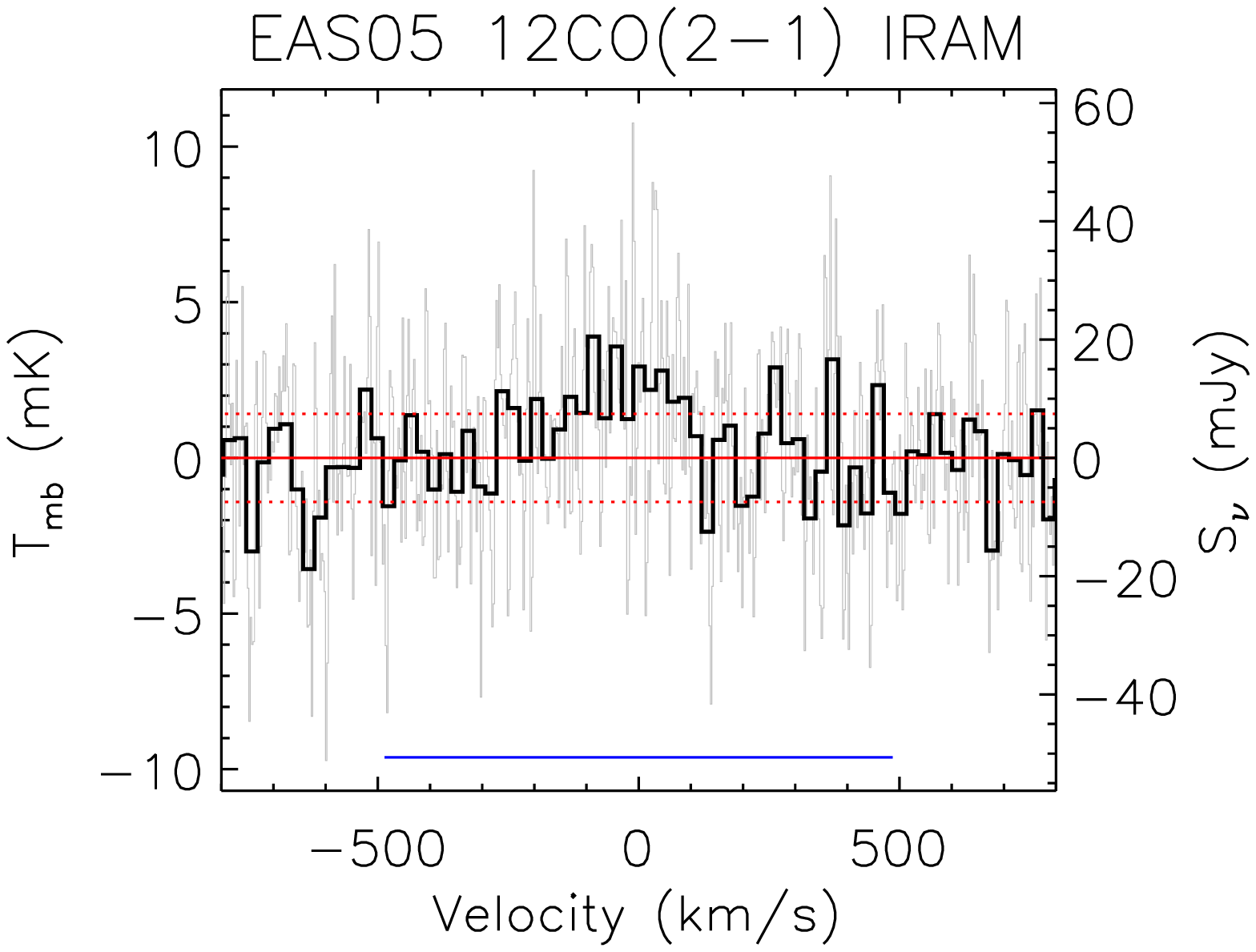}
\includegraphics[width=0.32\textwidth]{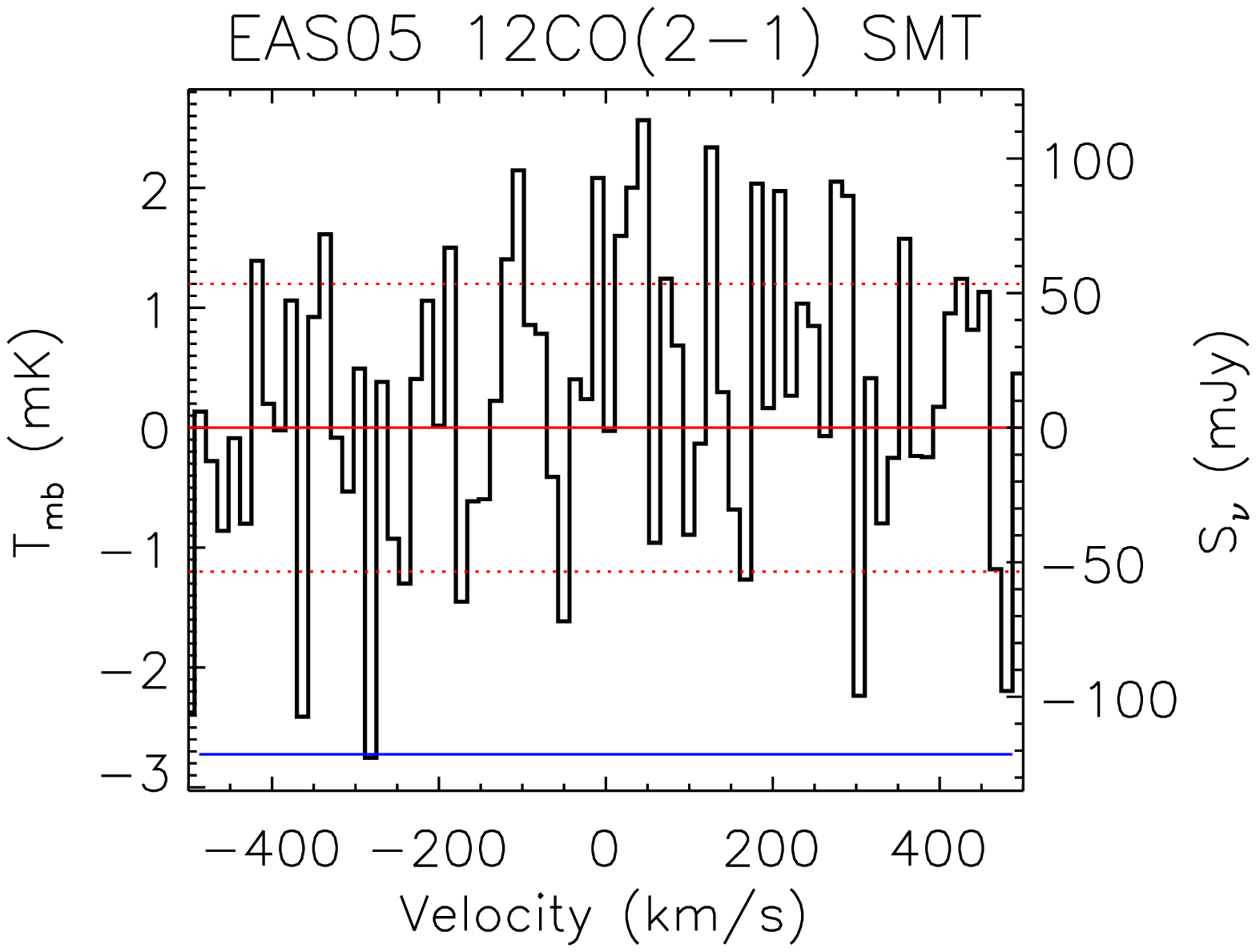}

\includegraphics[width=0.32\textwidth]{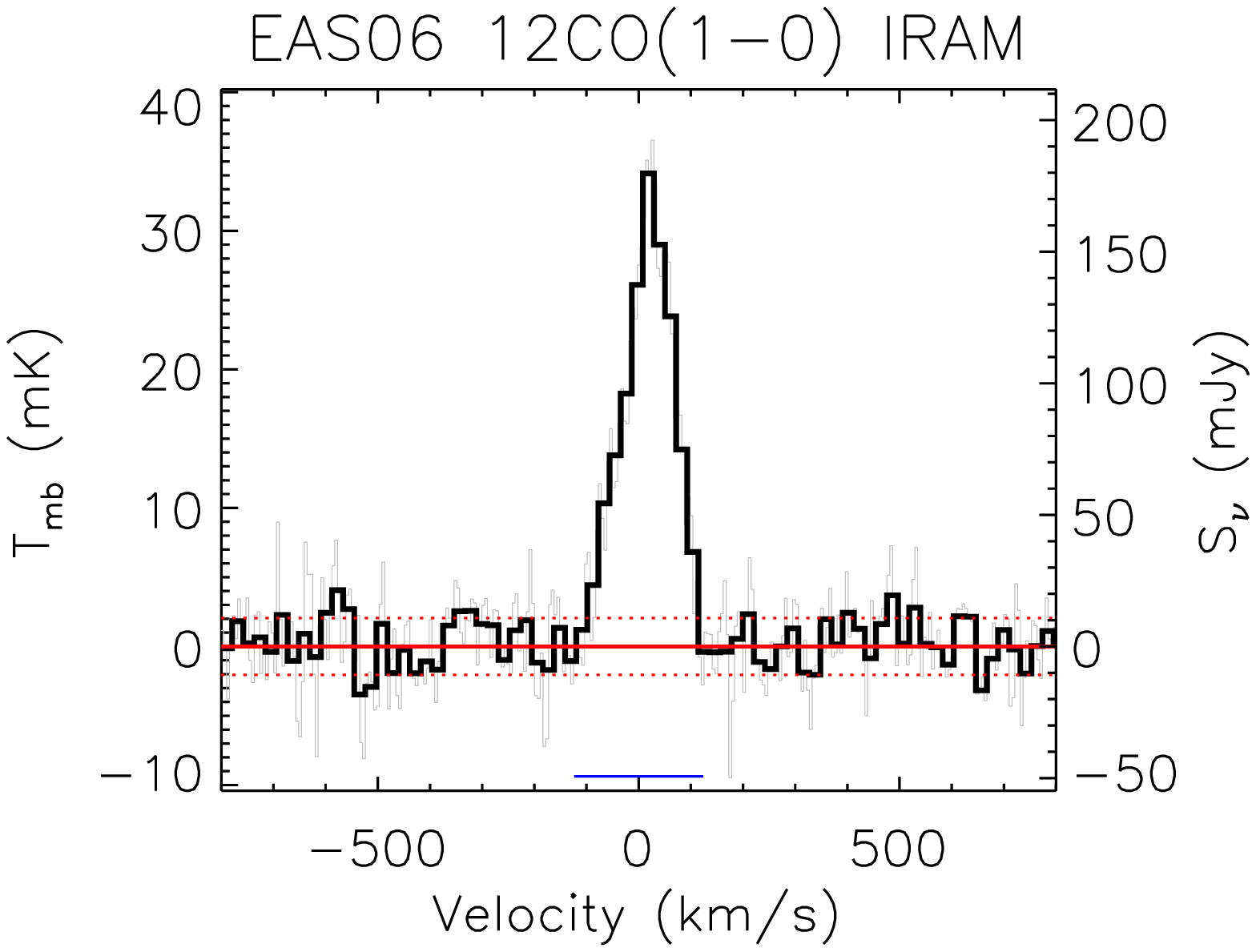}
\includegraphics[width=0.32\textwidth]{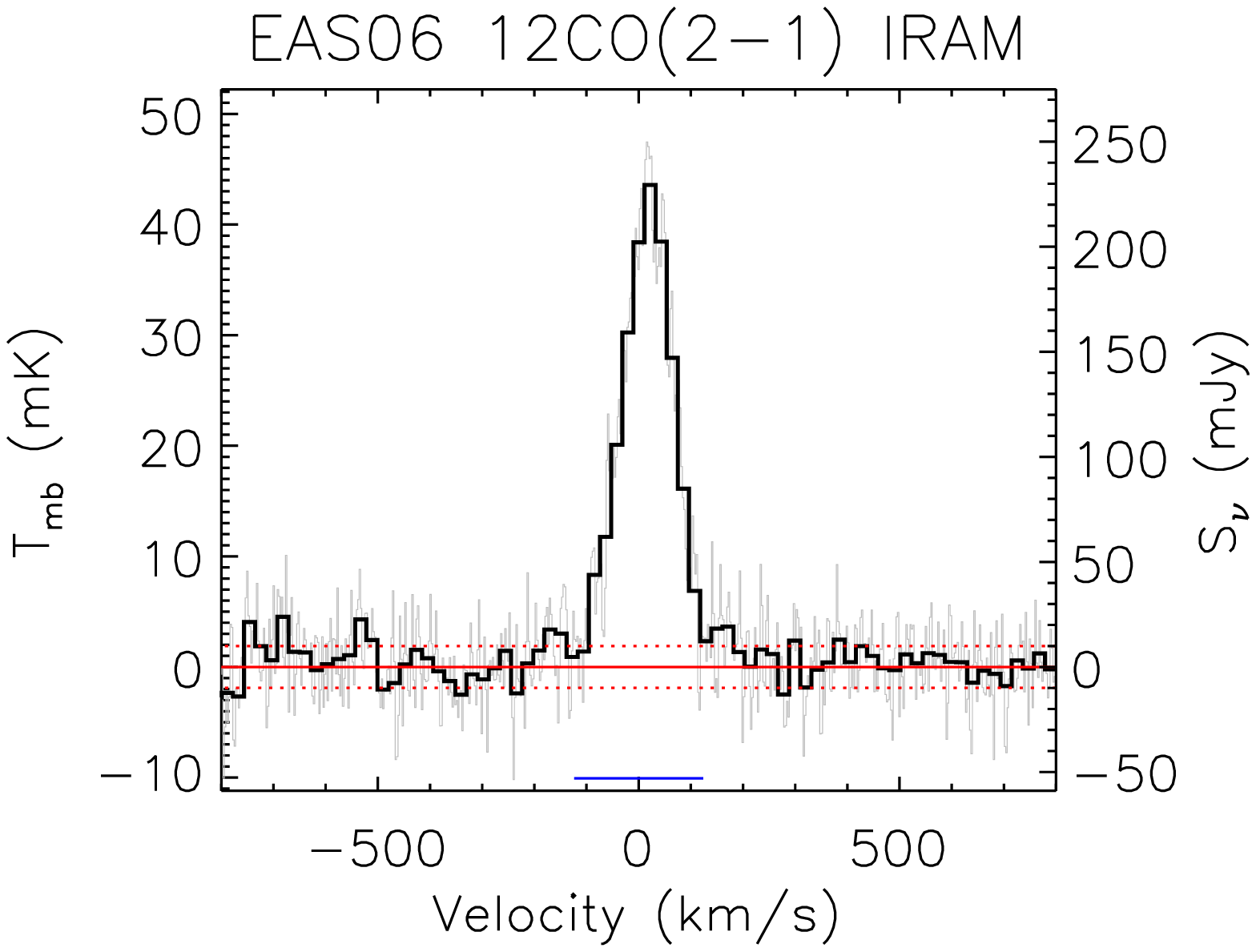}
\includegraphics[width=0.32\textwidth]{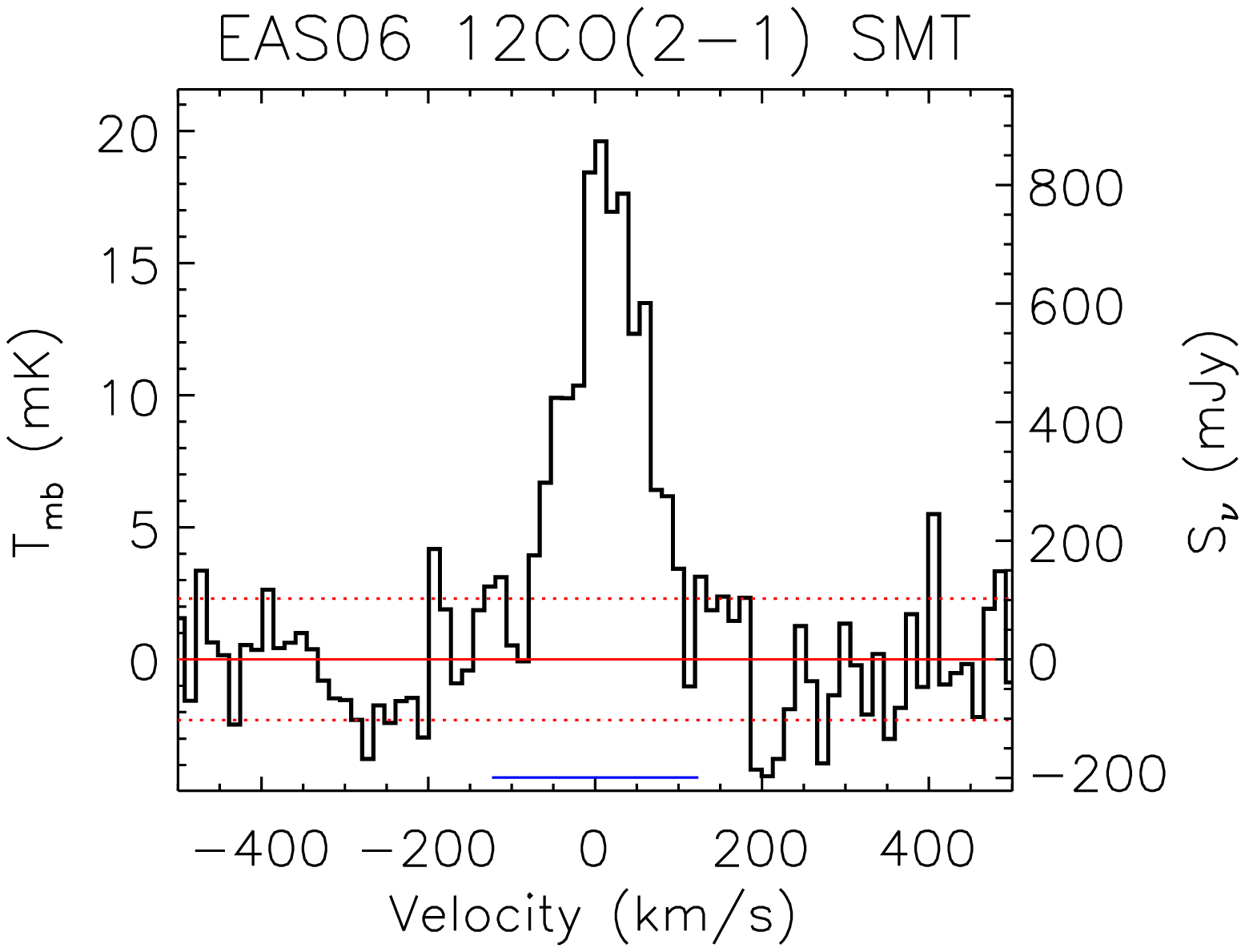}

\includegraphics[width=0.32\textwidth]{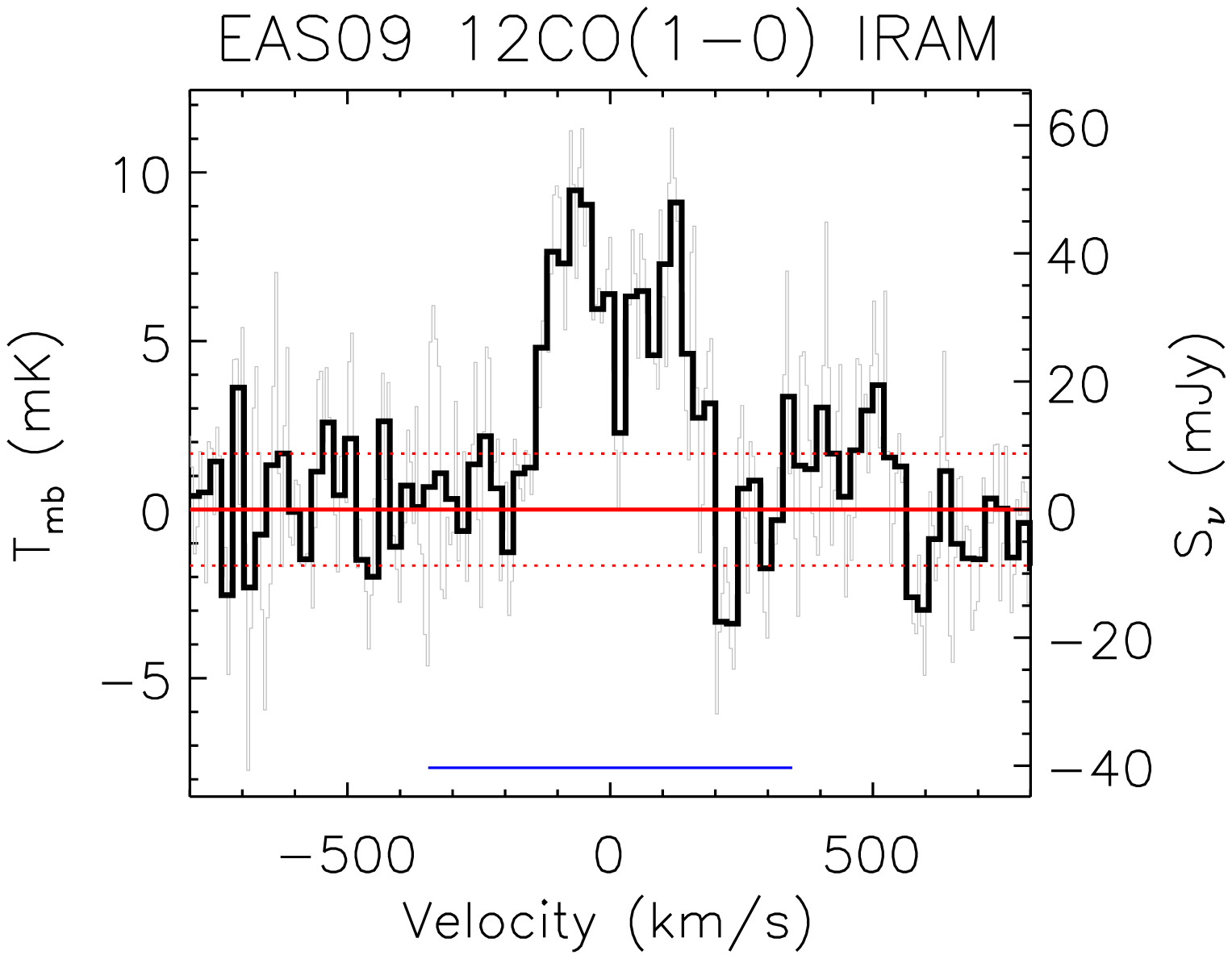}
\includegraphics[width=0.32\textwidth]{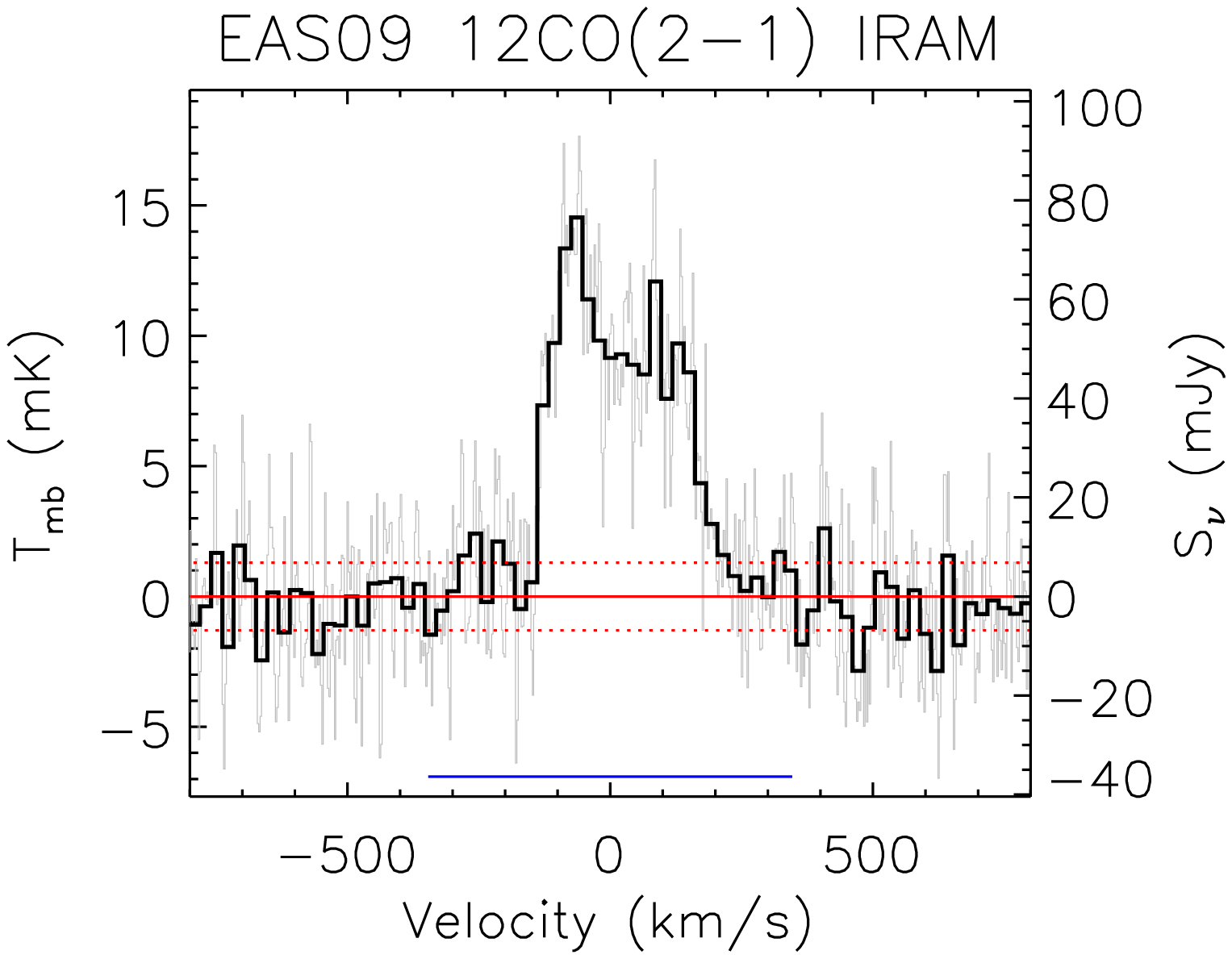}
\includegraphics[width=0.32\textwidth]{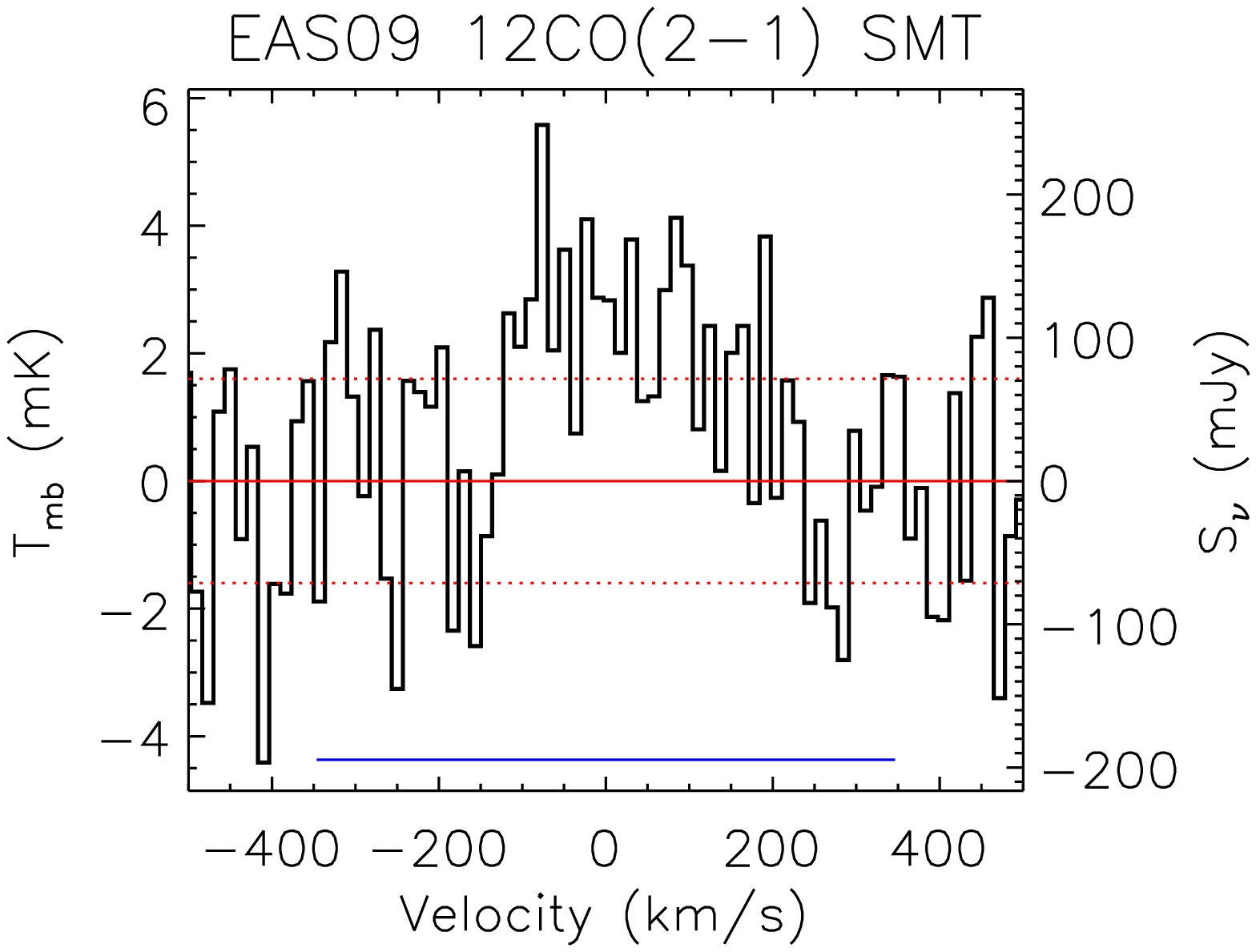}

\includegraphics[width=0.32\textwidth]{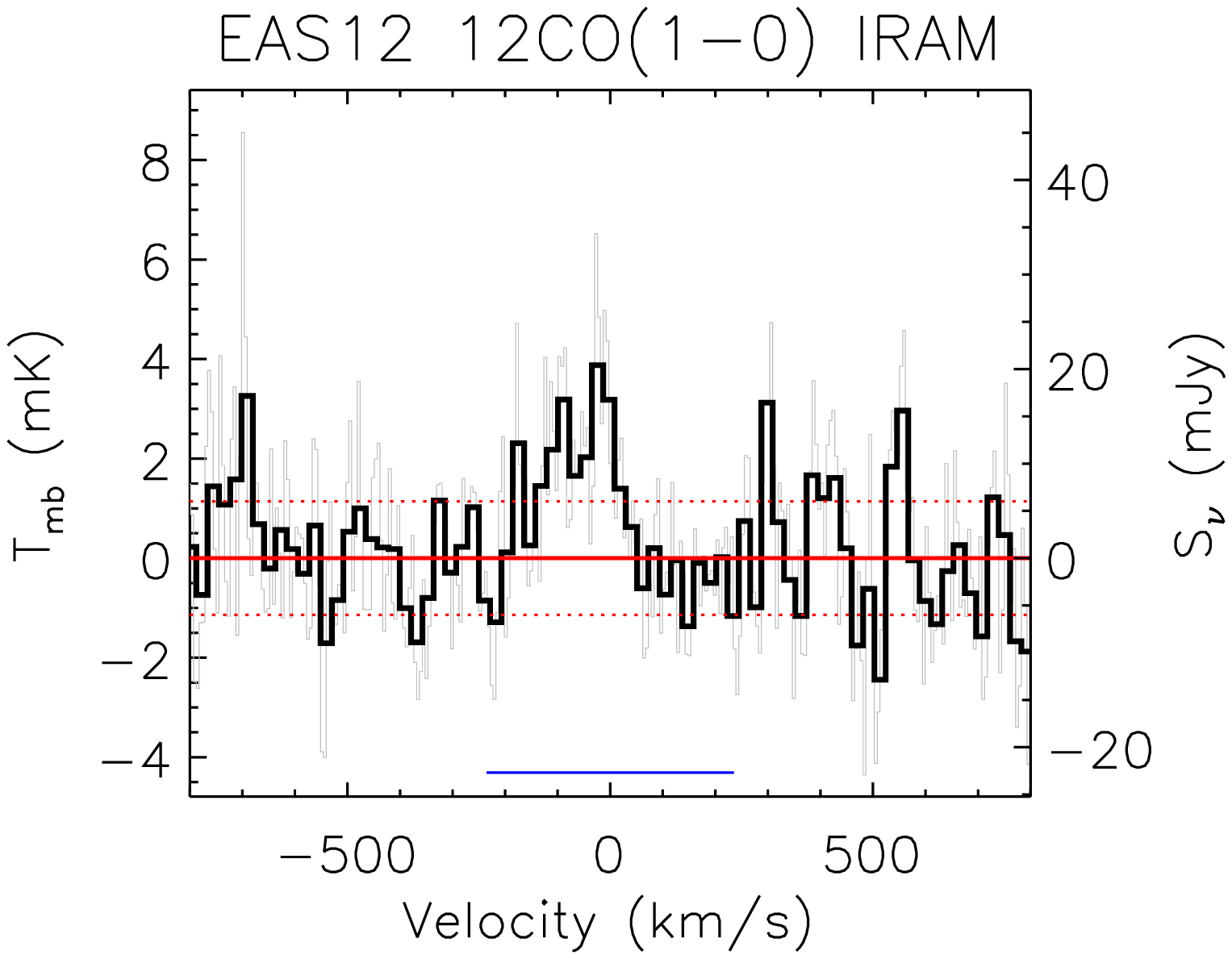}
\includegraphics[width=0.32\textwidth]{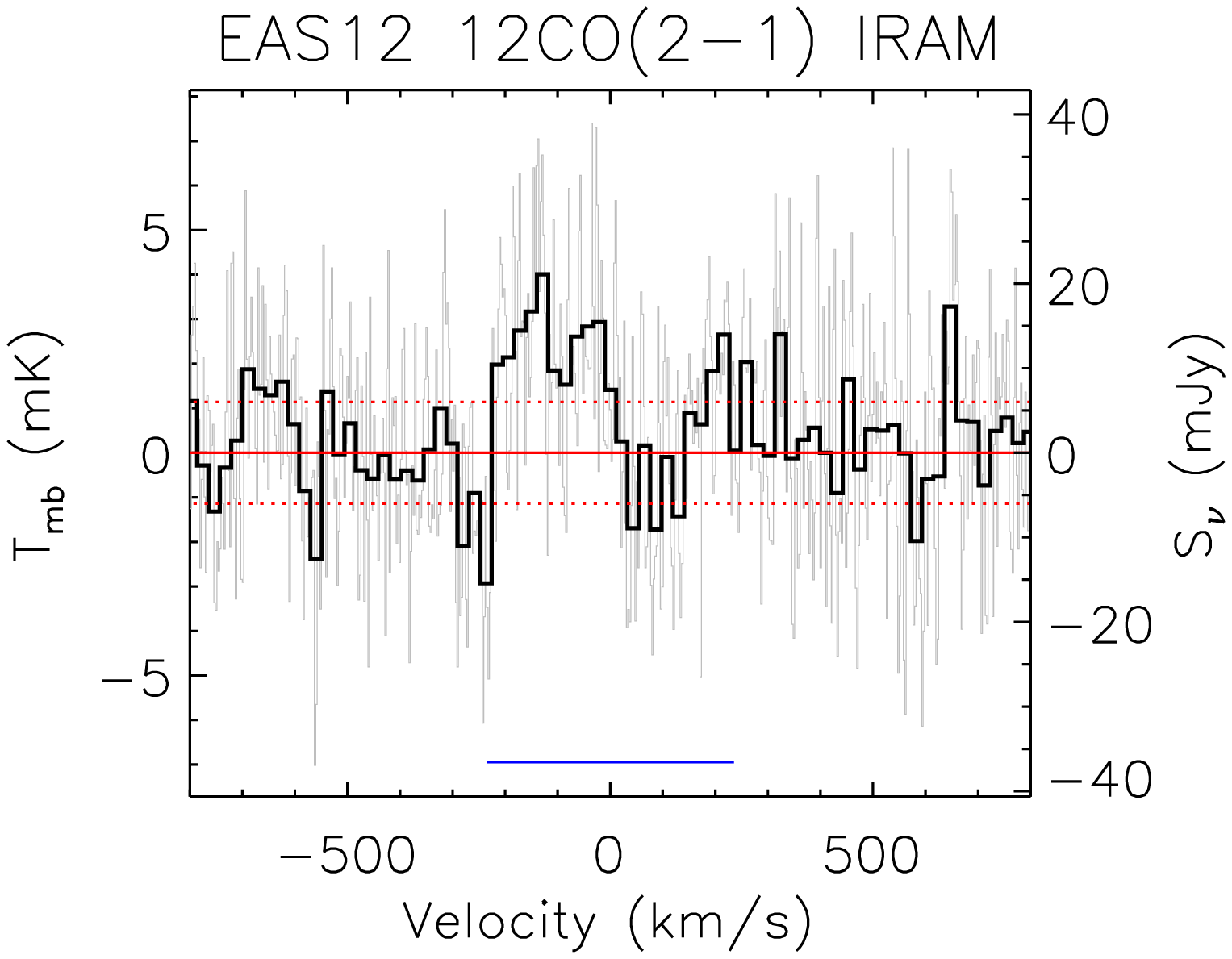}
\includegraphics[width=0.32\textwidth]{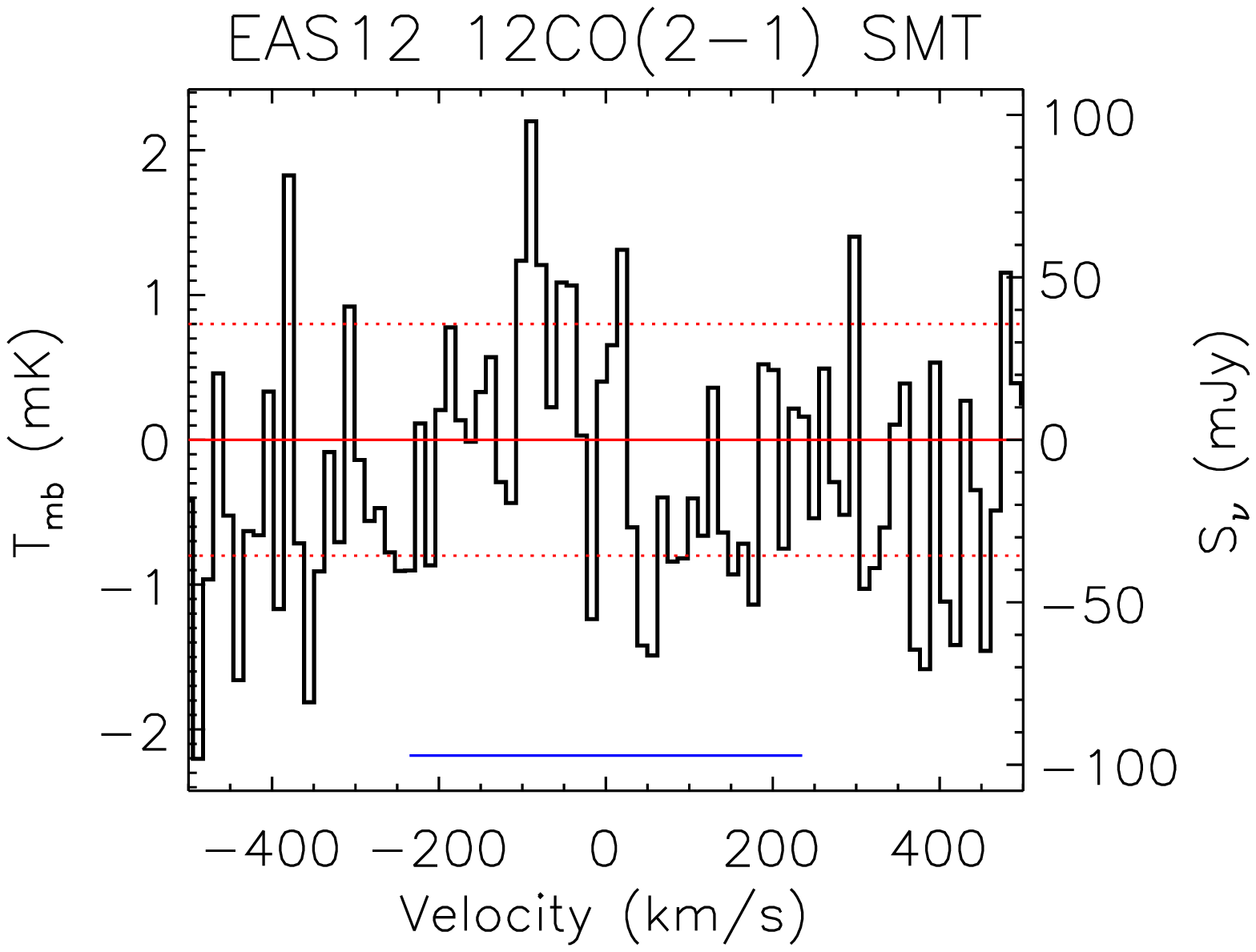}

\caption[]{continued}
\end{figure*}

\begin{figure*}
\ContinuedFloat

\includegraphics[width=0.32\textwidth]{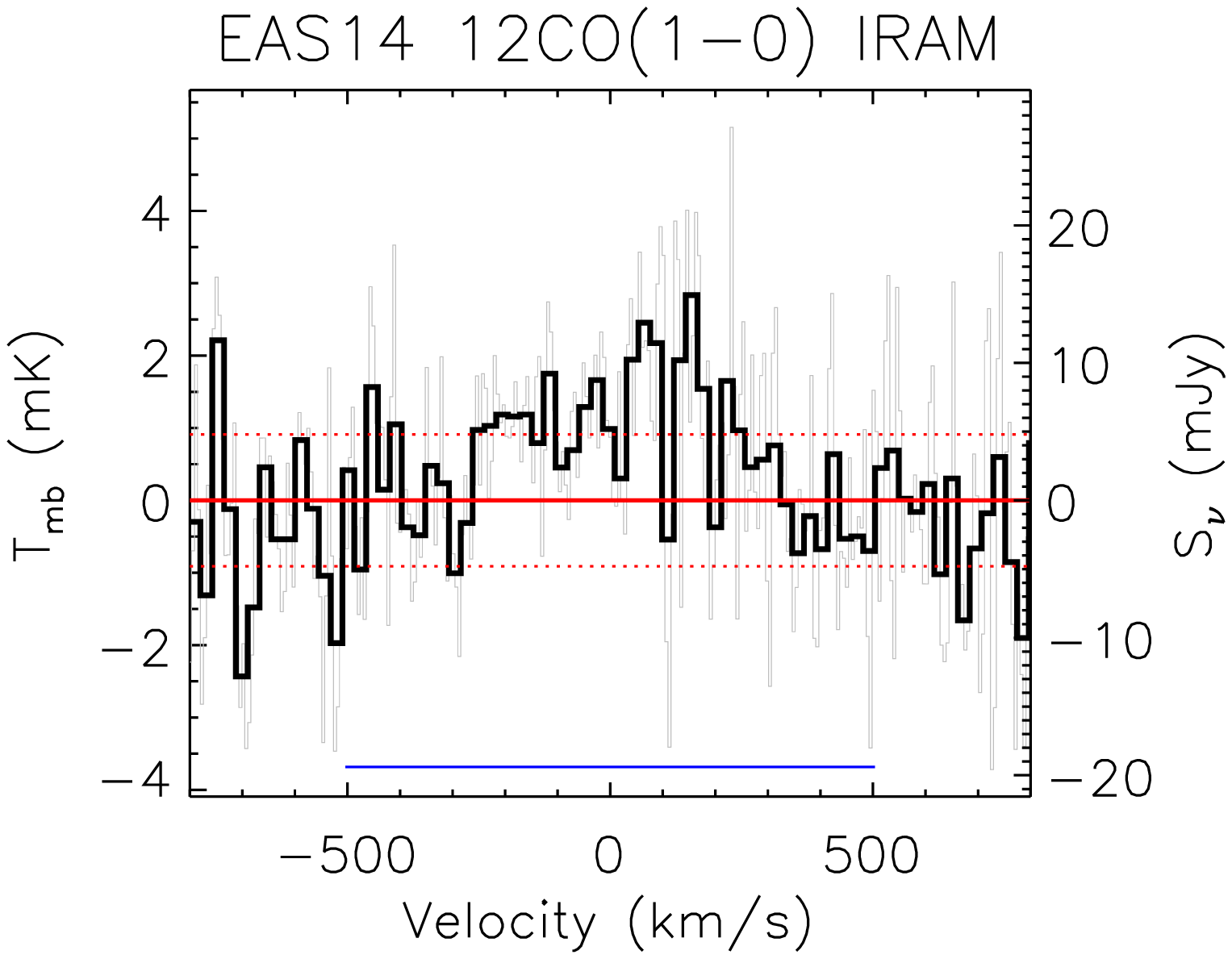}
\includegraphics[width=0.32\textwidth]{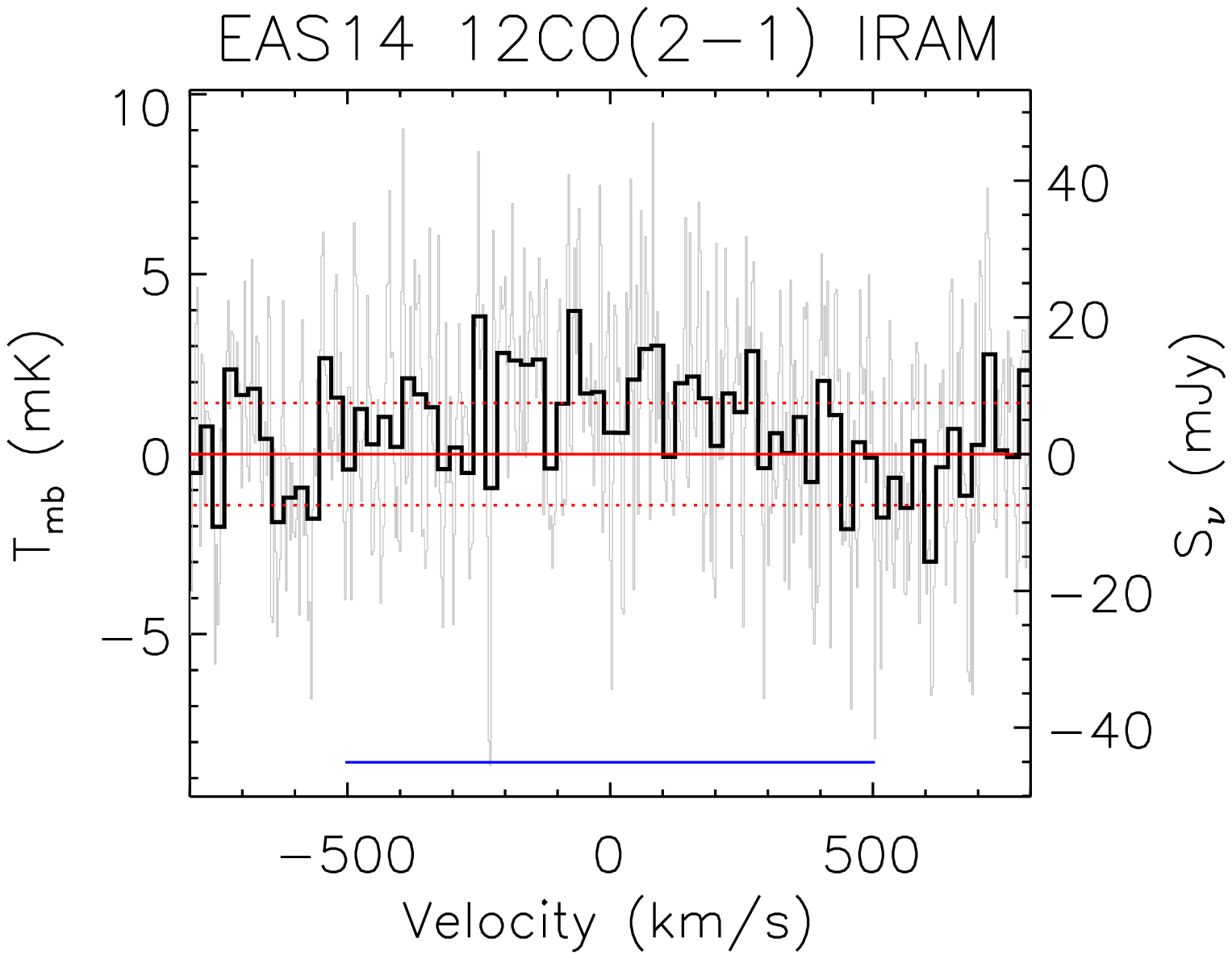}

\includegraphics[width=0.32\textwidth]{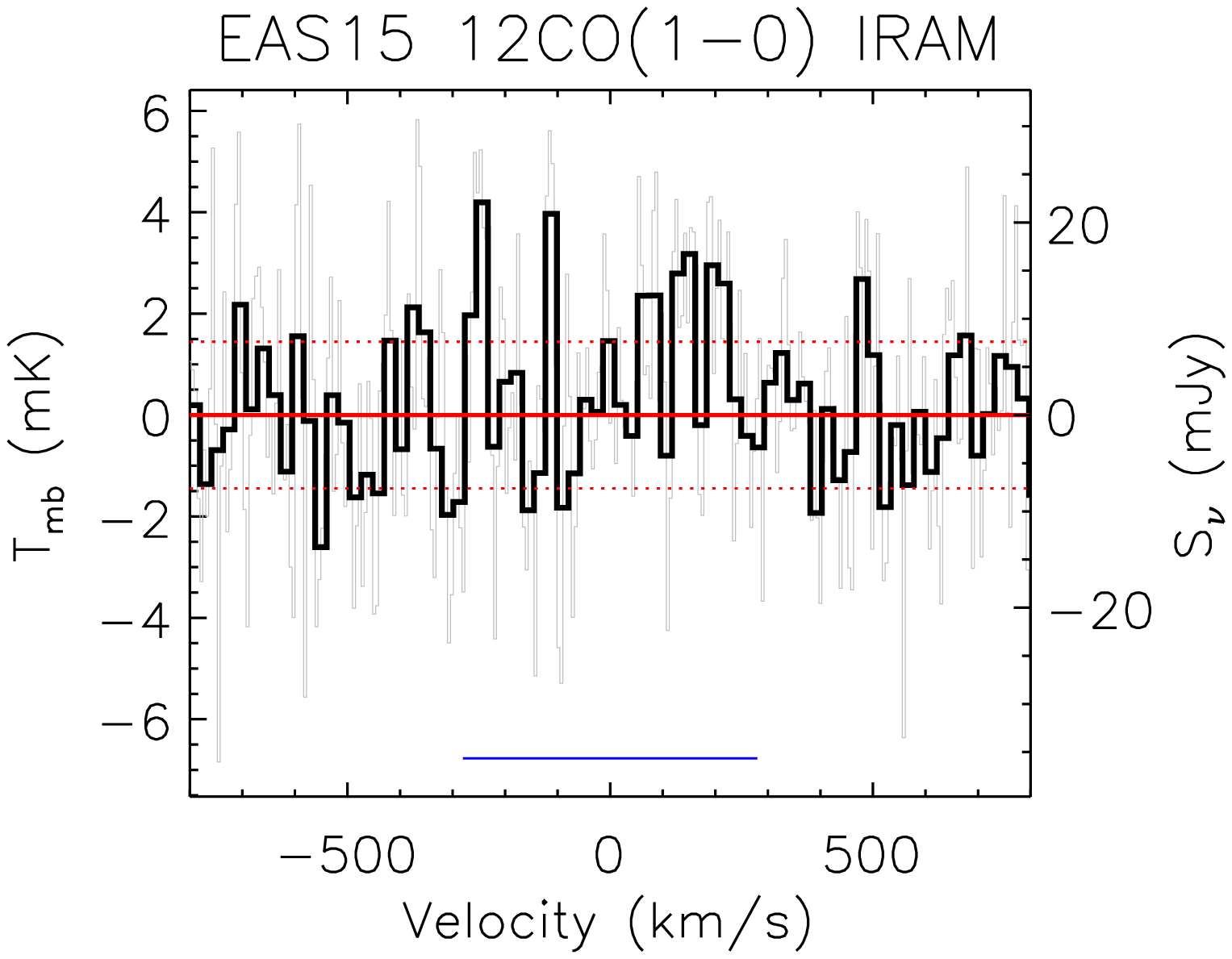}
\includegraphics[width=0.32\textwidth]{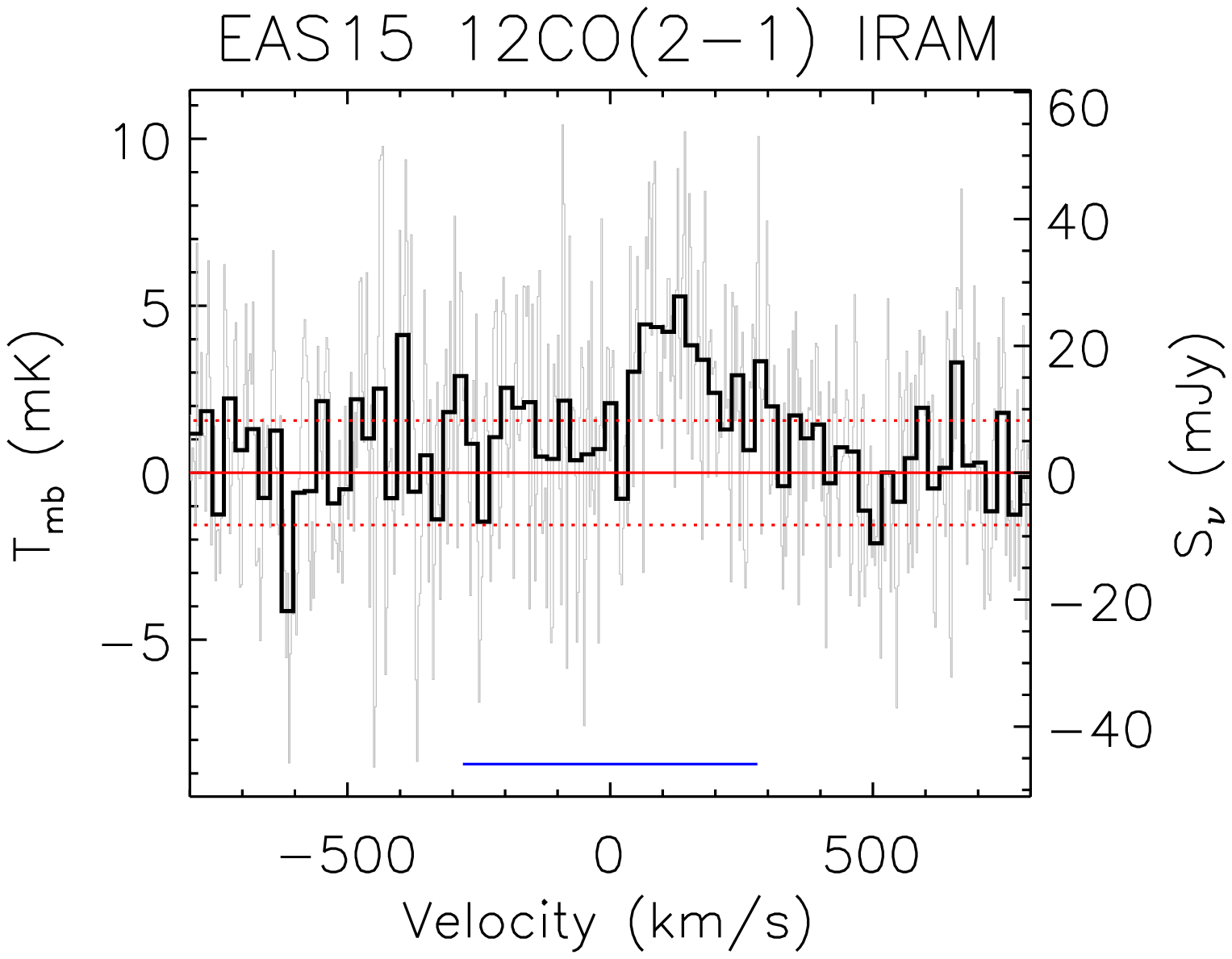}
\includegraphics[width=0.32\textwidth]{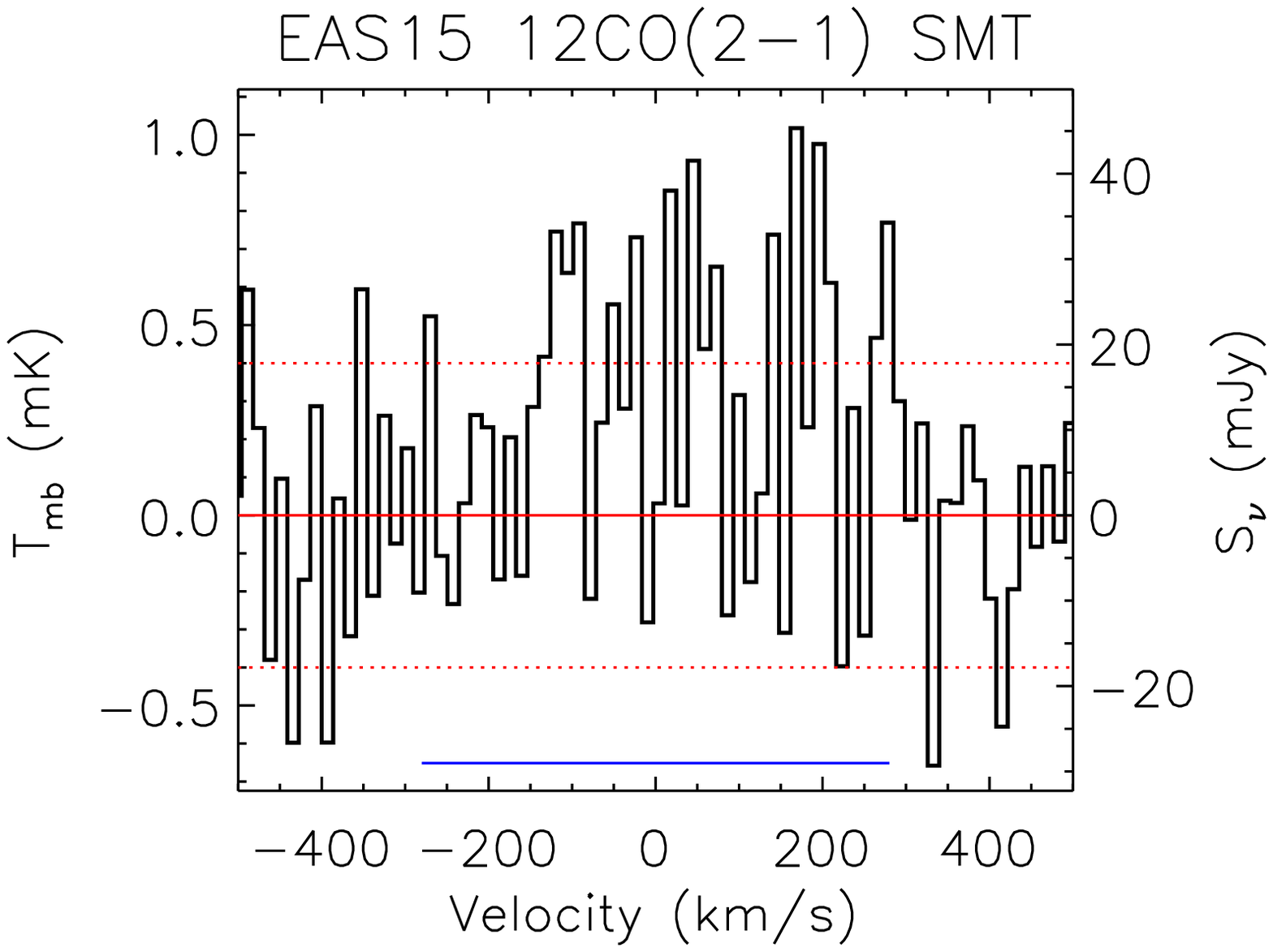}

\caption[]{continued}
\end{figure*}


\begin{figure*}

\includegraphics[width=0.32\textwidth]{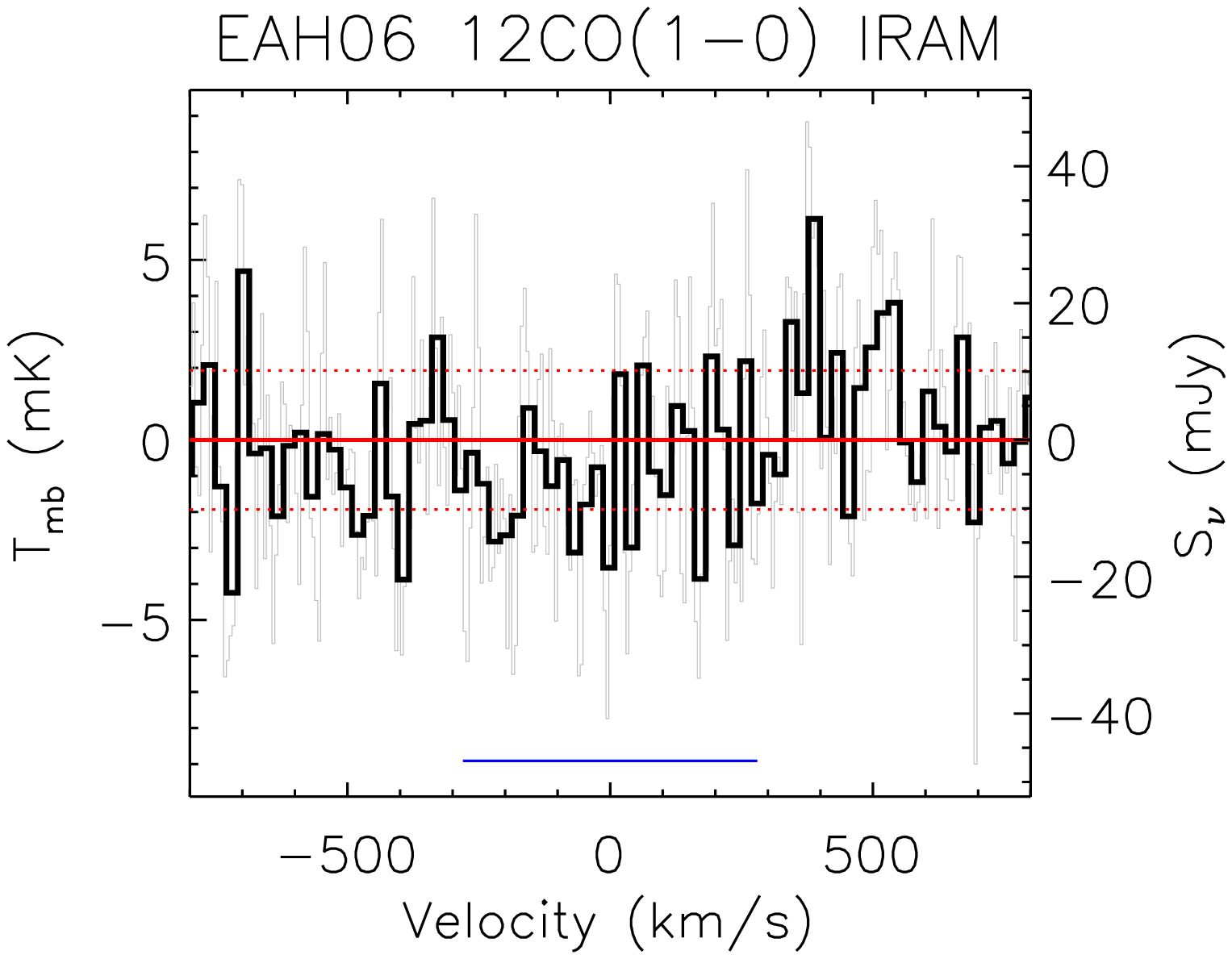}
\includegraphics[width=0.32\textwidth]{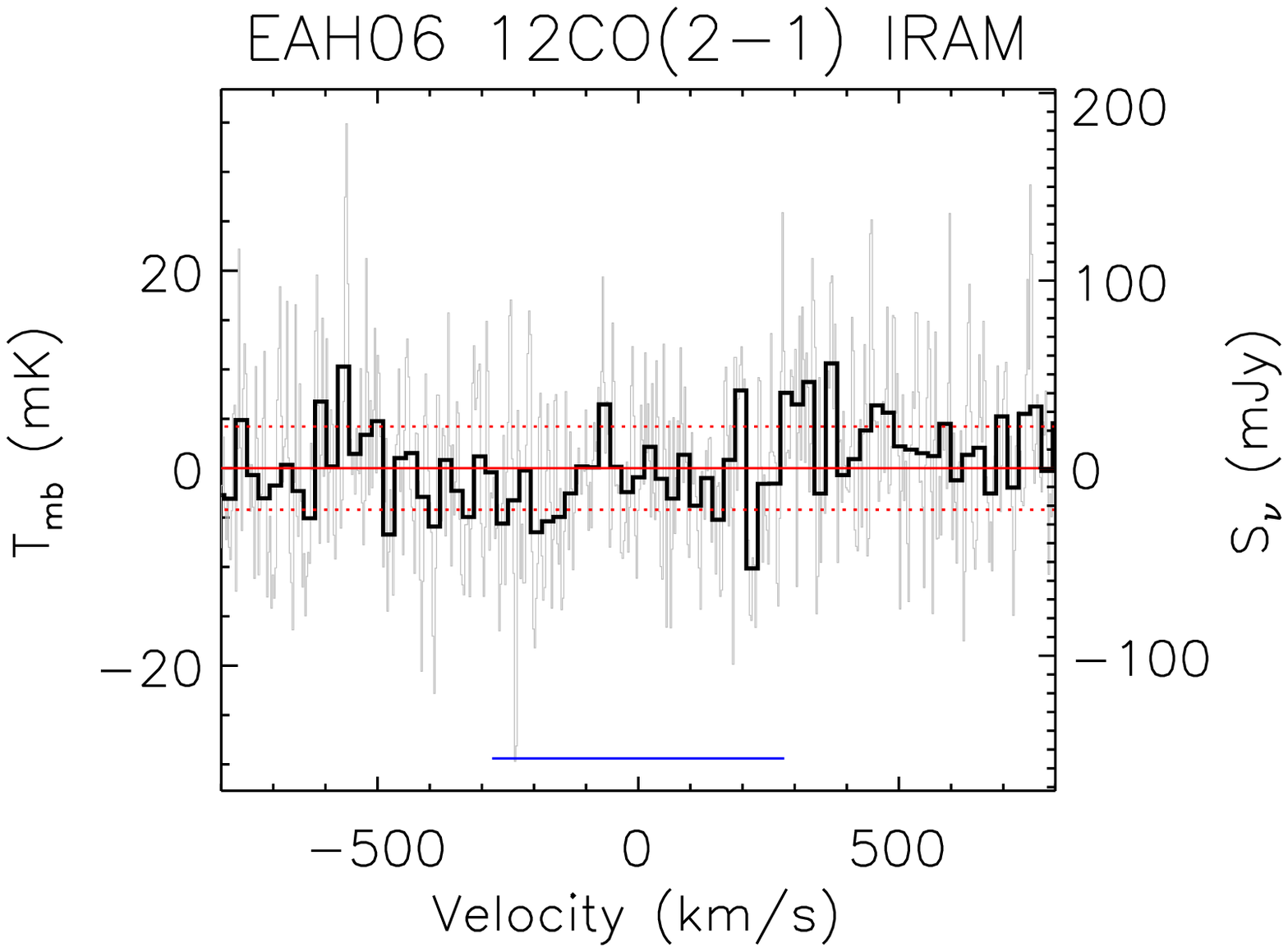}

\includegraphics[width=0.32\textwidth]{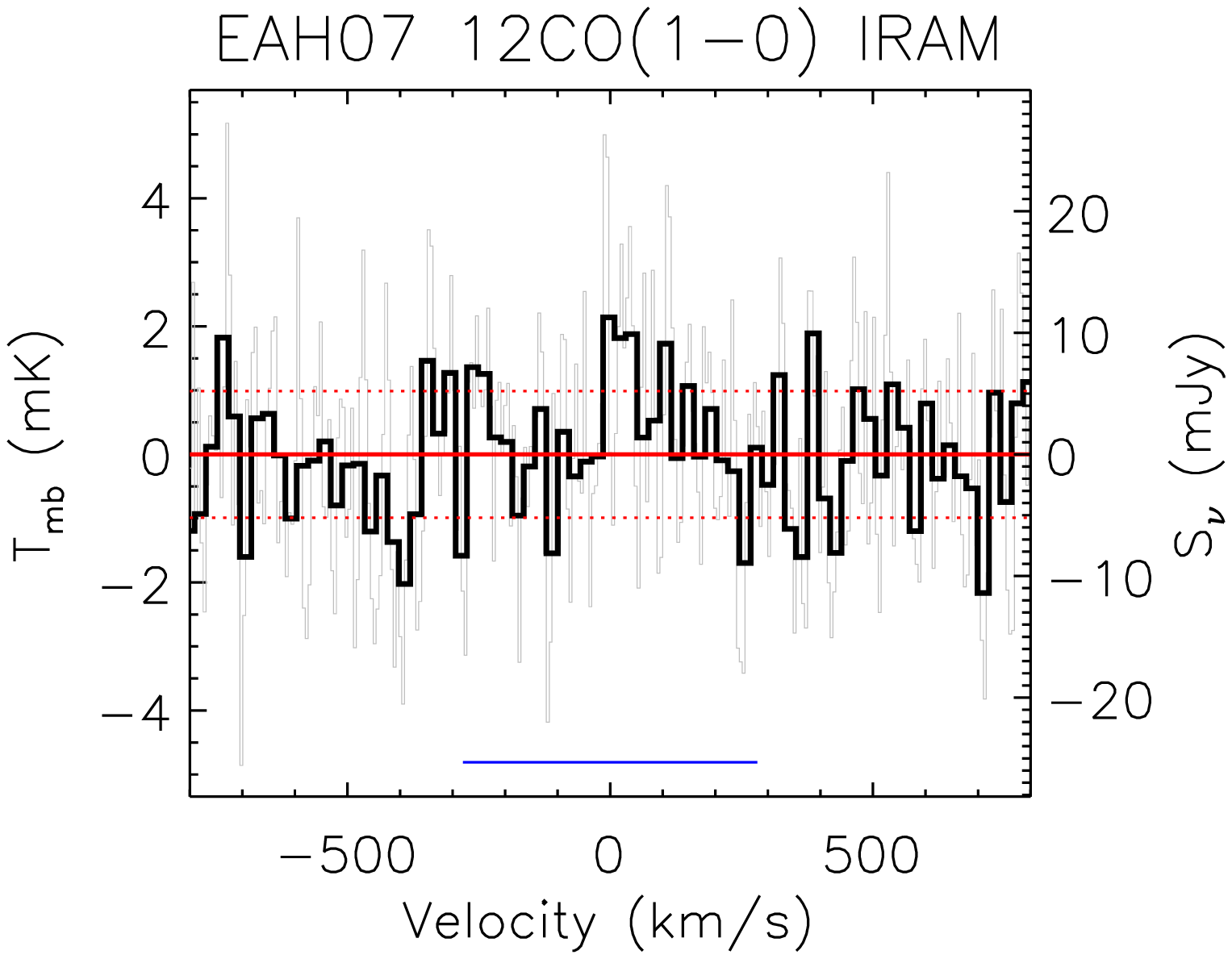}
\includegraphics[width=0.32\textwidth]{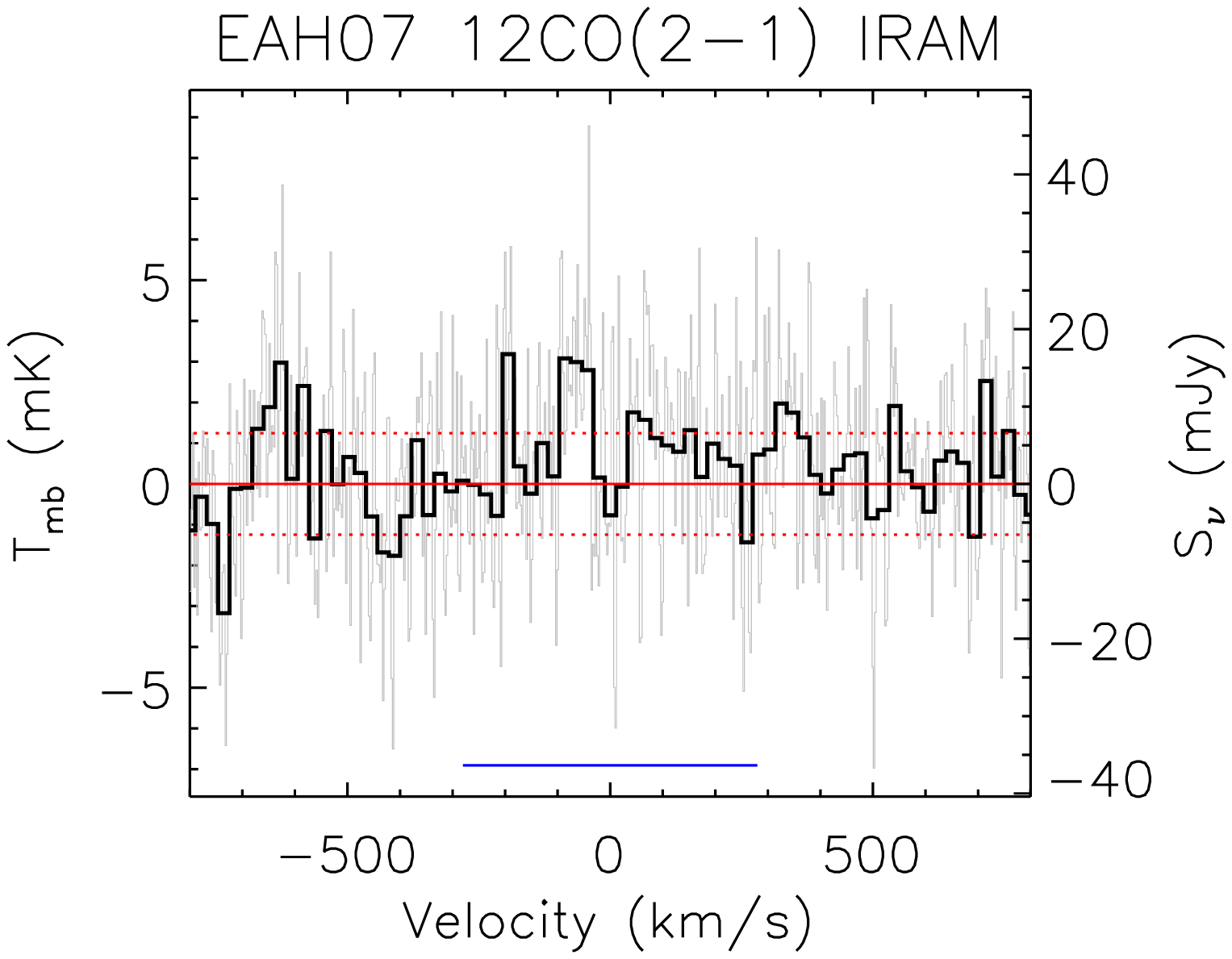}

\includegraphics[width=0.32\textwidth]{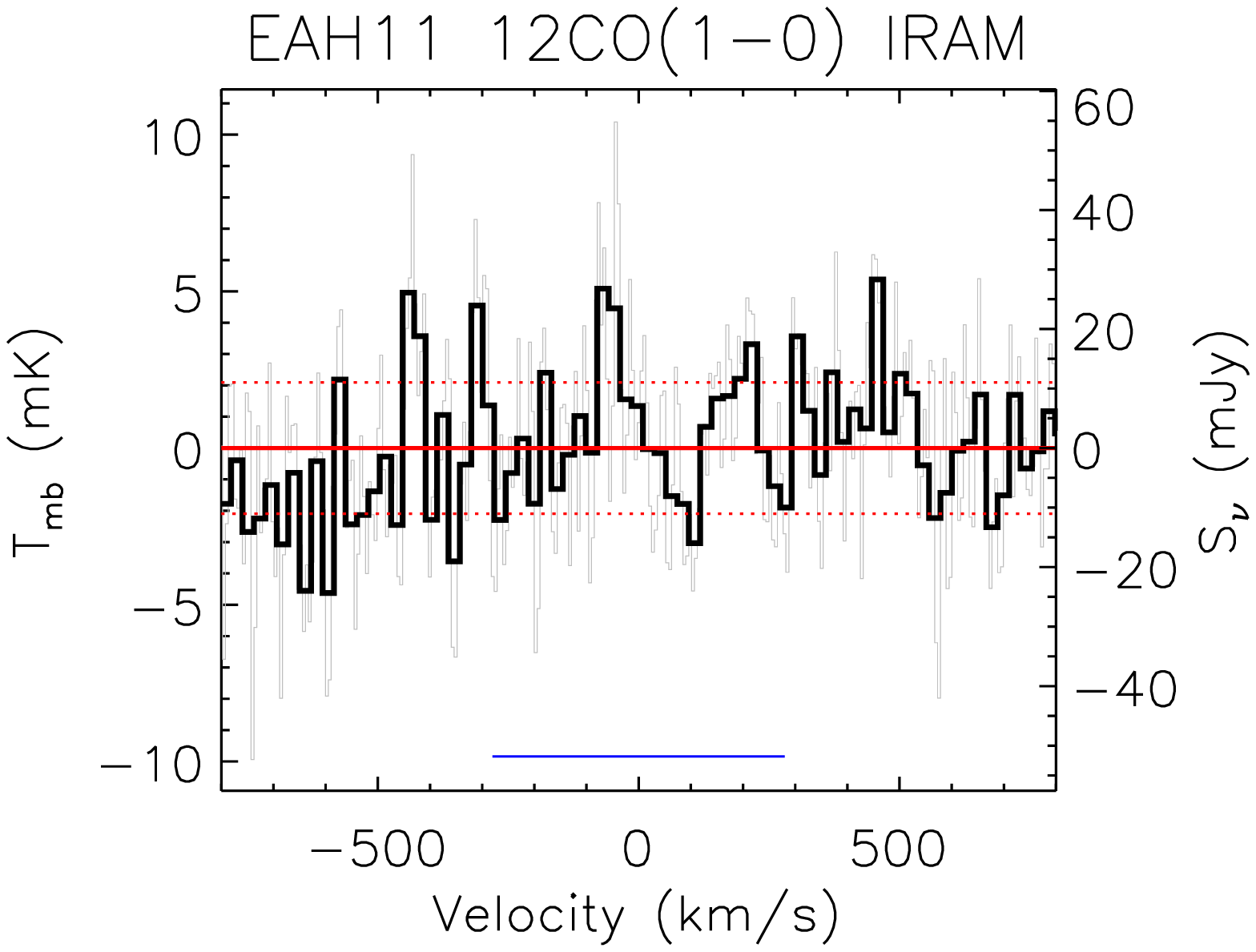}
\includegraphics[width=0.32\textwidth]{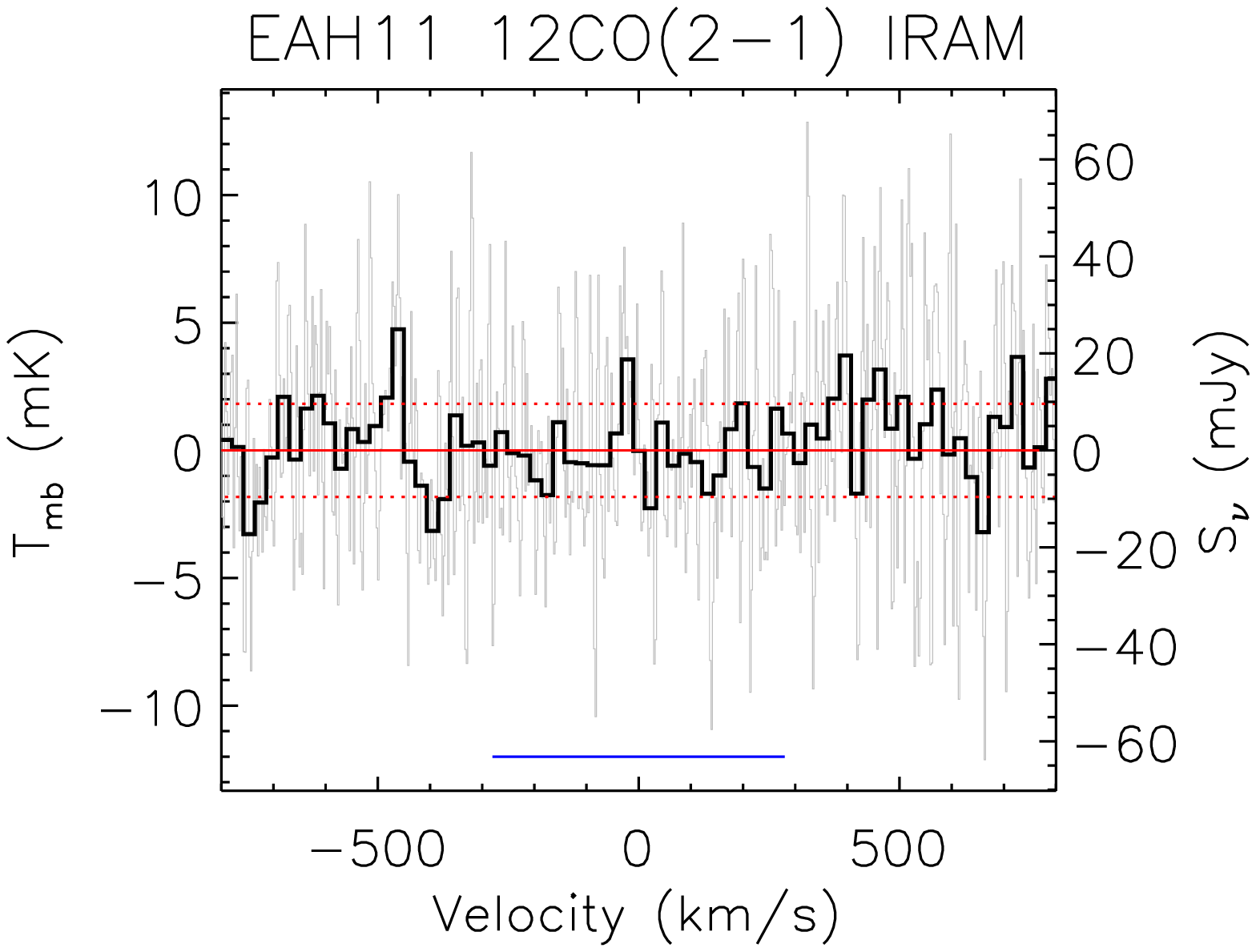}

\includegraphics[width=0.32\textwidth]{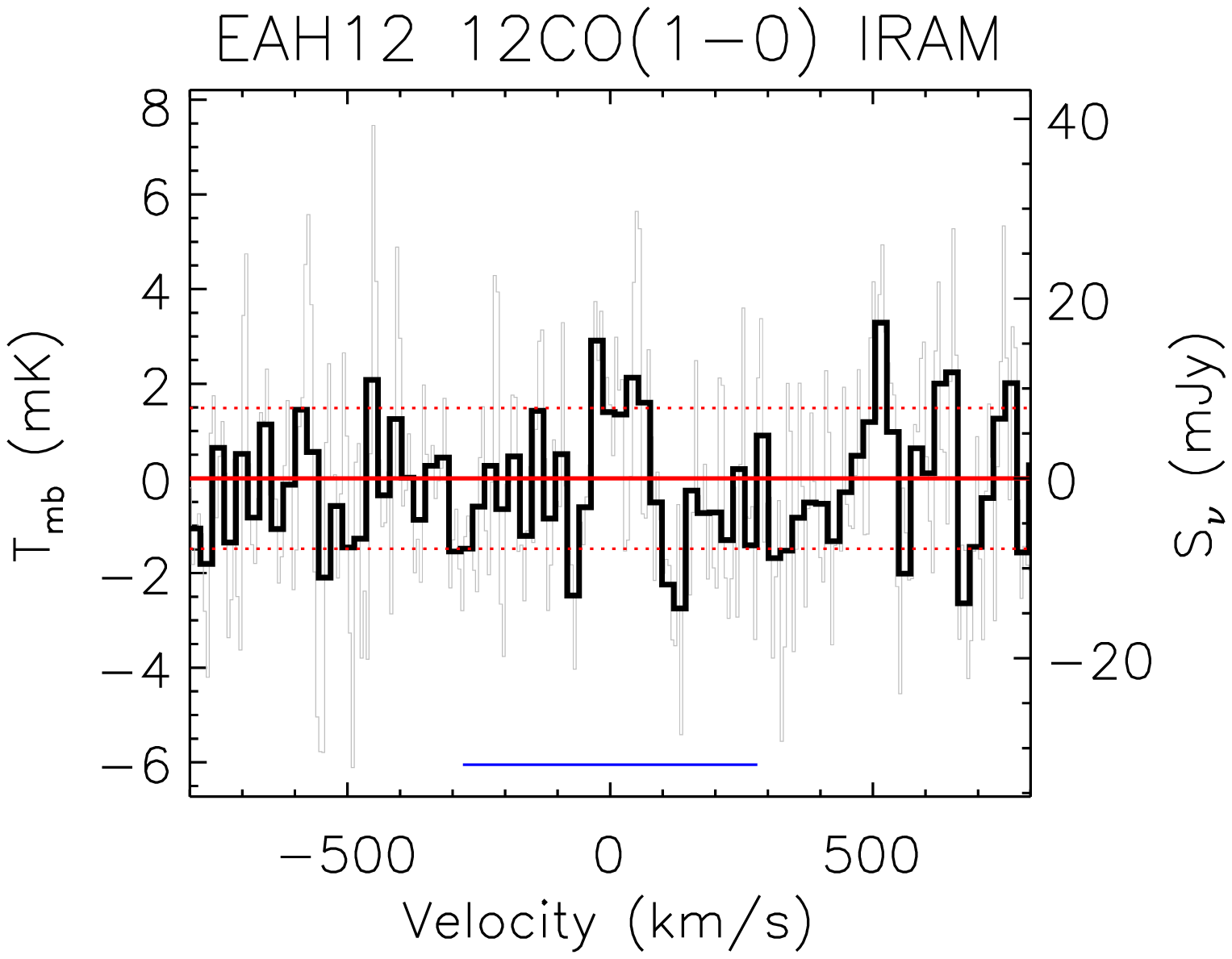}
\includegraphics[width=0.32\textwidth]{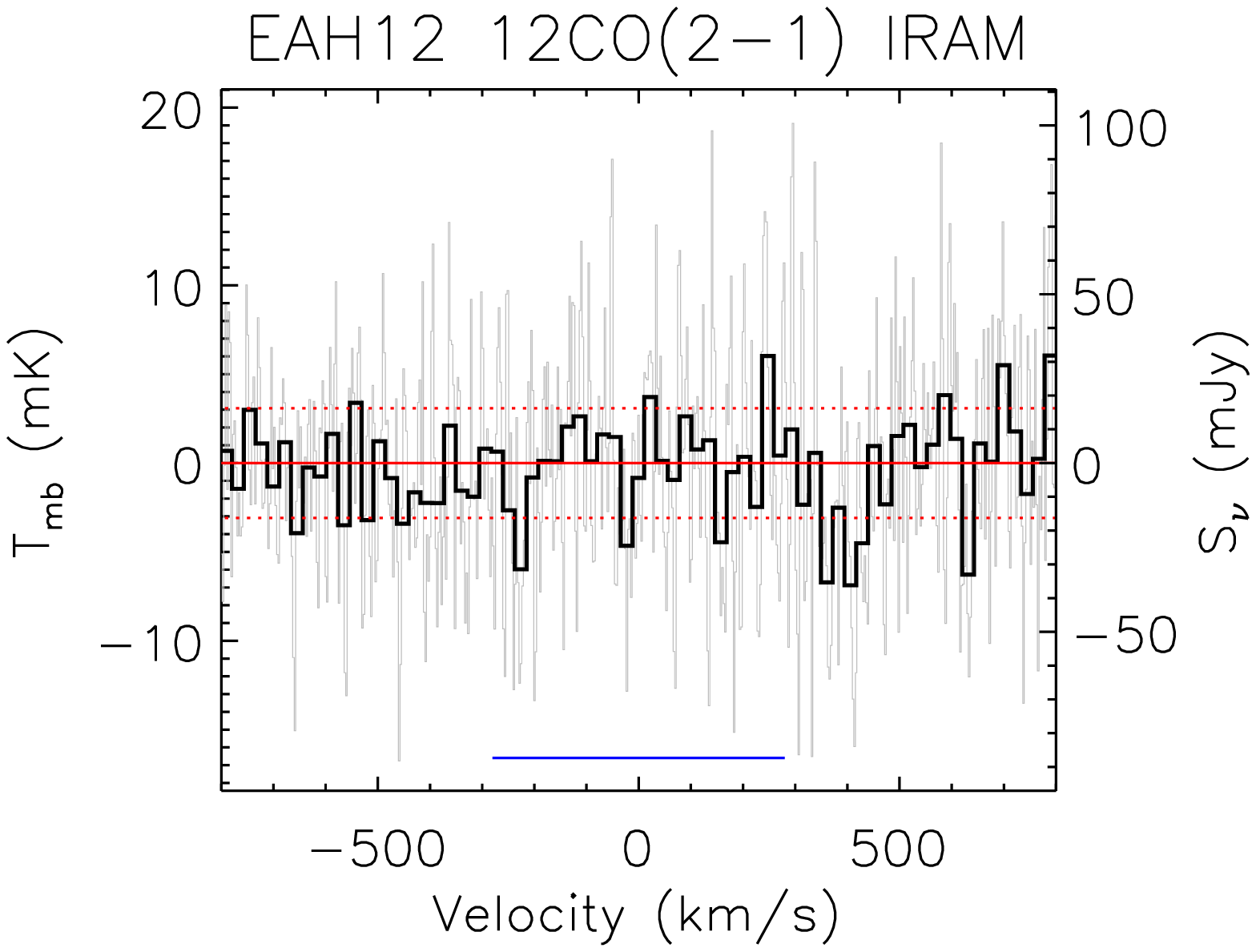}

\includegraphics[width=0.32\textwidth]{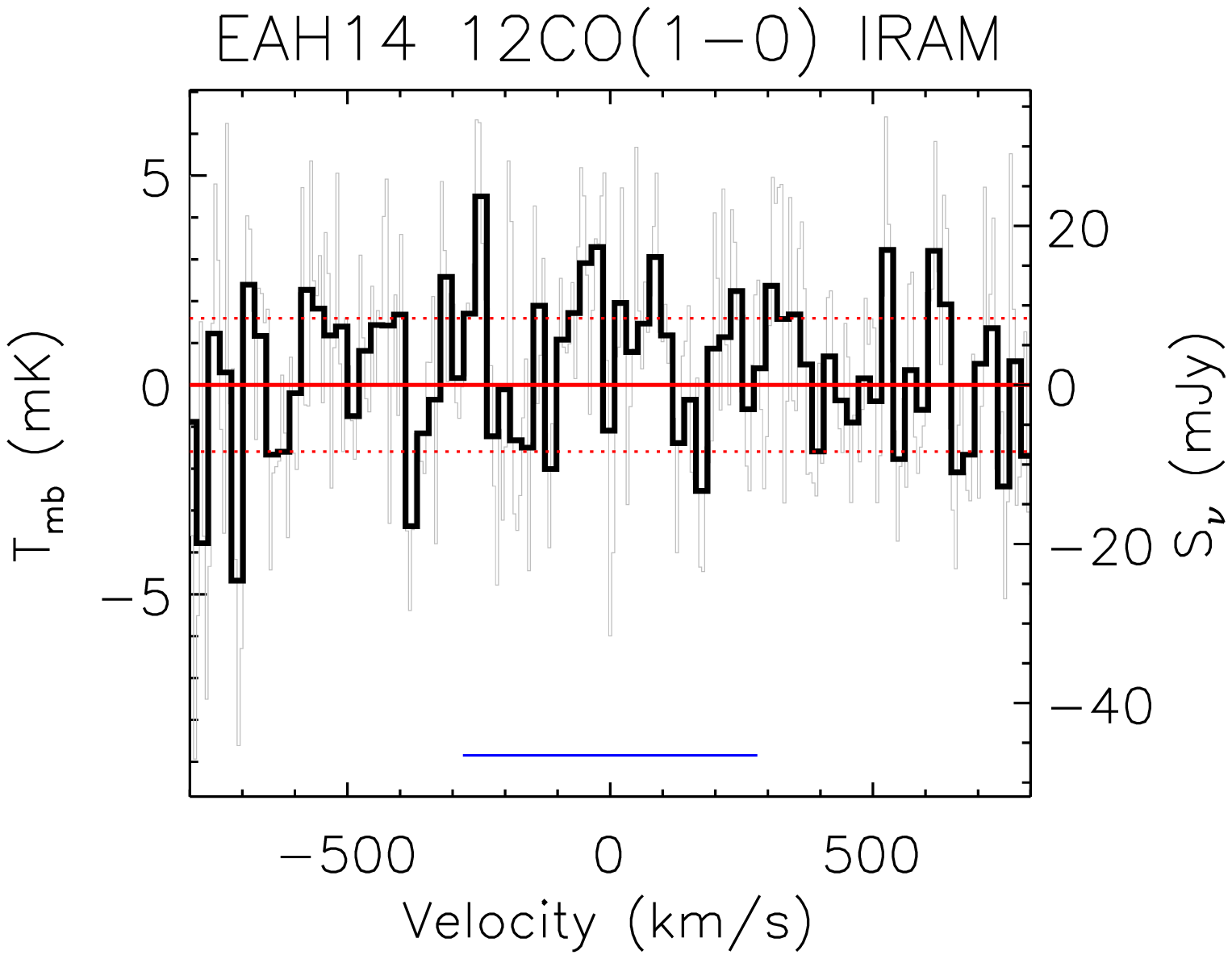}
\includegraphics[width=0.32\textwidth]{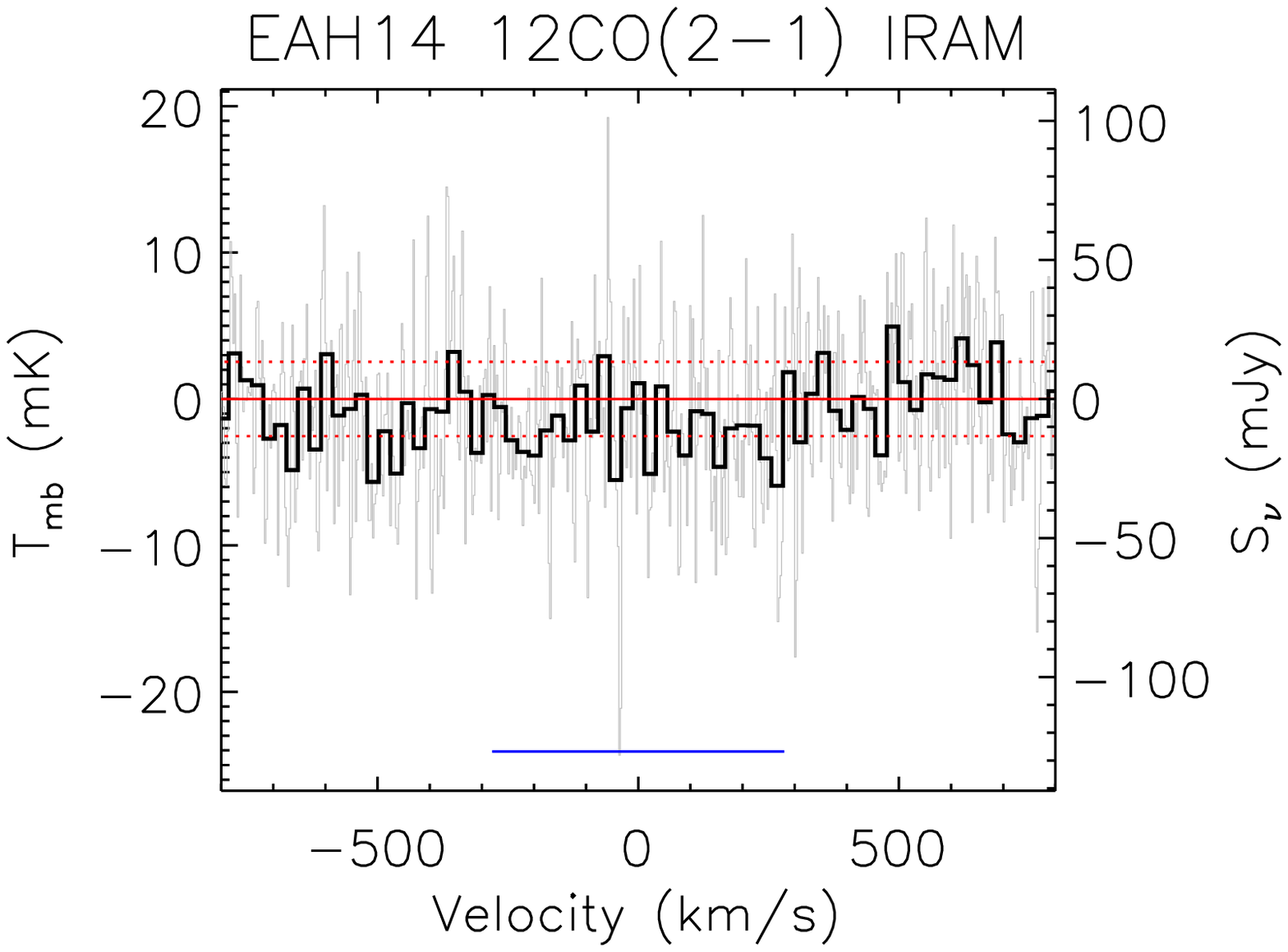}

\caption{CO(1-0) and CO(2-1) spectra from IRAM and SMT for galaxies with CO (1--0) not detected with the IRAM 30m. Spectra are shown in units of both main beam temperature T$_{mb}$ [mK] and S$_\nu$ [Jy]. Grey lines show the unbinned IRAM data for 5 km/s channels, and black lines show the data binned to 20 km/s. Dashed red lines represent the rms of the binned data. SMT data are shown in 13 km/s bins. Blue horizontal lines at bottom represent the integration intervals, as described in the text.}
\label{fig:CO2}

\end{figure*}

\begin{figure*}
\ContinuedFloat

\includegraphics[width=0.32\textwidth]{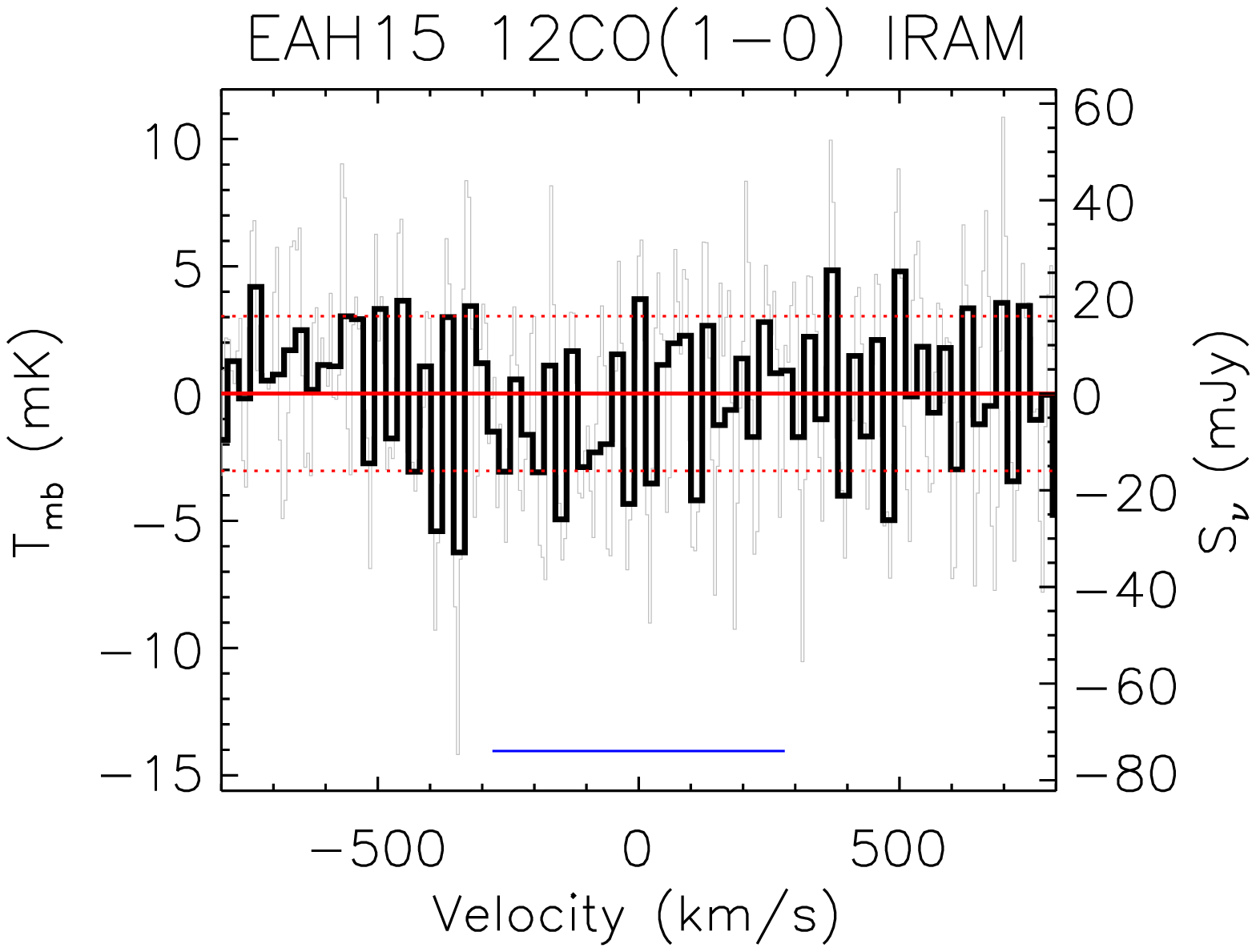}
\includegraphics[width=0.32\textwidth]{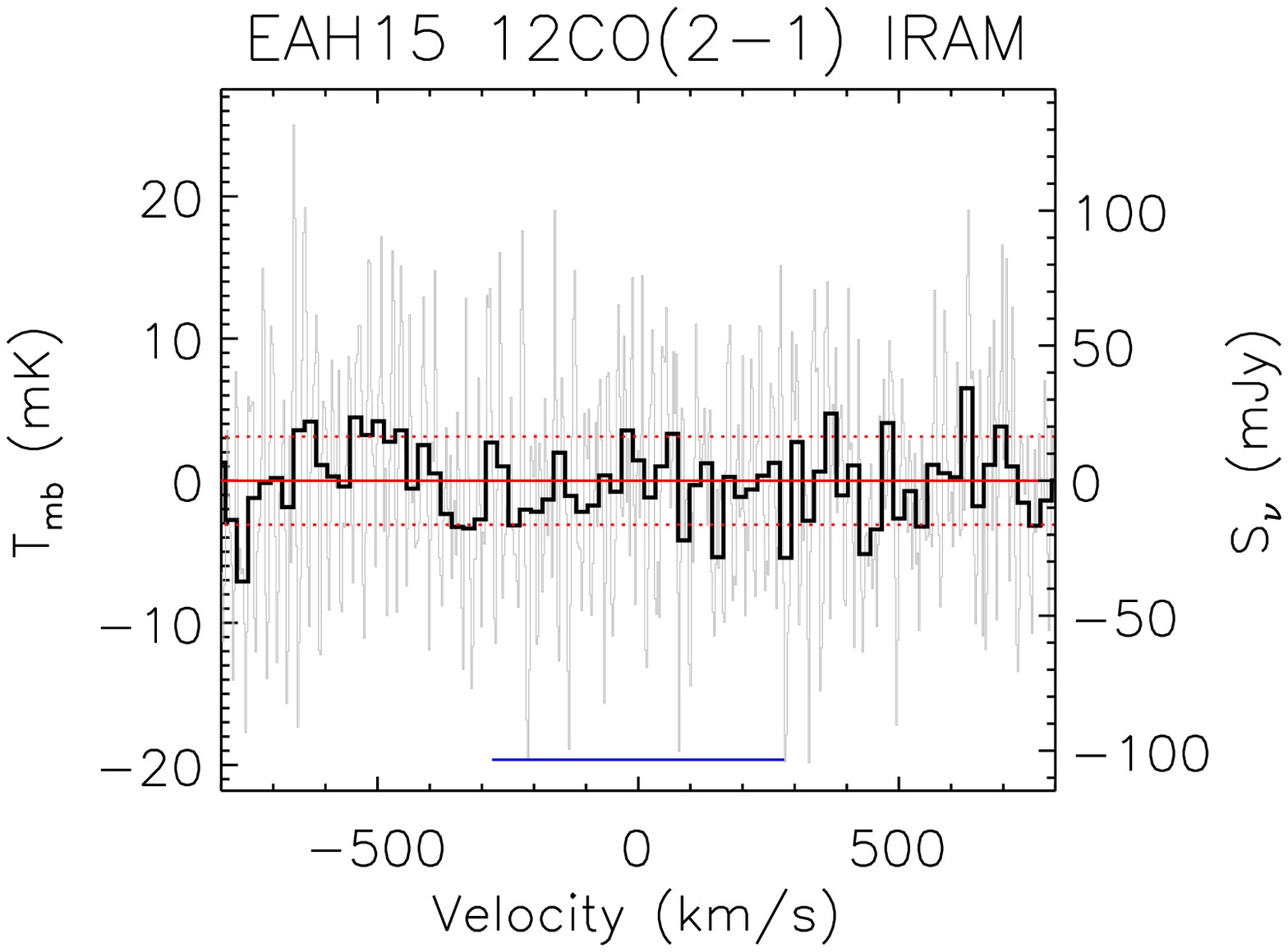}

\includegraphics[width=0.32\textwidth]{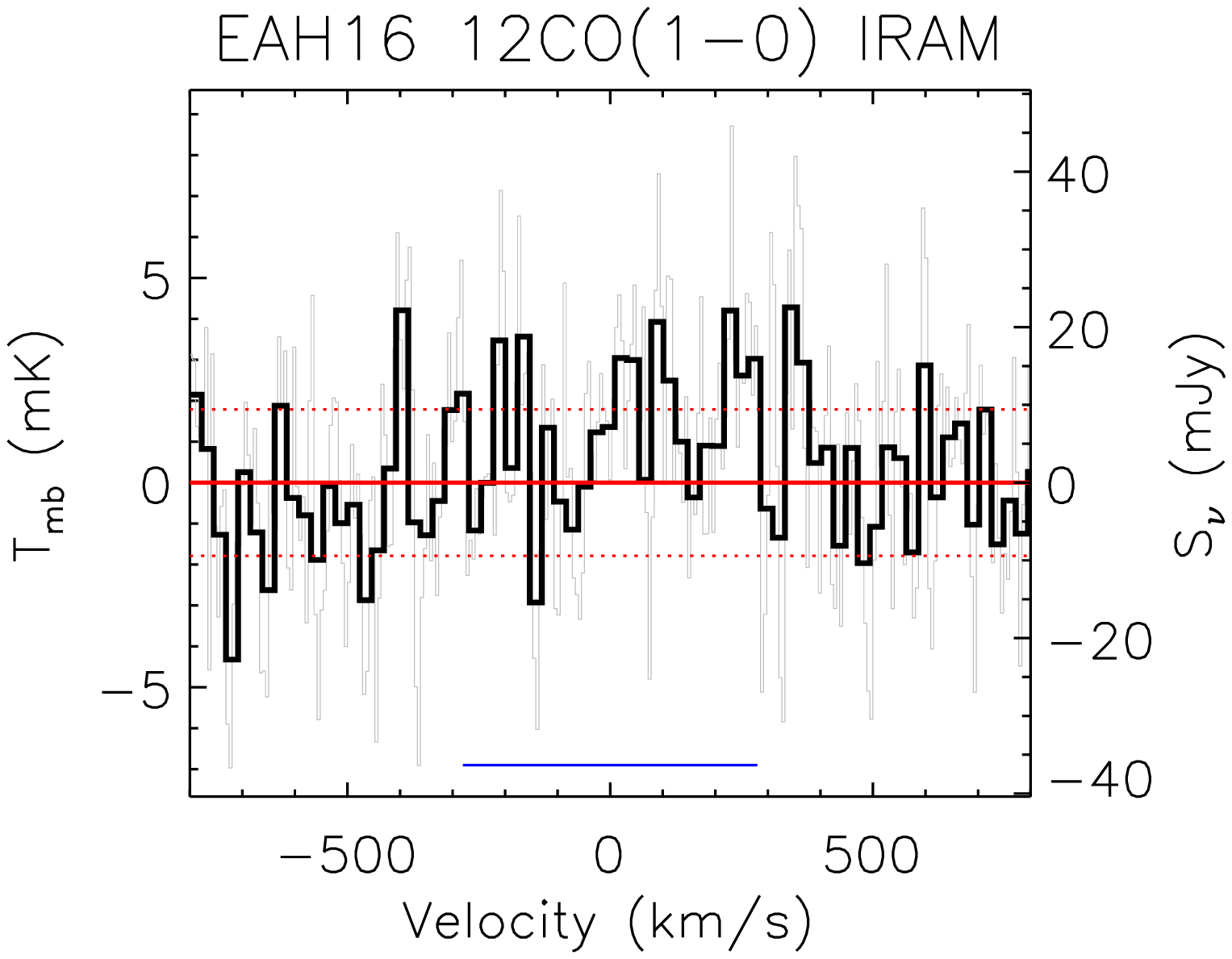}
\includegraphics[width=0.32\textwidth]{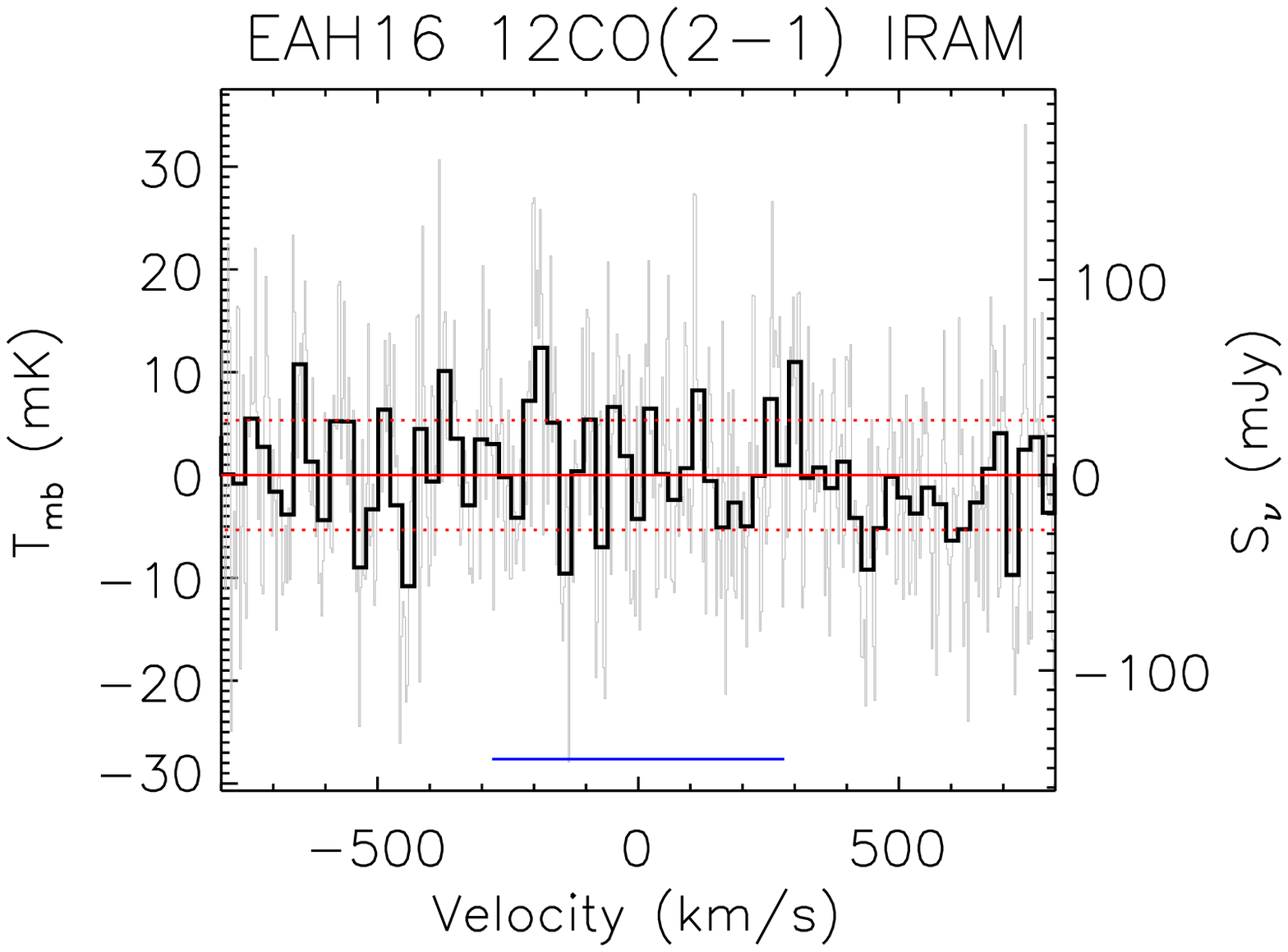}

\includegraphics[width=0.32\textwidth]{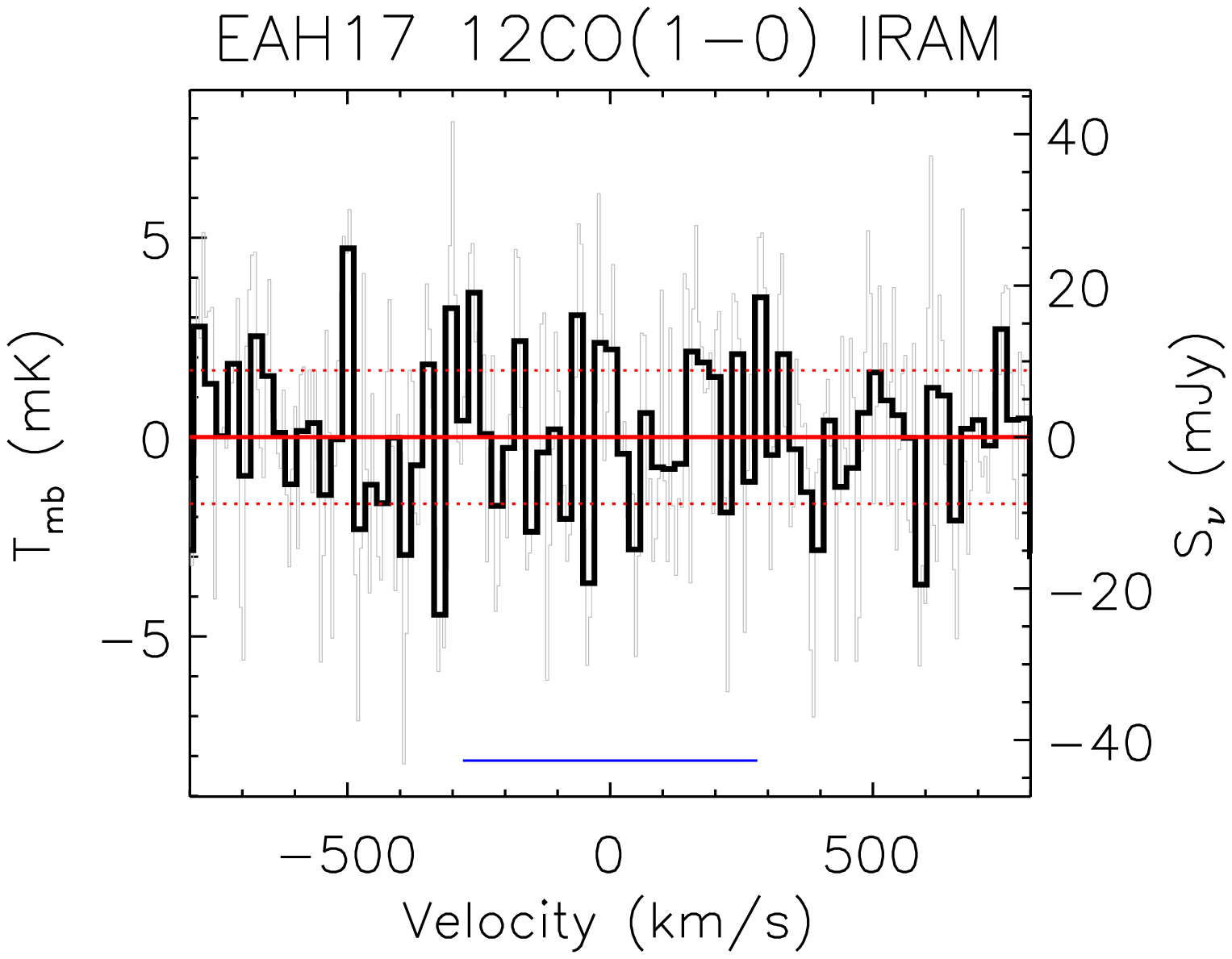}
\includegraphics[width=0.32\textwidth]{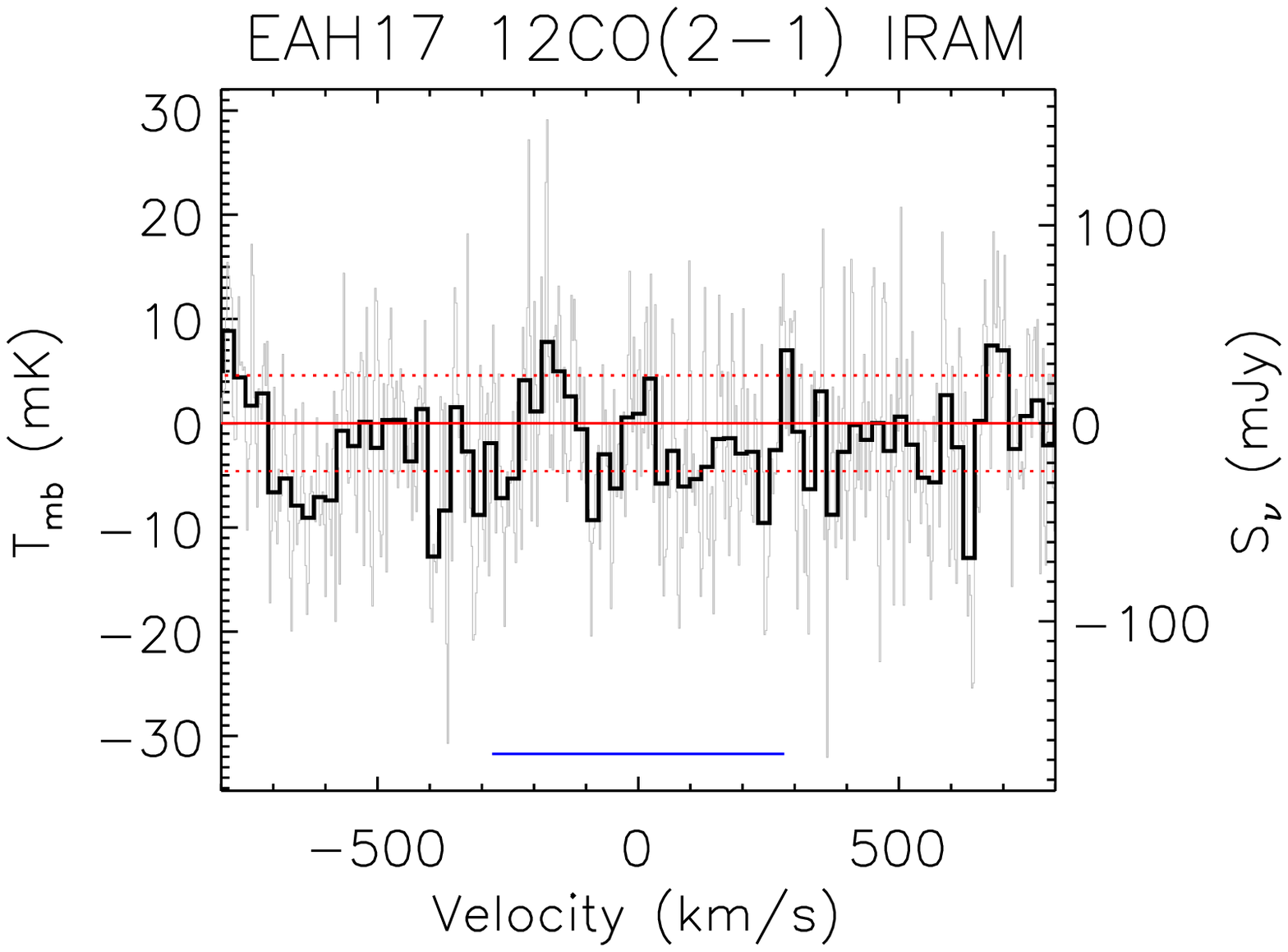}

\includegraphics[width=0.32\textwidth]{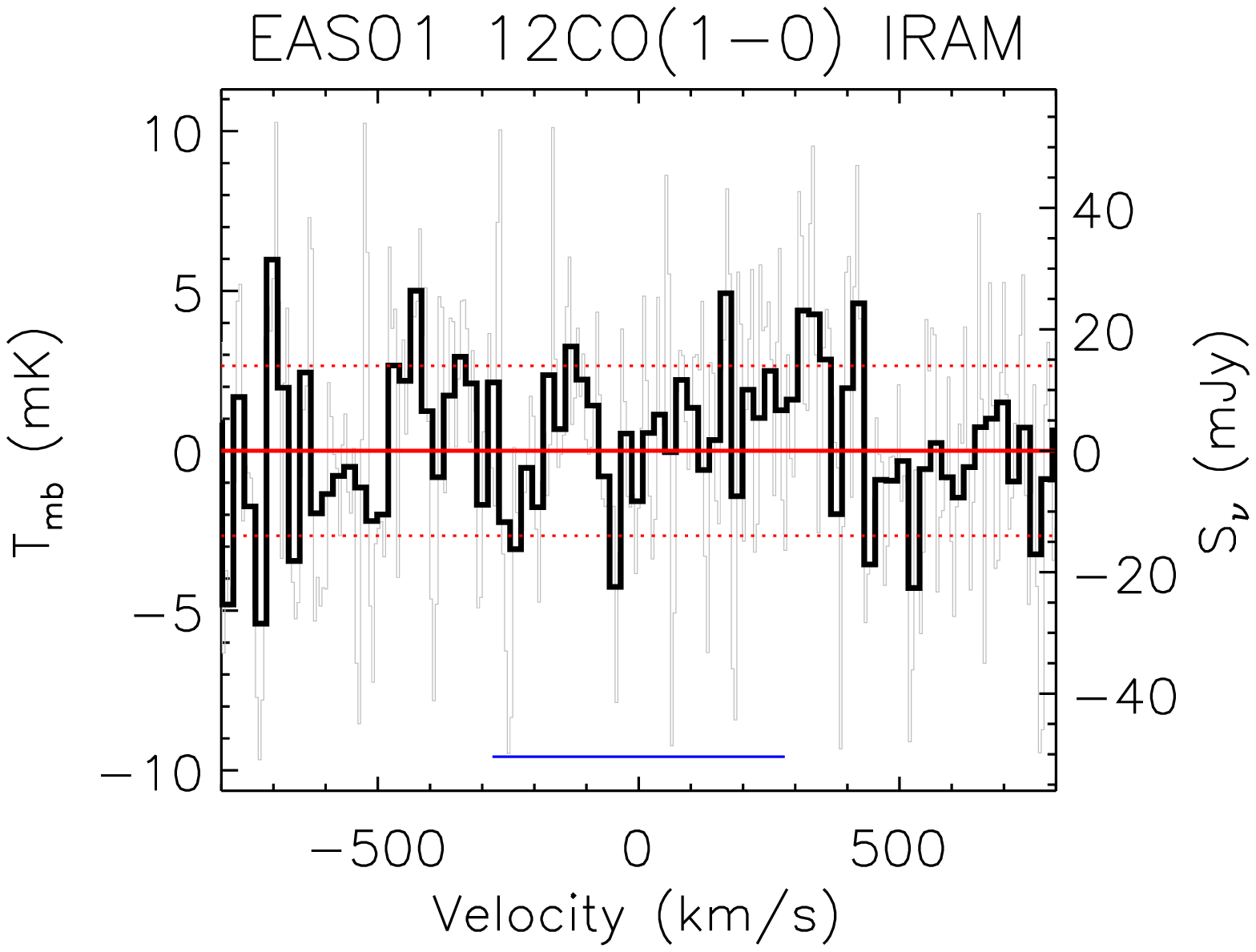}
\includegraphics[width=0.32\textwidth]{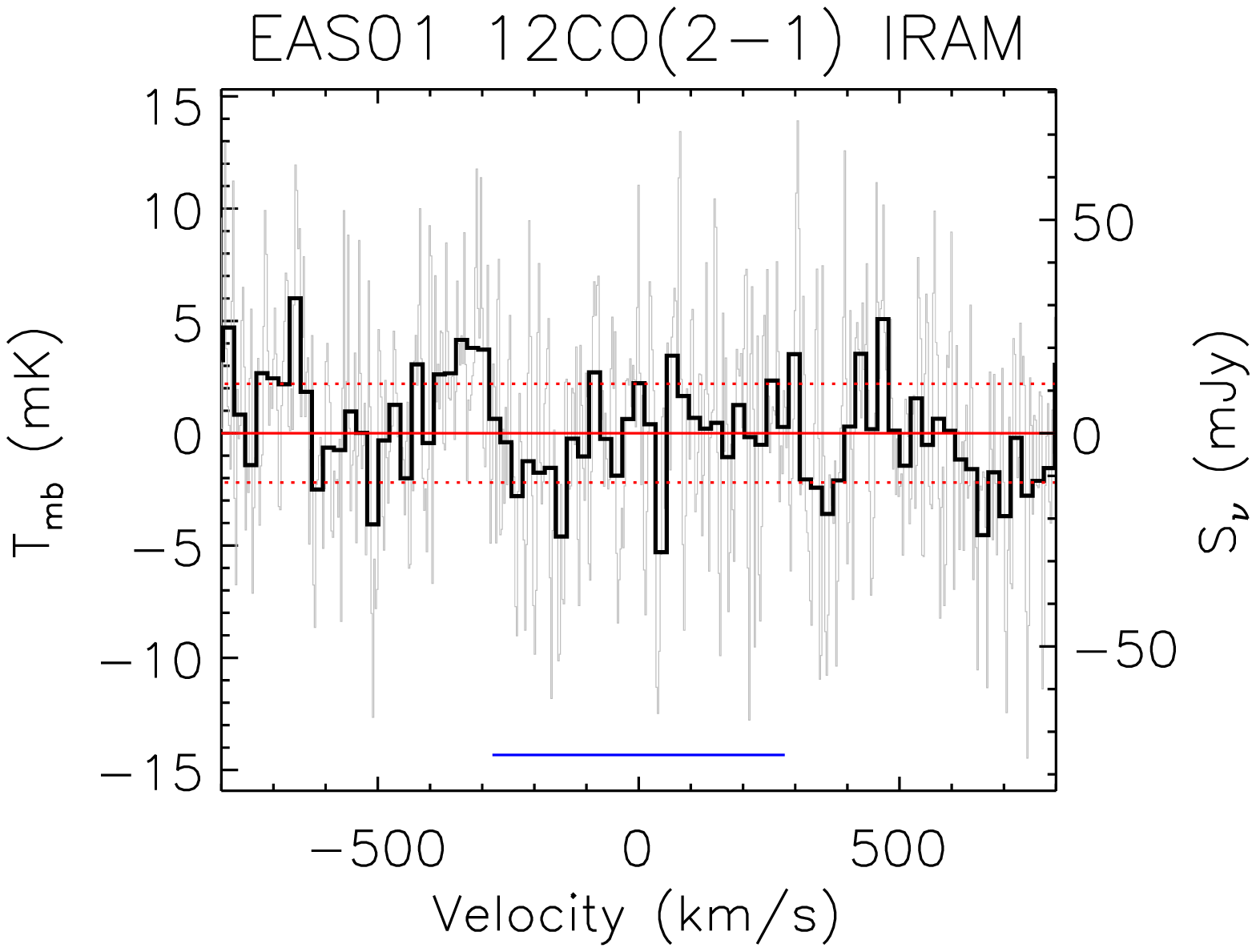}

\includegraphics[width=0.32\textwidth]{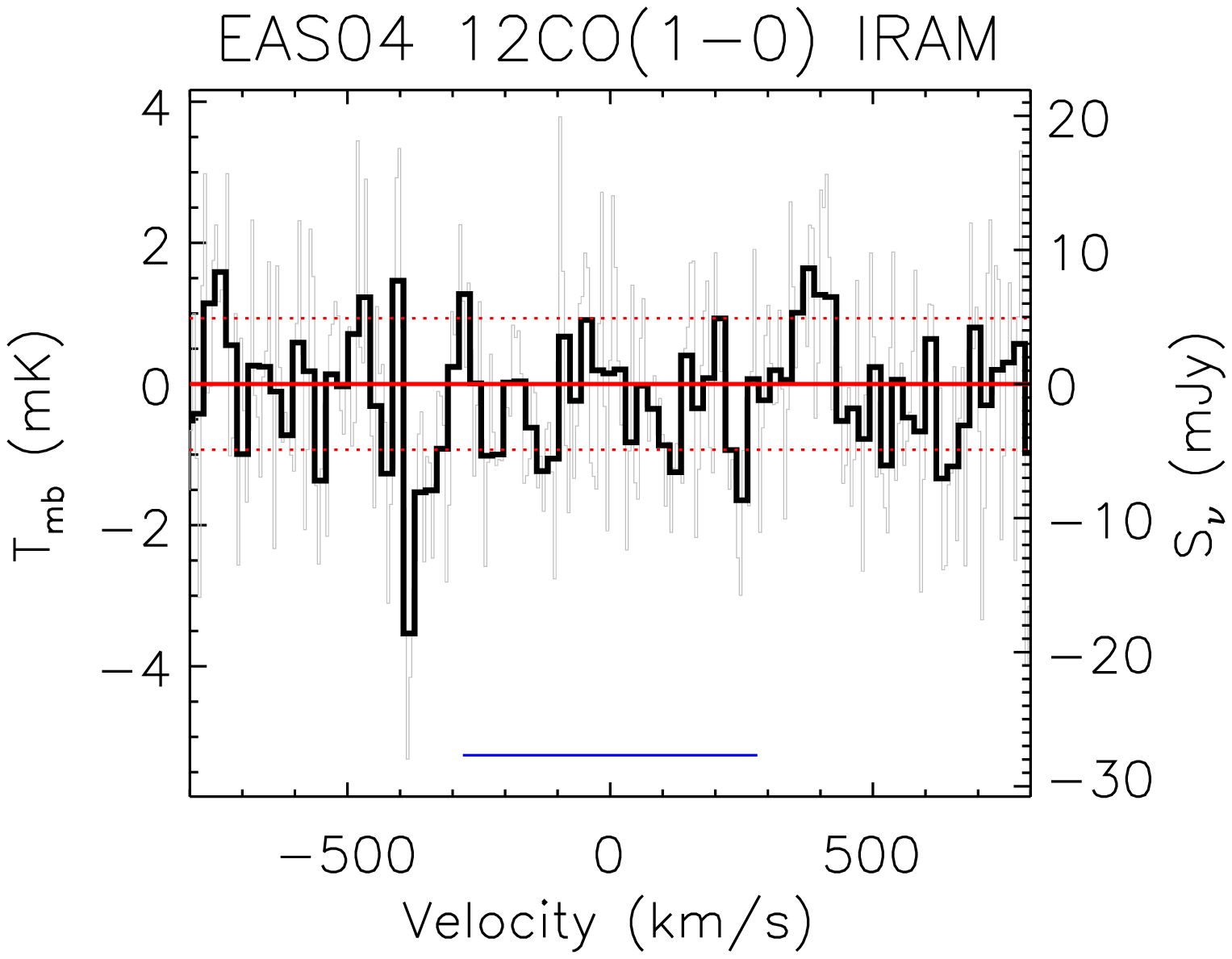}
\includegraphics[width=0.32\textwidth]{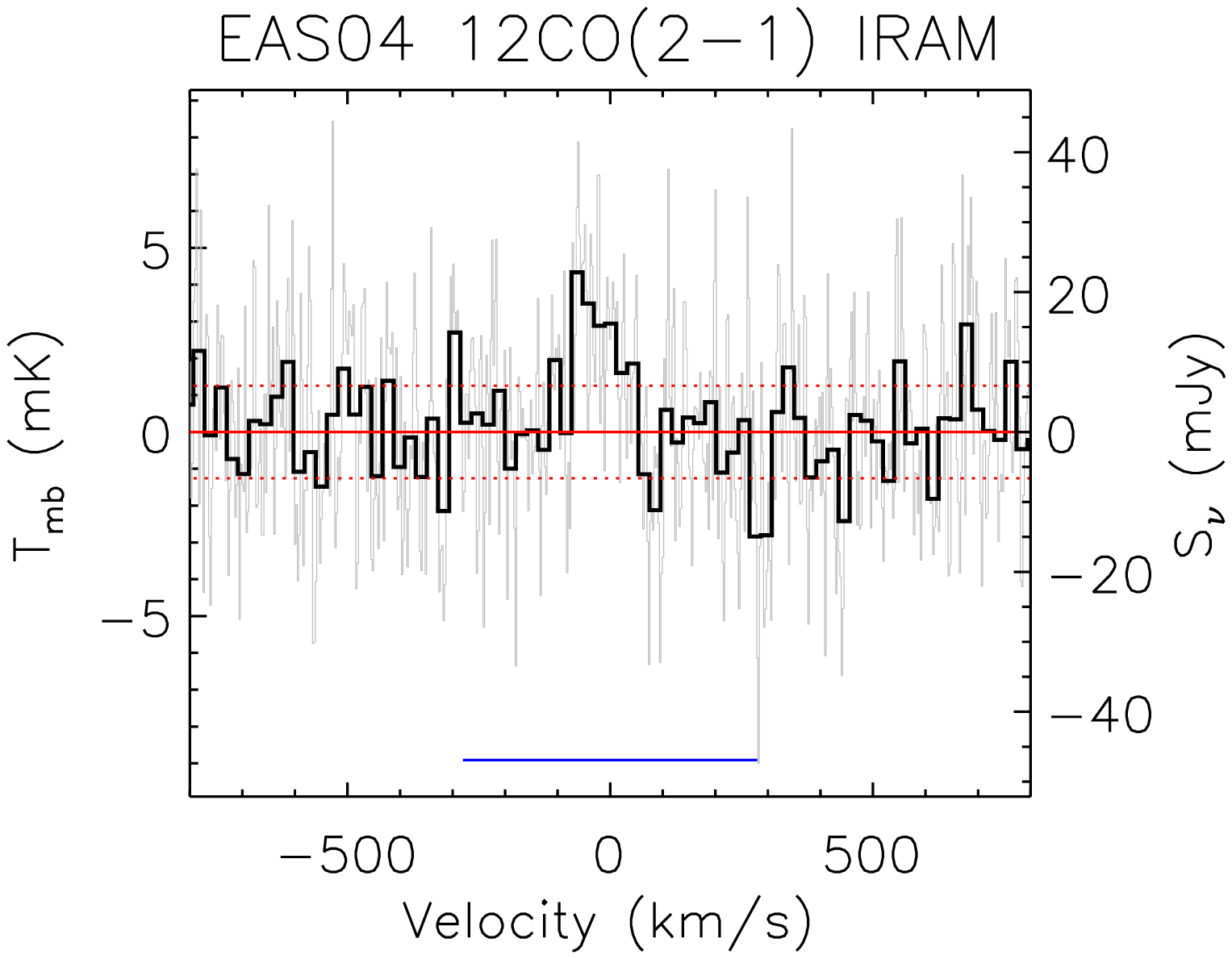}

\caption[]{continued}
\end{figure*}

\begin{figure*}
\ContinuedFloat

\includegraphics[width=0.32\textwidth]{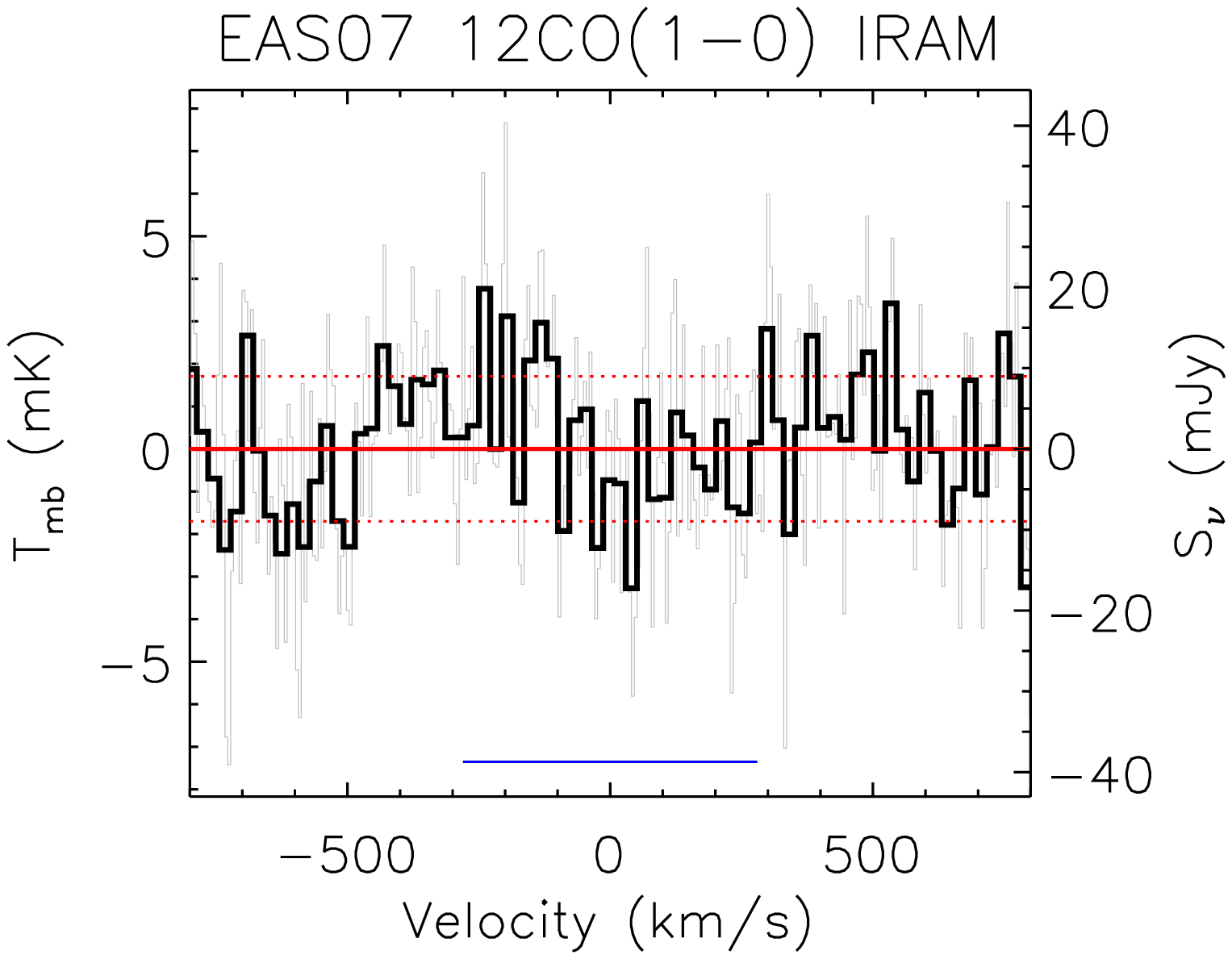}
\includegraphics[width=0.32\textwidth]{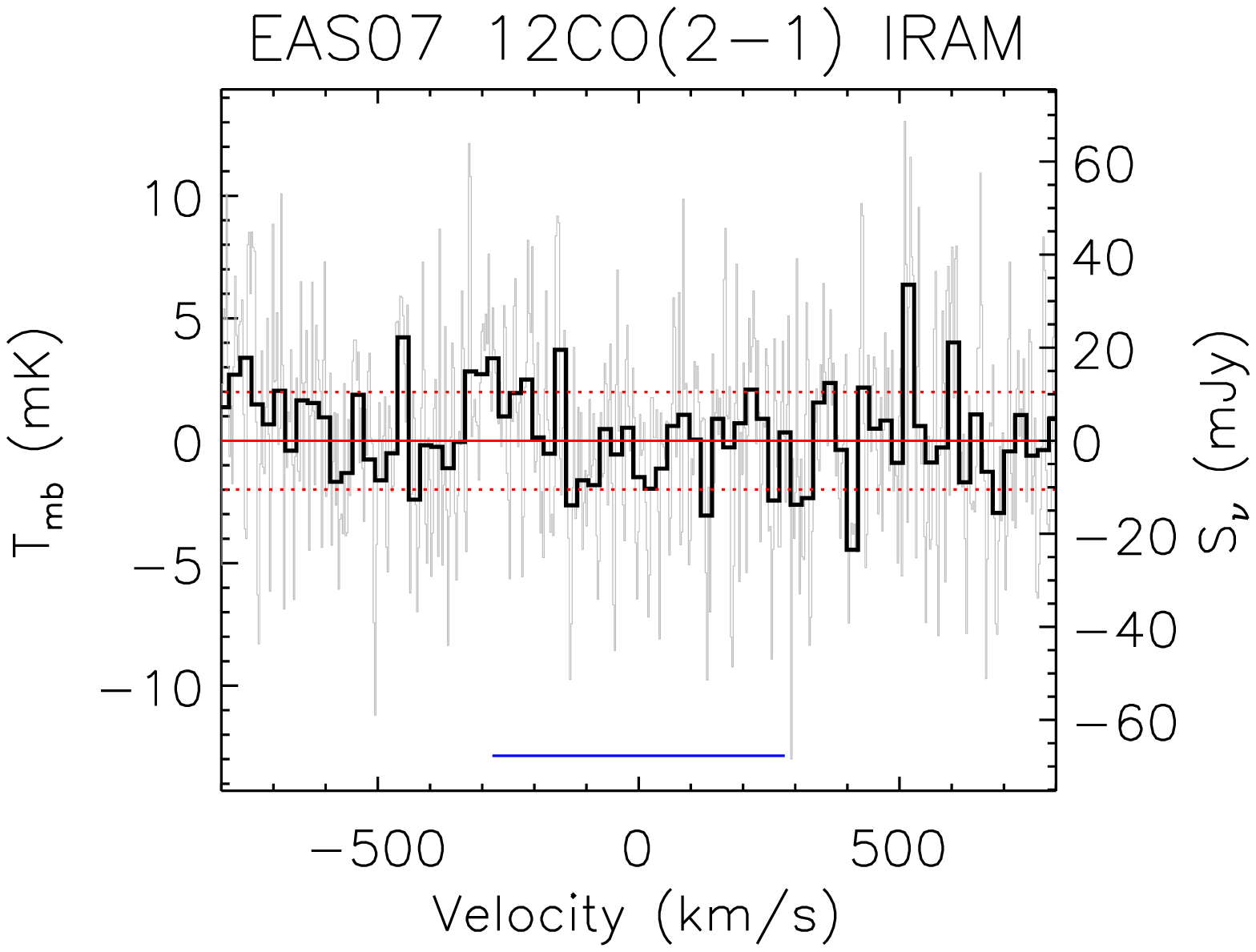}

\includegraphics[width=0.32\textwidth]{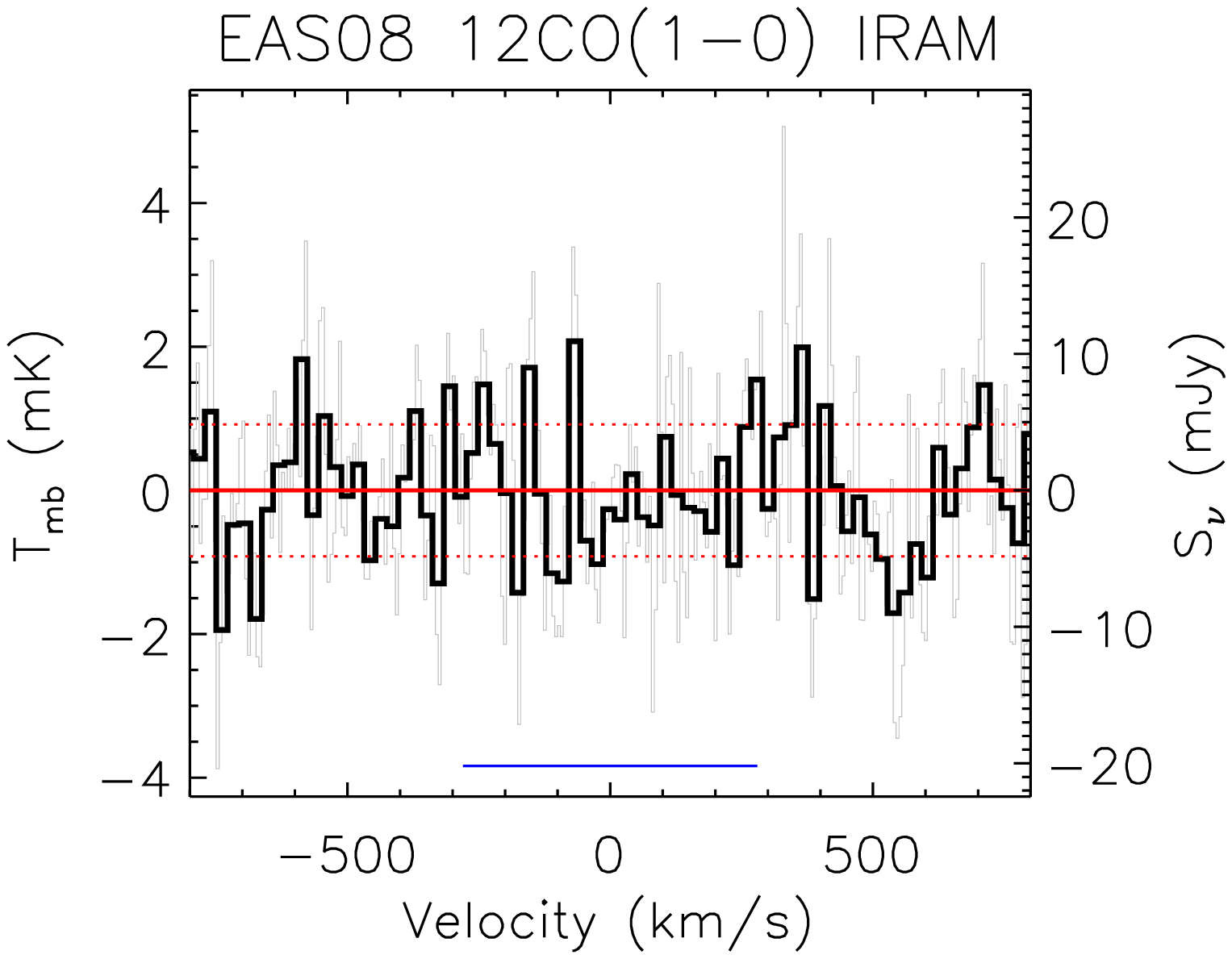}
\includegraphics[width=0.32\textwidth]{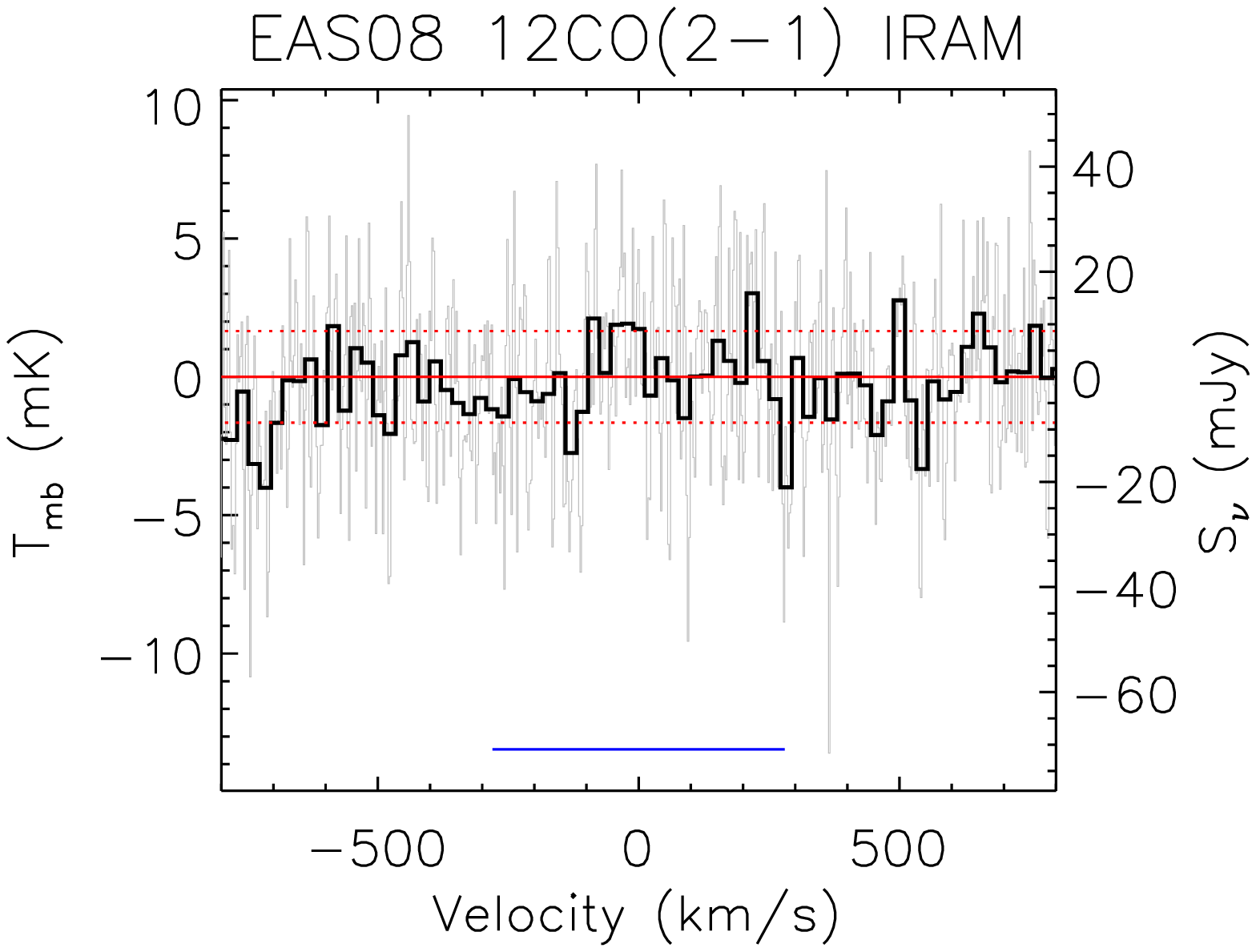}

\includegraphics[width=0.32\textwidth]{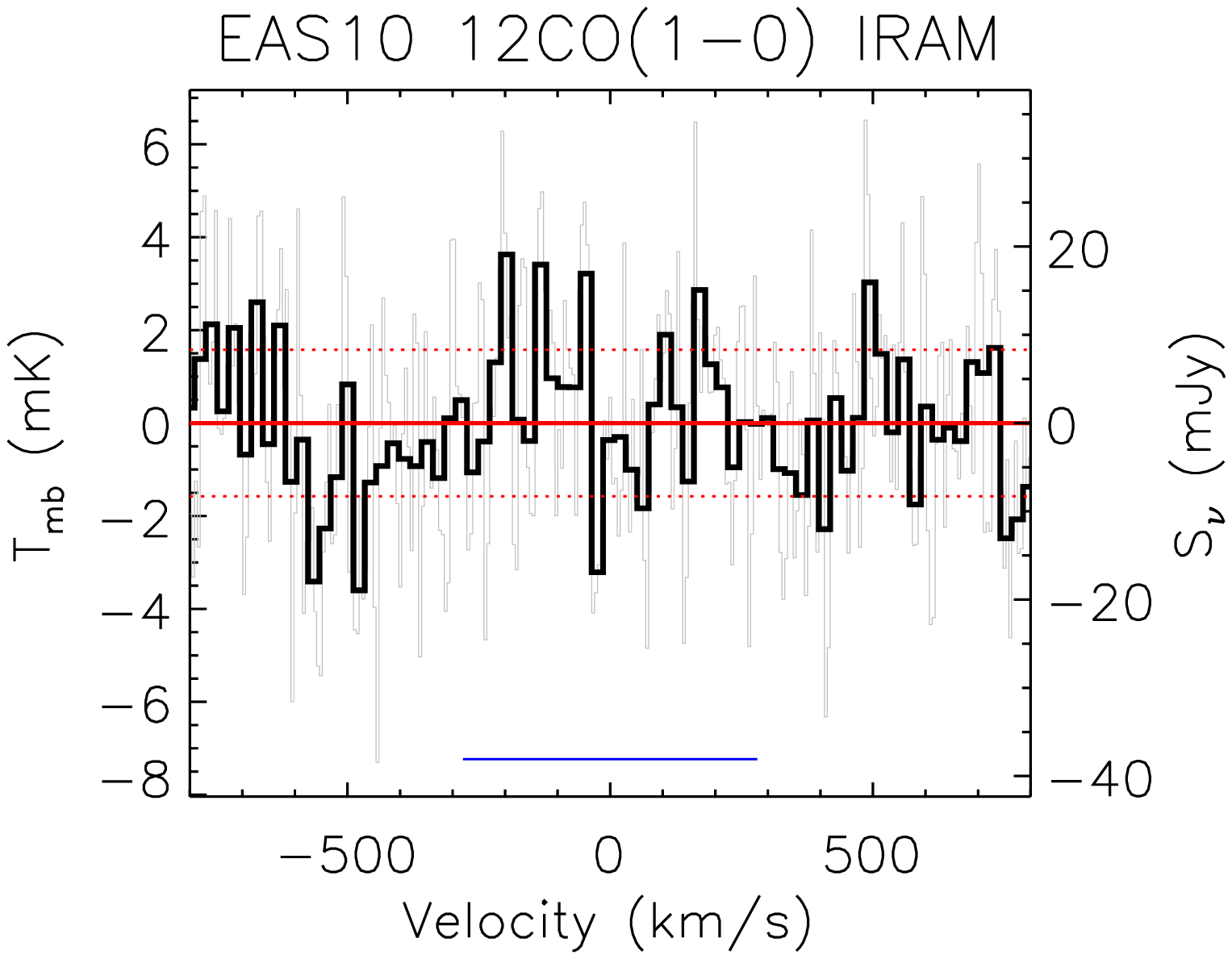}
\includegraphics[width=0.32\textwidth]{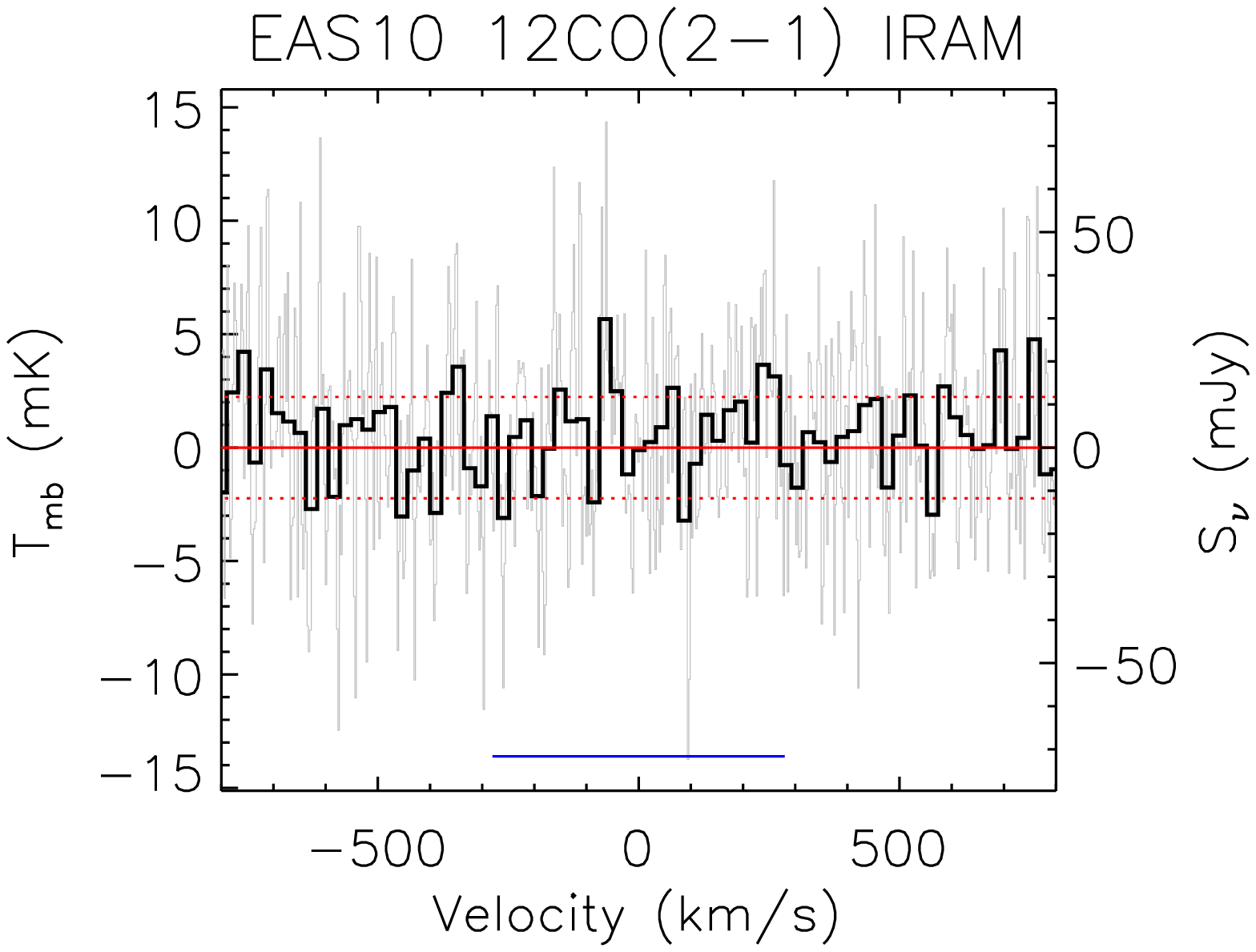}

\includegraphics[width=0.32\textwidth]{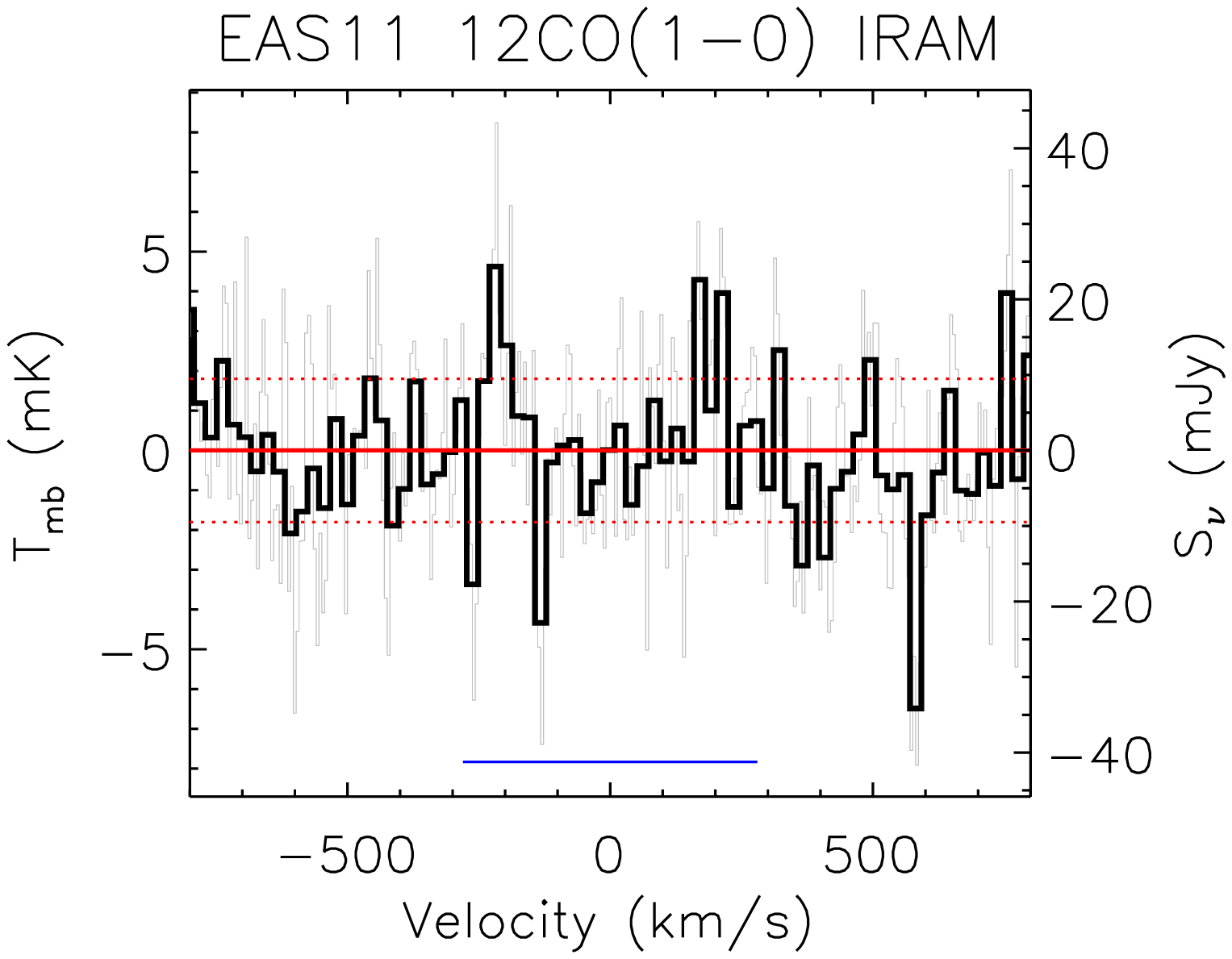}
\includegraphics[width=0.32\textwidth]{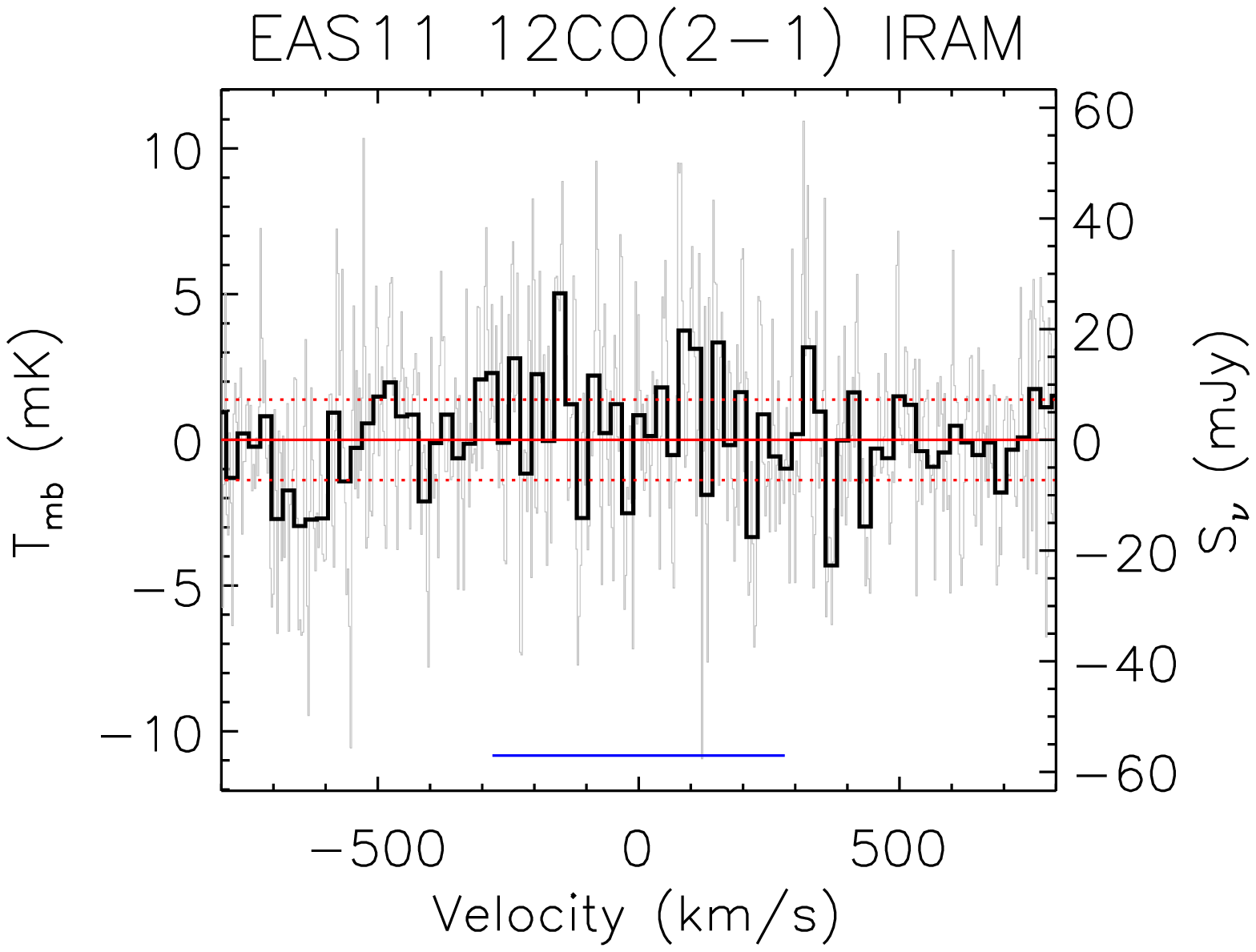}

\includegraphics[width=0.32\textwidth]{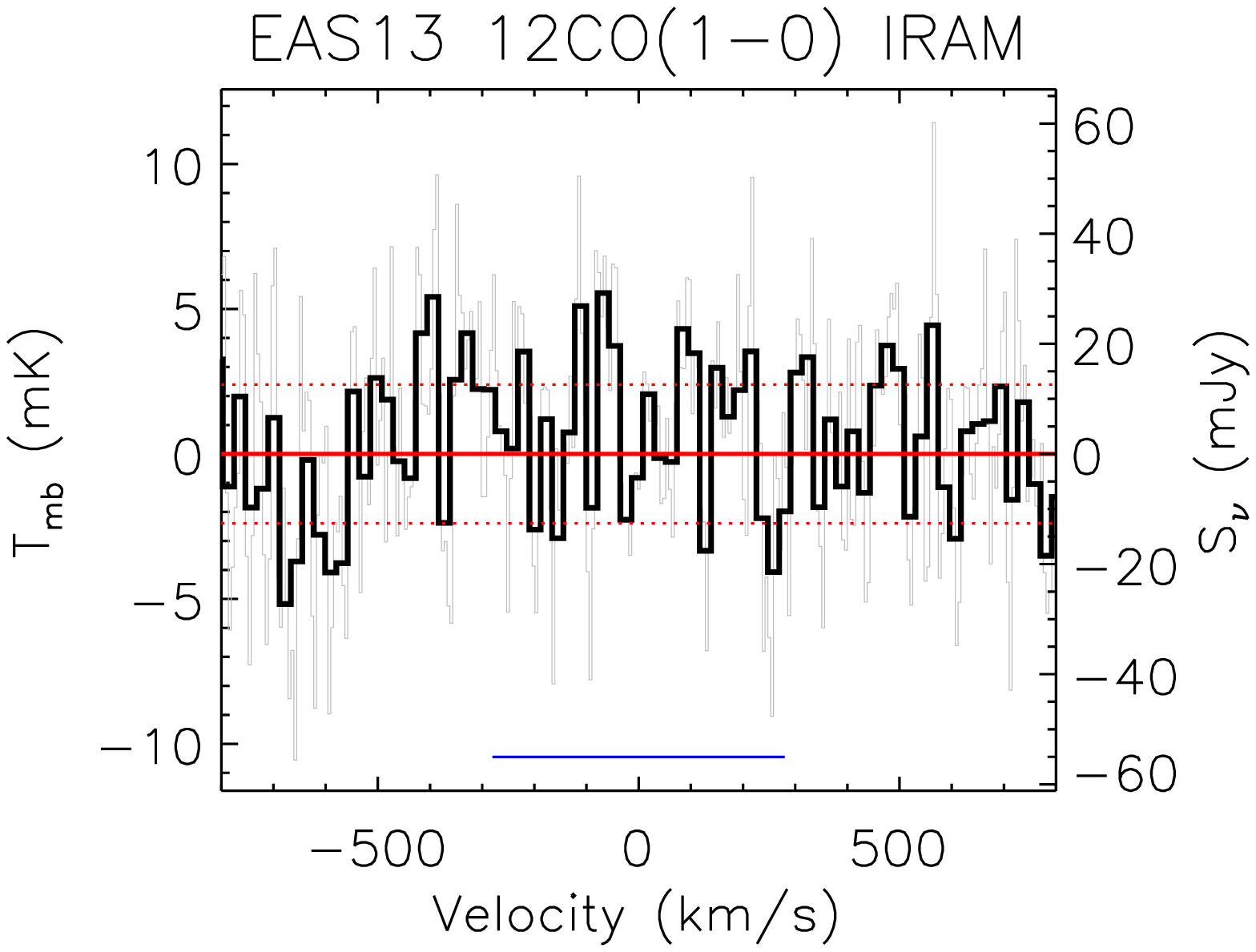}
\includegraphics[width=0.32\textwidth]{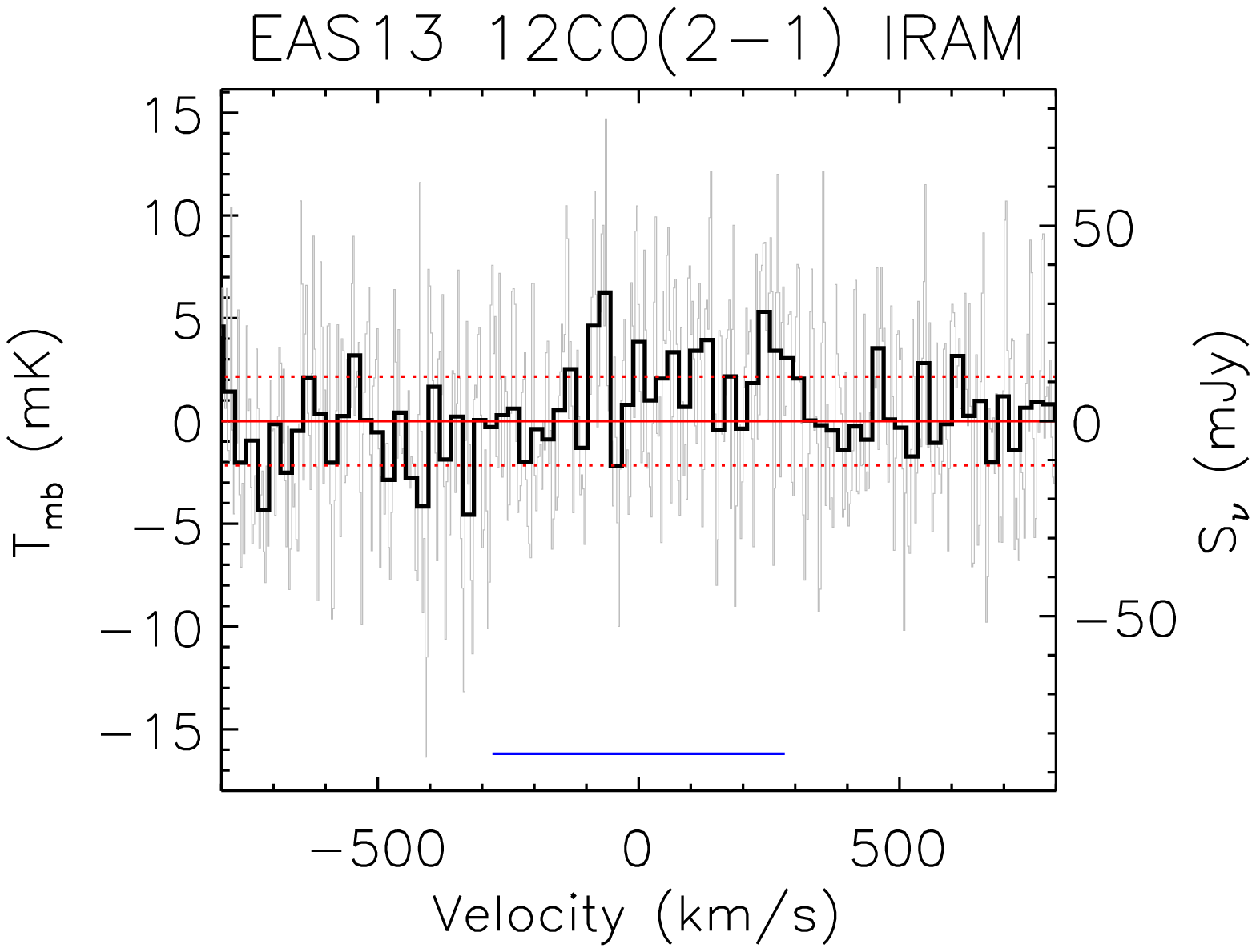}

\caption[]{continued}
\end{figure*}

\end{document}